\documentclass[amsmath,amssymb,prb]{revtex4}

\usepackage{graphicx}

\begin{document}

\title{Dynamical non-Condon effects in resonant tunneling}

\author{Kazuhiro Tsusaka}

\affiliation{Faculty of Science and Technology, Meijo University, Nagoya 468-8502, Japan}

\begin{abstract}
It is well known that the electron transfer (ET) reaction in
donor-bridge-acceptor systems is much influenced by the bridge
conformational changes.
However, the importance of the dynamical contribution to the ET rate is
still under debate.
In this study, we investigate the significance of the dynamical
non-Condon effect on the ET rate in the coherent resonant tunneling
regime.
To deal with the resonance, we generalize the time-dependent Fermi's
golden rule expression for the ET rate in the $T$-matrix framework.
The dynamical effect is thereby expressed in terms of the time
correlation function of the non-Hermitian $T$-matrix.
We discuss the property of the quantum time correlation function of the
non-Hermitian $T$-matrix and construct the semiclassical approximation.
Then we perform a computational study using classical equilibrium
molecular dynamics simulations and quantum chemical calculations.
It is shown that the ET rate can be significantly enhanced by the
dynamical non-Condon effects.
\end{abstract}

\maketitle

\section{\label{intro}Introduction}
Electron transfer (ET) is one of the most important elementary chemical
processes.\cite{Nitzan}
In many cases the simple picture of a direct transfer of an electron
from the donor (D) to the acceptor (A) does not apply.
Instead, bridge (B) molecules are involved.
The standard expression for the ET rate is based on the Condon
approximation, in which the dependence of the electronic coupling
between D and A on the nuclear configuration is disregarded.
However, molecules undergo structural fluctuations over a wide range of
timescales at thermal equilibrium.
If the B conformational changes influence the electronic coupling, the
system will experience a breakdown of the Condon approximation.
In the limit of slow electronic coupling fluctuations, the ET rate can
be expressed as the ensemble average of the instantaneous ET rate over
the static B configuration.
This expression captures the structural averaging effect of the B
conformation on the electronic coupling, while neglecting the coupling
between nuclear and electron dynamics.
The dynamical non-Condon effects may become significant as the
electronic coupling fluctuations occur on faster timescales.

There have been theoretical developments and computational studies
elucidating the dynamical non-Condon effect on the ET rate in the deep
(off-resonant) tunneling regime where the energy of the initial
electronic state is well separated from the eigenenergies of the
B subspace.\cite{Tang_JCP93,Goychuk_JCP95,Daizadeh_PNAS97,Medvedev_JCP97,Bixon_RJE03,Troisi_JCP03,Troisi_JACS04,Skourtis_PNAS05,Nishioka_JPCB05}
Starting with the Fermi's golden rule expression for the ET rate, the
dynamical effect is expressed in terms of the time autocorrelation
function of the electronic coupling.
In a quantum mechanical treatment, the electronic coupling is a
Hermitian operator, and its time autocorrelation function satisfies
detailed balance.
The calculation results in Ref.~\onlinecite{Medvedev_JCP97} show that
for the models of independently coupled bridge modes with coupling of
the tunneling electron to the bridge modes, the dynamical non-Condon
effects can result in substantial enhancement of the ET rate in the
activated inverted region.
The detailed balance property is reflected in the asymmetry of the ET
rate with respect to the activation energy.
On the other hand, the analogous classical time autocorrelation function
is a real symmetric function of time and thus does not satisfy detailed
balance.
Therefore, the classical approximation cannot reproduce the asymmetry of
the ET rate due to the dynamical non-Condon effects, but rather produces
the symmetric correction to the ET rate with respect to the activation
energy, assuming the classical Marcus expression for the Franck-Condon
factor.\cite{Troisi_JCP03}
The enhancement in the activated inverted region is much smaller than
that in the quantum mechanical treatment.
As pointed out in Ref.~\onlinecite{Troisi_JCP03}, it is possible to
restore detailed balance in the way that the Fourier transform of the
classical time autocorrelation function is multiplied by a quantum
correction factor.
Such semiclassical approximation schemes have been widely discussed over
the years.\cite{Schofield_PRL60,Egelstaff_AP62,Berne_JCP67,Berne_ACP70,An_JCP76,Oxtoby_ACP81,Berens_JCP81,Frommhold,Bader_JCP94,Skinner_JCP97,Egorov_CPL98,Egorov_JPCA99,Skinner_JPCB01,Kim_JPCB02}

The computational studies have been carried out using classical
equilibrium molecular dynamics (MD) simulations and quantum
chemical (QC) calculations to evaluate the classical and semiclassical
time autocorrelation functions of the electronic coupling.\cite{Daizadeh_PNAS97,Troisi_JACS04,Skourtis_PNAS05,Nishioka_JPCB05}
The results show that the characteristic timescales for the electronic
coupling fluctuations are too slow for the dynamical non-Condon effects
to be significant.
In the time domain, a comparison between the decay times of the
electronic coupling correlation function, and the time-dependent
Franck-Condon factor reveals the extent to which dynamical non-Condon
effects influence the ET rate.
The decay time of the time-dependent Franck-Condon factor is typically
$\sim 3-4$ fs.\cite{Lockwood_CPL01}
On the other hand, the decay time of the electronic coupling correlation
function is more than a few tens of
fs.\cite{Troisi_JACS04,Skourtis_PNAS05,Nishioka_JPCB05}
Therefore, the electronic coupling fluctuations are negligible on the
timescale of the decay of the time-dependent Franck-Condon factor.
A significantly short decay time of the electronic coupling correlation
function was observed only when the electron tunneling energy was
artificially brought into resonance with bridge-centered
eigenstates.\cite{Skourtis_PNAS05}
However, there has not ever been confirmation of this statement for a
natural system in the resonant tunneling regime.

The purpose of this work is to investigate the timescale of the
electronic coupling fluctuations and the possibility of the dynamical
non-Condon effects in the resonant tunneling regime through a
computational study.
We assume that transport is fully coherent, that is, the electron
transfers from D to A in a single quantum mechanical process.
In the resonant tunneling regime, another contribution comes from
sequential tunneling where an electron first tunnels into B and then,
after losing its phase memory, tunnels out of B.
The dynamical non-Condon effects can substantially increase the
contribution from the coherent tunneling, particularly if the B and A
electronic states are off resonance.
We will consider such situations.

The choice of materials are largely motivated by the previous work for
solvent-mediated ET, where B is the solvent\cite{Troisi_JACS04}.
Instead of the C-clamp molecule, we use unlinked D-A systems.
In fact, the curved saturated bridge in the C-clamp molecule plays an
important role in keeping the distance between D and A.
Instead, in this work the D and A molecules are held fixed during the
simulation.
As an example of the resonant tunneling we consider the ET from the
LUMO of naphthalene to the LUMO of tetracyanoethylene (TCNE) in
benzonitrile (PhCN).
This system lies in the highly activated inverted region (experimental
data available for the case of acetonitrile solvent\cite{Rehm_IJC70}).
Therefore, if the electronic coupling fluctuations occur on timescale as
short as the decay time of the time-dependent Franck-Condon factor, the
dynamical non-Condon effects may cause substantial enhancement of the
ET rate.

The Fermi's golden rule expression refers to the second-order
perturbation theory result for the ET rate.
In the resonant tunneling regime, higher-order tunneling processes can
come into play.
In Section \ref{tmatrix} we generalize the Fermi's golden rule
expression to include all high-order tunneling processes in the
$T$-matrix framework.
The time autocorrelation function of the electronic coupling is thereby
replaced by a time correlation function of the $T$-matrix.
In practice we will limit our calculations to a small part of a large
system, while the effect of the surroundings on the relevant system will
be included by broadening the energy levels of the system.
In this case the $T$-matrix should be treated as non-Hermitian.
In Section \ref{correlation} we discuss the properties of the quantum time
correlation functions of the $T$-matrix and then describe their
semiclassical approximations, which will be used to evaluate the time
correlation functions.
In Section \ref{methods} we give the details of the computational
methods.
In Section \ref{broadening} we first show that disregarding the level
broadening leads to unacceptable results even in the deep tunneling
regime.
Then we estimate the level broadening for the charge separation (CS)
from the LUMO of anthracene to the LUMO of TCNE and the charge
recombination (CR) from the LUMO of TCNE to the HOMO of naphthalene in
PhCN, which are in the deep tunneling regime, by comparing the full
$T$-matrix results with the second-order perturbation results.
Considering the similarities in chemical structure, we expect a similar
level broadening for the above-mentioned resonant tunneling process.
Taking this into consideration we evaluate the time correlation
functions for the resonant tunneling process in Section \ref{resonant}.
Then we discuss the significance of the dynamical non-Condon effect on
the ET rate.
In Section \ref{con} we conclude the paper.
In Appendix \ref{A} we sketch the derivation of a level broadening term
in the T-matrix formalism.

\section{\label{tmatrix}$T$-matrix formulation}
In the Born-Oppenheimer approximation, the system under consideration
can be described by the Hamiltonian
\begin{equation}
 \label{HS}
 \hat{H}_\mathrm{S} =
 (E_\mathrm{D} + \hat{H}_\mathrm{D})
 | \mathrm{D} \rangle \langle \mathrm{D} |
 + (E_\mathrm{A} + \hat{H}_\mathrm{A})
 | \mathrm{A} \rangle \langle \mathrm{A} |
 + \hat{H}_\mathrm{B}
 + (\hat{V}_\mathrm{AD} | \mathrm{A} \rangle \langle \mathrm{D} |
 + \mathrm{h.c.}),
\end{equation}
with
\begin{align}
 \label{HD}
 &\hat{H}_\mathrm{D} = \sum_v E_{\mathrm{D}v}
 | \mathrm{D}v \rangle \langle \mathrm{D}v |,
 \\
 \label{HA}
 &\hat{H}_\mathrm{A} = \sum_w E_{\mathrm{A}w}
 | \mathrm{A}w \rangle \langle \mathrm{A}w |,
 \\
 \label{HB}
 &\hat{H}_\mathrm{B} = \sum_a E_a | a \rangle \langle a |,
 \\
 \label{VAD}
 &\hat{V}_\mathrm{AD} =
 \sum_{v,a} \sum_{w,b} V_{\mathrm{A}wb,\mathrm{D}va}
 | b \rangle | \mathrm{A}w \rangle \langle \mathrm{D}v | \langle a |.
\end{align}
Here $| \mathrm{D} \rangle$ and $| \mathrm{A} \rangle$ are the D and A
electronic states, respectively, $| \mathrm{D}v \rangle$ and
$| \mathrm{A}w \rangle$ are vibrational states associated with the D and
A diabatic energy surfaces, respectively, and $| a \rangle$ and
$| b \rangle$ are B vibrational states that modulate only the tunneling
barrier between the electronic $| \mathrm{D} \rangle$ and
$| \mathrm{A} \rangle$ states.
The D and A electronic states depend parametrically on the nuclear
coordinates.
Because of the fact that the ET occurs only at a specific configuration
of the former vibrational modes, the Condon approximation is applicable
for them, assuming the local configuration regime.
On the other hand, since the latter vibrational modes affect strongly
the electronic coupling, the Condon approximation cannot be made.
Also, $E_\mathrm{D}$ and $E_\mathrm{A}$ are the diabatic energy surface
bottoms (referred to as the electronic origins) of the D and A states,
respectively, $E_{\mathrm{D}v}$ and $E_{\mathrm{A}w}$ are the energies
of vibrational states $| \mathrm{D}v \rangle$ and
$| \mathrm{A}w \rangle$ measured from the electronic origins, $E_a$ is
the energy of vibrational state $| a \rangle$, and
$V_{\mathrm{D}va,\mathrm{A}wb}$ denotes effective coupling matrix
elements between two vibronic states
$| \mathrm{D} \rangle | \mathrm{D}v \rangle | a \rangle$ and
$| \mathrm{A} \rangle | \mathrm{A}w \rangle | b \rangle$.

The Fermi's golden rule based expression for the ET rate is given by
\begin{equation}
 \label{kAwbDva1}
 k_{\mathrm{A}w,b \leftarrow \mathrm{D}v,a}
 = \frac{2 \pi}{\hslash}
 |\langle \mathrm{A}w,b | \hat{V}_\mathrm{AD}
 | \mathrm{D}v,a \rangle|^2
 \delta(E_\mathrm{D} + E_{\mathrm{D}v} + E_a
 - E_\mathrm{A} - E_{\mathrm{A}w} - E_b),
\end{equation}
where
$| \mathrm{D}v,a \rangle = | a \rangle | \mathrm{D}v \rangle$,
$| \mathrm{A}w,b \rangle = | b \rangle | \mathrm{A}w \rangle$.
The Condon approximation for the $| \mathrm{D}v \rangle$ and
$| \mathrm{A}w \rangle$ states allows us to write the effective coupling
matrix elements between two vibronic states
$| \mathrm{D} \rangle | \mathrm{D}v,a \rangle$
and
$| \mathrm{A} \rangle | \mathrm{A}w,b \rangle$
as a product of the electronic coupling element
$\langle b | \hat{V}_\mathrm{AD} | a \rangle$ and the nuclear overlap
function $\langle \mathrm{A}w | \mathrm{D}v \rangle$, that is,
\begin{equation}
 \label{COND}
 \langle \mathrm{A}w,b | \hat{V}_\mathrm{AD} | \mathrm{D}v,a \rangle
 = \langle b | \hat{V}_\mathrm{AD} | a \rangle
 \langle \mathrm{A}w | \mathrm{D}v \rangle.
\end{equation}

The electronic states can be constructed by using the Hartree-Fock
orbitals at fixed nuclear coordinates.
The full orbital space is divided into subspaces: the D subspace
containing the D orbitals, the A subspace containing the A orbitals, and
the B subspace containing the B orbitals.
The partitioning Fock matrix is
\begin{equation}
 \label{FOCK}
 \mathbf{F} =
 \begin{pmatrix}
 \mathbf{F}_\mathrm{DD}&\mathbf{F}_\mathrm{DA}&\mathbf{F}_\mathrm{DB}\\
 \mathbf{F}_\mathrm{AD}&\mathbf{F}_\mathrm{AA}&\mathbf{F}_\mathrm{AB}\\
 \mathbf{F}_\mathrm{BD}&\mathbf{F}_\mathrm{BA}&\mathbf{F}_\mathrm{BB}
 \end{pmatrix}
 ,
\end{equation}
e.g., $\mathbf{F}_\mathrm{AD}$ is the submatrix of $\mathbf{F}$ with
elements $F_{ij}$, where $i$ belongs to the subspace D, and $j$ belongs
to the subspace A.

In the deep tunneling regime, within the INDO approximation that we will
use, the electronic coupling element between the D and A states is
expressed by the effective Fock matrix element defined as
\begin{equation}
 \label{FADeff}
 \hat{F}_\mathrm{AD}^\mathrm{eff} =
 \langle \phi_\mathrm{A} | \hat{F} | \phi_\mathrm{D} \rangle
 + \sum_{b_i b_j} \langle \phi_\mathrm{A} | \hat{F}
 | \phi_{b_i} \rangle \langle \phi_{b_i} |
 \frac{1}{E_\mathrm{tun} - \hat{F}}
 | \phi_{b_j} \rangle \langle \phi_{b_j} |
 \hat{F} | \phi_\mathrm{D} \rangle,
\end{equation}
where $\hat{F}$ is the Fock operator, $E_\mathrm{tun}$ is the electron
tunneling energy, $| \phi_\mathrm{D} \rangle$ and
$| \phi_\mathrm{A} \rangle$ are the D and A orbitals of the tunneling
electron, $| \phi_{b_i} \rangle$ denotes the intervening B molecular
orbitals, and $\sum_{b_i}$ refers to a summation over the B orbitals
$b_i$.
Strictly speaking, the effective Fock matrix element should be
multiplied by a constant factor dependent on the D and A electronic
configurations.
We will consider the electronic coupling only between closed-shell
Hartree-Fock ground states and singlet spin-adapted configurations that
arise as a result of single excitations from them.
While the constant factor is $1$ for the coupling between the singly
excited electronic configurations, it is $\sqrt{2}$ for between the
excited and ground state.

The first and second term in Eq.~(\ref{FADeff}) gives the through-space
and through-bridge contribution, respectively.
The through-space coupling becomes negligible at the large DA distance
discussed below.
The coupling matrix element is very much unaffected by the tunneling
energy $E_\mathrm{tun}$ as long as it remains in the deep
tunneling regime.
In this case the tunneling energy can be safely varied around the
average value of the D and A orbitals.
However, this ambiguity becomes a serious problem in the resonant
tunneling regime.
In particular, the setting, in which the energy denominator in
Eq.~(\ref{FADeff}) becomes zero, causes the divergence of the effective
Fock matrix element.
In the following, we reformulate the ET rate expression in the
$T$-matrix framework whereby the tunneling energy is fixed at the
initial state energy of the tunneling electron.

The procedure starts by writing the system Hamiltonian as
$\hat{H}_\mathrm{S} = \hat{H}_0 + \hat{V}$ with
\begin{align}
 \label{H0}
 &\mathbf{H}_0 =
 \begin{pmatrix}
 \mathbf{F}_\mathrm{DD}&            \mathbf{0}&            \mathbf{0}\\
             \mathbf{0}&\mathbf{F}_\mathrm{AA}&            \mathbf{0}\\
             \mathbf{0}&            \mathbf{0}&\mathbf{F}_\mathrm{BB}
 \end{pmatrix}
 ,
 \\
 \label{V}
 &\mathbf{V} =
 \begin{pmatrix}
             \mathbf{0}&\mathbf{F}_\mathrm{DA}&\mathbf{F}_\mathrm{DB}\\
 \mathbf{F}_\mathrm{AD}&            \mathbf{0}&\mathbf{F}_\mathrm{AB}\\
 \mathbf{F}_\mathrm{BD}&\mathbf{F}_\mathrm{BA}&            \mathbf{0}
 \end{pmatrix}
 ,
\end{align}
where $\mathbf{H}_0$ and $\mathbf{V}$ are the matrix representation of
the operators $\hat{H}_0$ and $\hat{V}$, respectively, and $\hat{V}$ is
treated as a perturbation to $\hat{H}_0$.
The eigenstates of the unperturbed Hamiltonian $\hat{H}_0$ can be used
to construct the D, A, and B orbitals.

Defining the $T$-matrix as
\begin{equation}
 \label{T}
 \hat{T} = \hat{V} + \hat{V}\,\hat{G}_\mathrm{S}(E_i)\,\hat{V},
\end{equation}
with the retarded Green's function
\begin{equation}
 \label{GS}
 \hat{G}_\mathrm{S}(E) = \frac{1}{E - \hat{H}_\mathrm{S} + i 0^+},
\end{equation}
one obtains the transition rate from the initial state
$| \phi_i \rangle$ to the final state $| \phi_f \rangle$ as
\begin{equation}
 \label{kfi}
 k_{f \leftarrow i}
 = \frac{2 \pi}{\hslash}
 |\langle \phi_f | \hat{T} | \phi_i \rangle|^2
 \delta(E_i - E_f).
\end{equation}
Here the two states $| \phi_i \rangle$ and $| \phi_f \rangle$ are
eigenstates of the unperturbed Hamiltonian $\hat{H}_0$, and $E_i$ and
$E_f$ are the eigenenergies of states $| \phi_i \rangle$ and
$| \phi_f \rangle$, respectively.
We will consider the situation that the initial and final electronic
state is described by the D and A orbital $| \phi_\mathrm{D} \rangle$
and $| \phi_\mathrm{A} \rangle$, respectively.
In this case $E_i$ and $E_f$ are the energies of the the D and A
orbitals that will be denoted by $\epsilon_\mathrm{D}$ and
$\epsilon_\mathrm{A}$, respectively.
The transition is induced by the perturbation $\hat{V}$ that couples D,
A, and B.

Strictly speaking, the nuclear kinetic energy terms should be added to
the system Hamiltonian $\hat{H}_\mathrm{S}$ because the Fock matrix only
contains a contribution from the potential surface for the nuclear
motion.
Then, the delta function in Eq.~(\ref{kfi}) can be rewritten in the same
form as that in Eq.~(\ref{kAwbDva1}).
Assume that the nuclear kinetic energy is the same in the initial and
intermediate states to keep the $T$-matrix independent of the nuclear
kinetic energies.
This assumption seems to be reasonable under the Born-Oppenheimer
approximation.
Then, keeping in mind that the $T$-matrix is an operator in the space of
the B vibrational states, the transition rate can be rewritten in the
form
\begin{equation}
 \label{kAwbDva2}
 k_{\mathrm{A}w,b \leftarrow \mathrm{D}v,a}
 = \frac{2 \pi}{\hslash}
 |\langle b | \hat{T}_\mathrm{AD} | a \rangle|^2
 |\langle \mathrm{A}w | \mathrm{D}v \rangle|^2
 \delta(E_\mathrm{D} + E_{\mathrm{D}v} + E_a
 - E_\mathrm{A} - E_{\mathrm{A}w} - E_b),
\end{equation}
with
$\hat{T}_\mathrm{AD} =
\langle \phi_\mathrm{A} | \hat{T} | \phi_\mathrm{D} \rangle$.
In the deep tunneling regime, expansion of the $T$-matrix element
$\hat{T}_\mathrm{AD}$ to second order in the perturbation $\hat{V}$
gives the effective Fock matrix element
$\hat{F}_\mathrm{AD}^\mathrm{eff}$ with substitution of the D orbital
energy $\epsilon_\mathrm{D}$ for the tunneling energy $E_\mathrm{tun}$.
The $T$-matrix includes all higher-order tunneling processes where the
initial and final states are coupled by multiple scatterings described
by the perturbation $\hat{V}$.

The observed ET rate is given by summing over final vibrational states
and thermal averaging over initial vibrational states as
\begin{equation}
 \label{kAD1}
 k_{\mathrm{A} \leftarrow \mathrm{D}}
 = \sum_{v,a} P_{\mathrm{D}v} P_a \sum_{w,b}
 k_{\mathrm{A}w,b \leftarrow \mathrm{D}v,a},
\end{equation}
where $P_{\mathrm{D}v}$ and $P_a$ are the Boltzmann distributions over the
vibrational levels $| \mathrm{D}v \rangle$ and $| a \rangle$,
respectively.
We denote the thermal average over the vibrational states associated
with the D diabatic energy surface by
$\langle \ldots \rangle_\mathrm{D}$, that is,
$\langle \ldots \rangle_\mathrm{D} =
\sum_v P_{\mathrm{D}v}
\langle \mathrm{D}v | \ldots | \mathrm{D}v \rangle =
\mathrm{Tr}_\mathrm{D}[e^{- \beta \hat{H}_\mathrm{D}} \ldots] /
\mathrm{Tr}_\mathrm{D}[e^{- \beta \hat{H}_\mathrm{D}}]$,
where $\beta = (k_B T)^{-1}$ and $\mathrm{Tr}_\mathrm{D}$ denotes a
trace over the eigenstates of $\hat{H}_\mathrm{D}$, and the thermal
average over the B vibrational states by
$\langle \ldots \rangle_\mathrm{B}$, that is,
$\langle \ldots \rangle_\mathrm{B} =
\sum_a P_a \langle a | \ldots | a \rangle =
\mathrm{Tr}_\mathrm{B}[e^{- \beta \hat{H}_\mathrm{B}} \ldots] /
\mathrm{Tr}_\mathrm{B}[e^{- \beta \hat{H}_\mathrm{B}}]$,
where $\mathrm{Tr}_\mathrm{B}$ denotes a trace over the eigenstates of
$\hat{H}_\mathrm{B}$.
Then, writing the delta function in Eq.~(\ref{kAwbDva2}) as a Fourier
transform, we get
\begin{equation}
 \label{kAD2}
 k_{\mathrm{A} \leftarrow \mathrm{D}}
 = \frac{1}{\hslash^2} \int_{-\infty}^\infty d\epsilon\,
 \tilde{\rho}_\mathrm{FC}(E_\mathrm{D} - E_\mathrm{A} - \epsilon)\,
 \tilde{C}_T(\epsilon / \hslash),
\end{equation}
with the thermally averaged Franck-Condon weighted density of states,
$\tilde{\rho}_\mathrm{FC}$, given by the Fourier transform of the
thermally averaged time-dependent Franck-Condon factor
\begin{equation}
 \label{rhoFCdef}
 \rho_\mathrm{FC}(t) \equiv
 \langle e^{i \hat{H}_\mathrm{D} t / \hslash}
 e^{- i \hat{H}_\mathrm{A} t / \hslash} \rangle_\mathrm{D},
\end{equation}
such that
\begin{equation}
 \label{trhoFC1}
 \tilde{\rho}_\mathrm{FC}(\epsilon)
 = \frac{1}{2 \pi \hslash} \int_{-\infty}^\infty dt\,
 e^{i \epsilon t / \hslash} \rho_\mathrm{FC}(t),
\end{equation}
and with the spectral density $\tilde{C}_T$ given by the Fourier
transform of the time correlation function
\begin{equation}
 \label{CTdef}
 C_T(t) \equiv \langle \hat{T}_\mathrm{AD}^\dagger(t)\,
 \hat{T}_\mathrm{AD}(0) \rangle_\mathrm{B},
\end{equation}
where
$\hat{T}_\mathrm{AD}(t) = e^{i \hat{H}_\mathrm{B} t / \hslash}
\hat{T}_\mathrm{AD} e^{- i \hat{H}_\mathrm{B} t / \hslash}$,
such that
\begin{equation}
 \label{tildeCT1}
 \tilde{C}_T(\epsilon / \hslash)
 = \int_{-\infty}^\infty dt\,
 e^{i \epsilon t / \hslash} C_T(t).
\end{equation}
Applying the convolution theorem for Fourier transforms,
Eq.~(\ref{kAD2}) can be written as
\begin{equation}
 \label{kAD3}
 k_{\mathrm{A} \leftarrow \mathrm{D}}
 = \frac{1}{\hslash^2} \int_{-\infty}^\infty dt\,
 e^{i (E_\mathrm{D} - E_\mathrm{A}) t / \hslash}
 \rho_\mathrm{FC}(t)\, C_T(t).
\end{equation}
Note that the $T$-matrix is non-Hermitian due to the infinitesimal term
$i 0^+$.
In the deep tunneling regime it is treated as a Hermitian matrix by
disregarding the $i 0^+$ term.
However, we will see that in the resonant tunneling regime it is
necessary to modify the eigenvalues of $\hat{H}_\mathrm{S}$ in the
Green's function (\ref{GS}) to include a finite imaginary term, which
represents the effect of the surroundings on the relevant system.
Then it is important to deal with the non-Hermitian $T$-matrix.
In this case the operator $\hat{T}_\mathrm{AD}$ is also non-Hermitian.

\section{\label{correlation}Time correlation functions of $T$-matrix}
The non-Hermitian operator $\hat{T}_\mathrm{AD}$ can be decomposed into
the sum of two Hermitian operators as
\begin{equation}
 \label{TADdec}
 \hat{T}_\mathrm{AD} = \hat{A} + i \hat{B},
\end{equation}
where $\hat{A}$ and $\hat{B}$ are Hermitian operators given by
\begin{align}
 \label{A}
 &\hat{A} =
 \frac{1}{2} (\hat{T}_\mathrm{AD} + \hat{T}_\mathrm{AD}^\dagger),
 \\
 \label{B}
 &\hat{B} =
 \frac{1}{2i} (\hat{T}_\mathrm{AD} - \hat{T}_\mathrm{AD}^\dagger).
\end{align}
Then the quantum time correlation function $C_T(t)$ can be separated
into the autocorrelation and the cross-correlation part, that is,
\begin{equation}
 \label{CTdec}
 C_T(t) = C_\mathrm{A}(t) + i C_\mathrm{C}(t),
\end{equation}
with
\begin{align}
 \label{CA}
 &C_\mathrm{A}(t) = C_{AA}(t) + C_{BB}(t),
 \\
 \label{CC}
 &C_\mathrm{C}(t) = C_{AB}(t) - C_{BA}(t),
 \\
 \label{CAA}
 &C_{AA}(t) = \langle \hat{A}(t)\, \hat{A}(0) \rangle_\mathrm{B},
 \\
 \label{CBB}
 &C_{BB}(t) = \langle \hat{B}(t)\, \hat{B}(0) \rangle_\mathrm{B},
 \\
 \label{CAB}
 &C_{AB}(t) = \langle \hat{A}(t)\, \hat{B}(0) \rangle_\mathrm{B},
 \\
 \label{CBA}
 &C_{BA}(t) = \langle \hat{B}(t)\, \hat{A}(0) \rangle_\mathrm{B},
\end{align}
where
\begin{align}
 \label{At}
 &\hat{A}(t) =
 \frac{1}{2} (\hat{T}_\mathrm{AD}(t) + \hat{T}_\mathrm{AD}^\dagger(t)),
 \\
 \label{Bt}
 &\hat{B}(t) =
 \frac{1}{2i} (\hat{T}_\mathrm{AD}(t) - \hat{T}_\mathrm{AD}^\dagger(t)).
\end{align}
These two parts have the following properties:
\begin{align}
 \label{CAprop}
 &C_\mathrm{A}(-t) = C_\mathrm{A}^\ast(t) =
 C_\mathrm{A}(t - i \beta \hslash),
 \\
 \label{CCprop}
 &C_\mathrm{C}(-t) = -C_\mathrm{C}^\ast(t) =
 -C_\mathrm{C}(t - i \beta \hslash).
\end{align}
Thus, the Fourier transform
\begin{equation}
 \label{CAomega}
 \tilde{C}_\mathrm{A}(\omega) =
 \int_{-\infty}^\infty dt\, e^{i \omega t} C_\mathrm{A}(t)
\end{equation}
satisfies the detailed balance relation
\begin{equation}
 \label{CAdbr}
 \tilde{C}_\mathrm{A}(\omega) =
 e^{\beta \hslash \omega} \tilde{C}_\mathrm{A}(-\omega),
\end{equation}
while the Fourier transform
\begin{equation}
 \label{CComega}
 \tilde{C}_\mathrm{C}(\omega) =
 \int_{-\infty}^\infty dt\, e^{i \omega t} C_\mathrm{C}(t)
\end{equation}
does not satisfy detailed balance but
\begin{equation}
 \label{CCdbr}
 \tilde{C}_\mathrm{C}(\omega) =
 - e^{\beta \hslash \omega} \tilde{C}_\mathrm{C}(-\omega).
\end{equation}
The cross-correlation part represents the nonequilibrium properties of
the system.
In the deep tunneling regime, $\hat{T}_\mathrm{AD}$ is often treated as
a Hermitian operator, assuming that the D and A electronic wave
functions are real, so that the cross-correlation part vanishes.
From Eqs.~(\ref{CAprop})$-$(\ref{CComega}) it follows that
$\tilde{C}_\mathrm{A}(\omega)$ and $\tilde{C}_\mathrm{C}(\omega)$ are
real and imaginary, respectively.
While $\tilde{C}_\mathrm{A}(\omega) \ge 0$,
$i \tilde{C}_\mathrm{C}(\omega)$ can also be negative.
Since the trace is invariant to cyclic permutation of the operators
inside it, and the equilibrium density operator
$e^{- \beta \hat{H}_\mathrm{B}}$ and the time evolution operator
$e^{- i \hat{H}_\mathrm{B} t / \hslash}$ commute, the quantum time
correlation functions are stationary, that is,
$\langle \hat{A}(t + t')\, \hat{A}(t') \rangle_\mathrm{B} =
\langle \hat{A}(t)\, \hat{A}(0) \rangle_\mathrm{B}$, etc.

It will be convenient to decompose $C_\mathrm{A}(t)$ and
$C_\mathrm{C}(t)$ into the real parts
\begin{align}
 \label{CA+}
 &C_\mathrm{A}^{(+)}(t) =
 \frac{1}{2} (C_\mathrm{A}(t) + C_\mathrm{A}^\ast(t)),
 \\
 \label{CC+}
 &C_\mathrm{C}^{(+)}(t) =
 \frac{1}{2} (C_\mathrm{C}(t) + C_\mathrm{C}^\ast(t)),
\end{align}
and the imaginary parts
\begin{align}
 \label{CA-}
 &C_\mathrm{A}^{(-)}(t) =
 \frac{1}{2} (C_\mathrm{A}(t) - C_\mathrm{A}^\ast(t)),
 \\
 \label{CC-}
 &C_\mathrm{C}^{(-)}(t) =
 \frac{1}{2} (C_\mathrm{C}(t) - C_\mathrm{C}^\ast(t)).
\end{align}
From Eqs.~(\ref{CAprop})$-$(\ref{CComega}) the Fourier transforms can be
written as
\begin{align}
 \label{CA+-}
 &\tilde{C}_\mathrm{A}^{(\pm)}(\omega) =
 \frac{1}{2}
 (\tilde{C}_\mathrm{A}(\omega) \pm \tilde{C}_\mathrm{A}(-\omega)),
 \\
 \label{CC+-}
 &\tilde{C}_\mathrm{C}^{(\pm)}(\omega) =
 \frac{1}{2}
 (\tilde{C}_\mathrm{C}(\omega) \mp \tilde{C}_\mathrm{C}(-\omega)).
\end{align}
While $\tilde{C}_\mathrm{A}^{(+)}(\omega)$ and
$\tilde{C}_\mathrm{C}^{(-)}(\omega)$ are symmetric in $\omega$,
$\tilde{C}_\mathrm{A}^{(-)}(\omega)$ and
$\tilde{C}_\mathrm{C}^{(+)}(\omega)$ are antisymmetric in $\omega$.
Substitution of Eqs.~(\ref{CAdbr}) and (\ref{CCdbr}) into
Eqs.~(\ref{CA+-}) and (\ref{CC+-}) leads to
\begin{align}
 \label{CACA+-}
 &\tilde{C}_\mathrm{A}(\omega) =
 \frac{2}{1 \pm e^{- \beta \hslash \omega}}
 \tilde{C}_\mathrm{A}^{(\pm)}(\omega),
 \\
 \label{CCCC+-}
 &\tilde{C}_\mathrm{C}(\omega) =
 \frac{2}{1 \pm e^{- \beta \hslash \omega}}
 \tilde{C}_\mathrm{C}^{(\pm)}(\omega).
\end{align}

The quantum time correlation function $C_T(t)$ can be evaluated using
classical equilibrium MD simulations coupled with QC calculations.
This approach involves the replacement of the quantum time correlation
function by its classical counterpart.
We denote the classical time correlation function by
\begin{equation}
 \label{CTcl}
 C_T^\mathrm{cl}(t) =
 \langle T_\mathrm{AD}^\ast(t)\, T_\mathrm{AD}(0) \rangle_\mathrm{cl},
\end{equation}
with the classical variables
\begin{align}
 \label{TADcl}
 &T_\mathrm{AD}(t) =
 \langle \phi_\mathrm{A}(\mathbf{Q}_\mathrm{B}(t)) |
 \hat{T}(\mathbf{Q}_\mathrm{B}(t))
 | \phi_\mathrm{D}(\mathbf{Q}_\mathrm{B}(t)) \rangle,
 \\
 \label{TADast}
 &T_\mathrm{AD}^\ast(t) =
 \langle \phi_\mathrm{D}(\mathbf{Q}_\mathrm{B}(t)) |
 \hat{T}^\dagger(\mathbf{Q}_\mathrm{B}(t))
 | \phi_\mathrm{A}(\mathbf{Q}_\mathrm{B}(t)) \rangle.
\end{align}
Here, the $T$-matrix $\hat{T}(\mathbf{Q}_\mathrm{B}(t))$, and the D and
A orbitals $| \phi_\mathrm{D}(\mathbf{Q}_\mathrm{B}(t)) \rangle$ and
$| \phi_\mathrm{A}(\mathbf{Q}_\mathrm{B}(t)) \rangle$ are parametric
functions of the B nuclear coordinates $\mathbf{Q}_\mathrm{B}(t)$ at
time $t$ that move along the classical trajectories obtained by the MD
simulations, and the classical thermal average
$\langle \ldots \rangle_\mathrm{cl}$ should be understood as the time
average over the MD trajectory, as implied by the ergodic theorem of
statistical mechanics.

Using the classical counterparts of $C_\mathrm{A}(t)$ and
$C_\mathrm{C}(t)$, the classical correlation function
$C_T^\mathrm{cl}(t)$ can be written as
\begin{equation}
 \label{CTcldec}
 C_T^\mathrm{cl}(t) =
 C_\mathrm{A}^\mathrm{cl}(t) + i C_\mathrm{C}^\mathrm{cl}(t),
\end{equation}
with
\begin{align}
 \label{CAcl}
 &C_\mathrm{A}^\mathrm{cl}(t) =
 C_{AA}^\mathrm{cl}(t) + C_{BB}^\mathrm{cl}(t),
 \\
 \label{CCcl}
 &C_\mathrm{C}^\mathrm{cl}(t) =
 C_{AB}^\mathrm{cl}(t) - C_{BA}^\mathrm{cl}(t),
 \\
 \label{CAAcl}
 &C_{AA}^\mathrm{cl}(t) = \langle A(t)\, A(0) \rangle_\mathrm{cl},
 \\
 \label{CBBcl}
 &C_{BB}^\mathrm{cl}(t) = \langle B(t)\, B(0) \rangle_\mathrm{cl},
 \\
 \label{CABcl}
 &C_{AB}^\mathrm{cl}(t) = \langle A(t)\, B(0) \rangle_\mathrm{cl},
 \\
 \label{CBAcl}
 &C_{BA}^\mathrm{cl}(t) = \langle B(t)\, A(0) \rangle_\mathrm{cl},
\end{align}
where
\begin{align}
 \label{Acl}
 &A(t) = \frac{1}{2} (T_\mathrm{AD}(t) + T_\mathrm{AD}^\ast(t)),
 \\
 \label{Bcl}
 &B(t) = \frac{1}{2i} (T_\mathrm{AD}(t) - T_\mathrm{AD}^\ast(t)).
\end{align}
From the time reversal symmetry of the classical equations of motion, it
follows that
\begin{align}
 \label{CAcltsym}
 &C_\mathrm{A}^\mathrm{cl}(t) =
 C_\mathrm{A}^\mathrm{cl}(-t),
 \\
 \label{CCcltsym}
 &C_\mathrm{C}^\mathrm{cl}(t) =
 C_\mathrm{C}^\mathrm{cl}(-t).
\end{align}
Furthermore, the stationary property implies that
\begin{equation}
 \label{CCclsta}
 C_\mathrm{C}^\mathrm{cl}(-t) = - C_\mathrm{C}^\mathrm{cl}(t).
\end{equation}
Alternatively, taking the classical limit $\hslash \rightarrow 0$,
Eq.~(\ref{CCprop}) yields Eq.~(\ref{CCclsta}).
From Eqs.~(\ref{CCcltsym}) and (\ref{CCclsta}) it follows that
\begin{equation}
 \label{CCcl0}
 C_\mathrm{C}^\mathrm{cl}(t) = 0.
\end{equation}
Namely, the classical approximation disregards the contribution of the
cross-correlation part $C_\mathrm{C}(t)$ in the time correlation
function $C_T(t)$.

Using Eqs.~(\ref{CTcldec}), (\ref{CCcl0}), (\ref{CAcl}), (\ref{CAAcl}),
(\ref{CBBcl}), (\ref{Acl}), and (\ref{Bcl}) yields
\begin{align}
 \label{CReclCImcl}
 C_T^\mathrm{cl}(t)
 &= C_\mathrm{A}^\mathrm{cl}(t) = \frac{1}{2}
 \langle
 T_\mathrm{AD}(t)\, T_\mathrm{AD}^\ast(0) +
 T_\mathrm{AD}^\ast(t)\, T_\mathrm{AD}(0)
 \rangle_\mathrm{cl}
 \nonumber
 \\
 &= \langle \mathrm{Re}T_\mathrm{AD}(t)\,
 \mathrm{Re}T_\mathrm{AD}(0) \rangle_\mathrm{cl}
 + \langle \mathrm{Im}T_\mathrm{AD}(t)\,
 \mathrm{Im}T_\mathrm{AD}(0) \rangle_\mathrm{cl}
 \nonumber
 \\
 &\equiv C_\mathrm{Re}^\mathrm{cl}(t) + C_\mathrm{Im}^\mathrm{cl}(t),
\end{align}
where $C_\mathrm{Re}^\mathrm{cl}(t)$ and $C_\mathrm{Im}^\mathrm{cl}(t)$
define the autocorrelation functions of the real and imaginary parts of
$T_\mathrm{AD}$, respectively.
Since $C_\mathrm{A}^\mathrm{cl}(t)$ is a real symmetric function of $t$,
its Fourier transform
\begin{equation}
 \label{CAclFT}
 \tilde{C}_\mathrm{A}^\mathrm{cl}(\omega) =
 \int_{-\infty}^\infty dt\, e^{i \omega t} C_\mathrm{A}^\mathrm{cl}(t)
\end{equation}
is a real symmetric function of $\omega$, that is,
$\tilde{C}_\mathrm{A}^\mathrm{cl}(\omega) =
\tilde{C}_\mathrm{A}^\mathrm{cl}(-\omega)$.
This means that the classical time correlation function does not satisfy
the detailed balance condition (\ref{CAdbr}).
The simplest way to solve this problem is to identify
$C_\mathrm{A}^\mathrm{cl}(t)$ with $C_\mathrm{A}^{(+)}(t)$.
Using the relation (\ref{CACA+-}) we obtain
\begin{equation}
 \label{CACAcl}
 \tilde{C}_\mathrm{A}(\omega) =
 \tilde{Q}(\omega)\, \tilde{C}_\mathrm{A}^\mathrm{cl}(\omega),
\end{equation}
where $\tilde{Q}(\omega)$ is the quantum correction factor given by
\begin{equation}
 \label{Qst}
 \tilde{Q}(\omega) = \frac{2}{1 + e^{- \beta \hslash \omega}},
\end{equation}
which guarantees detailed balance.
There are other approximations involving quantum correction factors such
as
\begin{align}
 \label{Qh}
 &\tilde{Q}(\omega) =
 \frac{\beta \hslash \omega}{1 - e^{- \beta \hslash \omega}},
 \\
 \label{Qsc}
 &\tilde{Q}(\omega) = e^{\beta \hslash \omega / 2}.
\end{align}
More complex examples can be found in
Refs.~\onlinecite{Egorov_CPL98,Egorov_JPCA99,Skinner_JPCB01,Kim_JPCB02}

Employing such a semiclassical approximation, the ET rate (\ref{kAD2})
can be expressed as
\begin{equation}
 \label{kAD4}
 k_{\mathrm{A} \leftarrow \mathrm{D}}
 = \frac{1}{\hslash^2} \int_{-\infty}^\infty d\epsilon\,
 \tilde{\rho}_\mathrm{FC}(E_\mathrm{D} - E_\mathrm{A} - \epsilon)\,
 \tilde{Q}(\epsilon / \hslash)\,
 \tilde{C}_\mathrm{A}^\mathrm{cl}(\epsilon / \hslash).
\end{equation}
Here, from Eqs.~(\ref{CReclCImcl}) and (\ref{CAclFT}),
$\tilde{C}_\mathrm{A}^\mathrm{cl}(\omega) =
\tilde{C}_\mathrm{Re}^\mathrm{cl}(\omega) +
\tilde{C}_\mathrm{Im}^\mathrm{cl}(\omega)$.
The general expressions for $\tilde{\rho}_\mathrm{FC}(\epsilon)$ have
been widely discussed.\cite{Bixon_ACP99,Newton_ACP99}
Adopting the simplest classical expression
\begin{equation}
 \label{trhoFCM}
 \tilde{\rho}_\mathrm{FC}(\epsilon)
 = \sqrt{\frac{1}{4 \pi \lambda k_B T}}
 \exp\left(- \frac{(\epsilon - \lambda)^2}{4 \lambda k_B T}\right),
\end{equation}
where $\lambda$ is the reorganization energy, Eq.~(\ref{kAD4}) yields
\begin{equation}
 \label{kAD5}
 k_{\mathrm{A} \leftarrow \mathrm{D}}
 = \frac{1}{\hslash^2} \sqrt{\frac{1}{4 \pi \lambda k_B T}}
 \int_{-\infty}^\infty d\epsilon\,
 \tilde{Q}(\epsilon / \hslash)\,
 \tilde{C}_\mathrm{A}^\mathrm{cl}(\epsilon / \hslash)\,
 \exp\left(- \frac{(E_\mathrm{D} - E_\mathrm{A} - \lambda - \epsilon)^2}
 {4 \lambda k_B T}\right),
\end{equation}
which reduces to the form given in Ref.~\onlinecite{Nishioka_JPCB05} if
we replace $T_\mathrm{AD}$ by the real terms up to second order in
perturbation theory.
According to Eq.~(\ref{trhoFCM}), $\tilde{\rho}_\mathrm{FC}(\epsilon)$
has a maximum for $\epsilon = \lambda$, and the energy scale for the
decay is given by
$\Delta \epsilon_\mathrm{FC} = \sqrt{4 \lambda k_B T}$.
If $\tilde{C}_\mathrm{A}^\mathrm{cl}(\epsilon / \hslash)$ goes
asymptotically to zero for $|\epsilon|$ larger than the width
$\Delta \epsilon_\mathrm{A} \ll \Delta \epsilon_\mathrm{FC}$, then
$\tilde{\rho}_\mathrm{FC}(E_\mathrm{D} - E_\mathrm{A} - \epsilon)$ in
Eq.~(\ref{kAD5}) can be replaced by the value at $\epsilon = 0$ to yield
the static limit expression
\begin{equation}
 \label{kAD6}
 k_{\mathrm{A} \leftarrow \mathrm{D}}
 = \frac{1}{\hslash^2}
 \tilde{\rho}_\mathrm{FC}(E_\mathrm{D} - E_\mathrm{A})\,
 \int_{-\infty}^\infty d\epsilon\,
 \tilde{Q}(\epsilon / \hslash)\,
 \tilde{C}_\mathrm{A}^\mathrm{cl}(\epsilon / \hslash).
\end{equation}
The corresponding quantum expression is given by
\begin{equation}
 \label{kAD7}
 k_{\mathrm{A} \leftarrow \mathrm{D}}
 = \frac{2 \pi}{\hslash}
 \tilde{\rho}_\mathrm{FC}(E_\mathrm{D} - E_\mathrm{A})\, C_T(0).
\end{equation}
Taking this limit corresponds to replacing
$\delta(E_\mathrm{D} + E_{\mathrm{D}v} + E_a
- E_\mathrm{A} - E_{\mathrm{A}w} - E_b)$
in Eq.~(\ref{kAwbDva2}) by
$\delta(E_\mathrm{D} + E_{\mathrm{D}v}
- E_\mathrm{A} - E_{\mathrm{A}w})$.
In this case the initial and final vibrational energies of B are the
same.
Hence, in Eq.~(\ref{kAD7}) $C_T(0)$ should be replaced by
$\sum_a P_a \sum_b
|\langle b | \hat{T}_\mathrm{AD} | a \rangle|^2 \delta_{E_b,E_a}$.
When $\Delta \epsilon_\mathrm{A}$ is comparable to
$\Delta \epsilon_\mathrm{FC}$, the dynamical corrections become
significant.
As shown in Eqs.~(\ref{Qst})$-$(\ref{Qsc}), the quantum correction
factors $\tilde{Q}(\omega)$ are positive increasing functions.
Moreover, $\tilde{C}_\mathrm{A}^\mathrm{cl}(\omega) \ge 0$.
Thus, due to the convolution in Eq.~(\ref{kAD5}), the dynamical
non-Condon effects can result in substantial enhancement of the ET rate
in the activated inverted region.
If the dynamical non-Condon effects are significant, its resultant ET
rate is highly asymmetric with respect to the activation energy.
In Eq.~(\ref{kAD5}),
$\tilde{Q}(\epsilon / \hslash)\,
\tilde{C}_\mathrm{A}^\mathrm{cl}(\epsilon / \hslash)$
is the semiclassical approximation of the spectral density
$\tilde{C}_T(\epsilon / \hslash)$, which represents the electronic
transition rate between D and A, averaged over the thermal distribution
of B vibrational states, and $\epsilon$ denotes the energy difference
between the D and A diabatic vibronic states
$| \mathrm{D} \rangle | \mathrm{D}v \rangle$ and
$| \mathrm{A} \rangle | \mathrm{A}w \rangle$, that is,
$\epsilon =
E_\mathrm{D} + E_{\mathrm{D}v} - E_\mathrm{A} - E_{\mathrm{A}w}$,
which is equal to the energy transferred to/from B during the electronic
transition.
This can be understood by rewriting
$\tilde{\rho}_\mathrm{FC}(E_\mathrm{D} - E_\mathrm{A} - \epsilon)$
and $\tilde{C}_T(\epsilon / \hslash)$ in Eq.~(\ref{kAD2}) as
\begin{align}
 \label{trhoFC2}
 &\tilde{\rho}_\mathrm{FC}(E_\mathrm{D} - E_\mathrm{A} - \epsilon)
 = \sum_{v,w} P_{\mathrm{D}v}
 |\langle \mathrm{A}w | \mathrm{D}v \rangle|^2
 \delta(E_\mathrm{D} + E_{\mathrm{D}v}
 - E_\mathrm{A} - E_{\mathrm{A}w} - \epsilon),
 \\
 \label{tildeCT2}
 &\tilde{C}_T(\epsilon / \hslash)
 = 2 \pi \hslash \sum_{a,b} P_a
 |\langle b | \hat{T}_\mathrm{AD} | a \rangle|^2
 \delta(E_a - E_b + \epsilon).
\end{align}
According to the Franck-Condon factor, downward (i.e. $\epsilon > 0$)
electronic transitions are more probable than upward
(i.e. $\epsilon < 0$) electronic transitions in the inverted region,
while the opposite is true in the normal region.
The asymmetry of the ET rate reflects the detailed balance property that
downward transitions are more probable than upward transitions.
This is a feature of the autocorrelation part $C_\mathrm{A}(t)$ of
$C_T(t)$.
If the cross-correlation part $C_\mathrm{C}(t)$ contributes to the ET
rate, the dynamical non-Condon effects can result in a more complicated
dependence of the ET rate on the energy gap $E_\mathrm{D}- E_\mathrm{A}$
due to the minus sign in Eq.~(\ref{CCdbr}), which does not exist in the
detailed balance relation.
In any case, the dynamical non-Condon effects are accompanied by the
vibrational excitation or deexcitation of B and in this sense inelastic.

Within the semiclassical approximation Eq.~(\ref{kAD4}) can be formally
written as
\begin{equation}
 \label{kAD8}
 k_{\mathrm{A} \leftarrow \mathrm{D}}
 = \frac{1}{\hslash^2} \int_{-\infty}^\infty dt\,
 e^{i (E_\mathrm{D} - E_\mathrm{A}) t / \hslash}
 \rho_\mathrm{FC}(t)
 \int_{-\infty}^\infty dt'\,
 Q(t')\, C_\mathrm{A}^\mathrm{cl}(t - t'),
\end{equation}
with $Q(t)$ the inverse Fourier transform of the quantum correction
factor, defined by
\begin{equation}
 \label{Qt}
 Q(t) =
 \frac{1}{2 \pi} \int_{-\infty}^\infty d\omega\,
 e^{- i \omega t} \tilde{Q}(\omega).
\end{equation}
Performing the inverse Fourier transform of Eq.~(\ref{trhoFCM}), we
obtain
\begin{equation}
 \label{rhoFCM}
 \rho_\mathrm{FC}(t) =
 e^{- \lambda k_B T t^2 / \hslash^2} e^{- i \lambda t / \hslash}.
\end{equation}
From this the characteristic decay timescale is given by
\begin{equation}
 \label{tauFC}
 \tau_\mathrm{FC} = \hslash / \sqrt{\lambda k_B T}.
\end{equation}
If $\tau_\mathrm{FC}$ is much smaller than the shortest timescale
$\tau_\mathrm{A}$ on which $C_\mathrm{A}^\mathrm{cl}(t)$ changes, we can
replace $C_\mathrm{A}^\mathrm{cl}(t - t')$ in Eq.~(\ref{kAD8}) by
$C_\mathrm{A}^\mathrm{cl}(t')$ to get the expression (\ref{kAD6}).
When $\tau_\mathrm{A}$ is comparable to $\tau_\mathrm{FC}$, the
dynamical non-Condon effects will become significant.
In the following, we attempt to evaluate $C_\mathrm{A}^\mathrm{cl}(t)$
in the resonant tunneling regime, using MD simulations and QC
semiempirical calculations.

\section{\label{methods}Computational methods}
We consider the ET processes between anthracene and TCNE and between
naphthalene and TCNE in PhCN solvent, where the solvent molecules act as
B.

The MD simulations were carried out by using the program
NAMD.\cite{NAMD}
We also use the program WORDOM to analyze the MD
trajectories.\cite{WORDOM}
We employed the AMBER force field,\cite{AMBER,AMBER9} converted for use
in CHARMM, and used the Grabuleda et al. force
field\cite{Grabuleda_JCC00} for acetonitrile in place of that for the
cyano group of PhCN.
To determine atomic charges of the D, A, and solvent molecules,
B3LYP/6-31G(d) optimizations of individual molecules were performed, and
then the ESP fits were implemented using Gaussian 03\cite{GAUSSIAN}.
The temperature was maintained at $300$ K by means of Langevin dynamics.
Also, the particle-mesh Ewald method for full-system periodic
electrostatics was taken into account.
The D and A molecules were placed in the middle of a solvent cube and
face to face with a $7\, \mathrm{\AA}$ distance in such a manner that
the C(9)$-$C(10) line of anthracene or the central C$-$C single bond of
naphthalene and the central C$=$C double bond of TCNE were parallel to
each other, after energy minimization.
The number of solvent molecules was $550$.
To build the starting geometry for the production dynamics, the system
was equilibrated at $300$ K and $1$ atm with periodic boundary conditions.
The system was first energy minimized, and then a $100$ ps equilibration
dynamics led to a starting simulation box of c.a.
$46.2 \times 46.2 \times 46.2\, \mathrm{\AA}$.
The simulation was performed in the NVT ensemble where the number of
particles, N, the volume, V, and the temperature, T, are kept constant.
Hydrogen atoms in the solvent molecules were constrained using the SHAKE
algorithm, and also the D and A molecules were held fixed during the
simulation.
The remaining degrees of freedom were left flexible.
The integration time step was $1$ fs.
The length of each trajectory was in the $50$ ps range, and the
interval between QC calculations was in $1$ fs range.

The PM3 method\cite{MOPAC} in the MOPAC7 program was used to perform the
QC calculation.
For each snapshot of the MD trajectory, the atomic coordinates of the D,
A, and solvent fragments are extracted and used for the QC calculation.
The solvent molecules involved in the coupling was selected on a
geometrical basis as the ones entirely within or partially within a
$3.5\, \mathrm{\AA}$ radius from the center of mass of the D and A
molecules.

\section{\label{broadening}Level broadening}
We limit our calculations to the relevant subspace of an overall system.
Then we need to characterize the effect of the rest of the system.
This effect can be included by introducing self-energy terms in the
denominator of the Green's function (\ref{GS}) for the isolated relevant
system.
Considering the weak perturbation $\hat{V}$, we expect that the
second-order perturbation theory provides sufficient accuracy in the
deep tunneling regime where the energy of the initial electronic state
is well separated from the eigenenergies of the B subspace.
Then the self-energy terms can be safely disregarded.

\begin{figure}[!htbp]
\centering\includegraphics[width=3.8in, angle=-90]{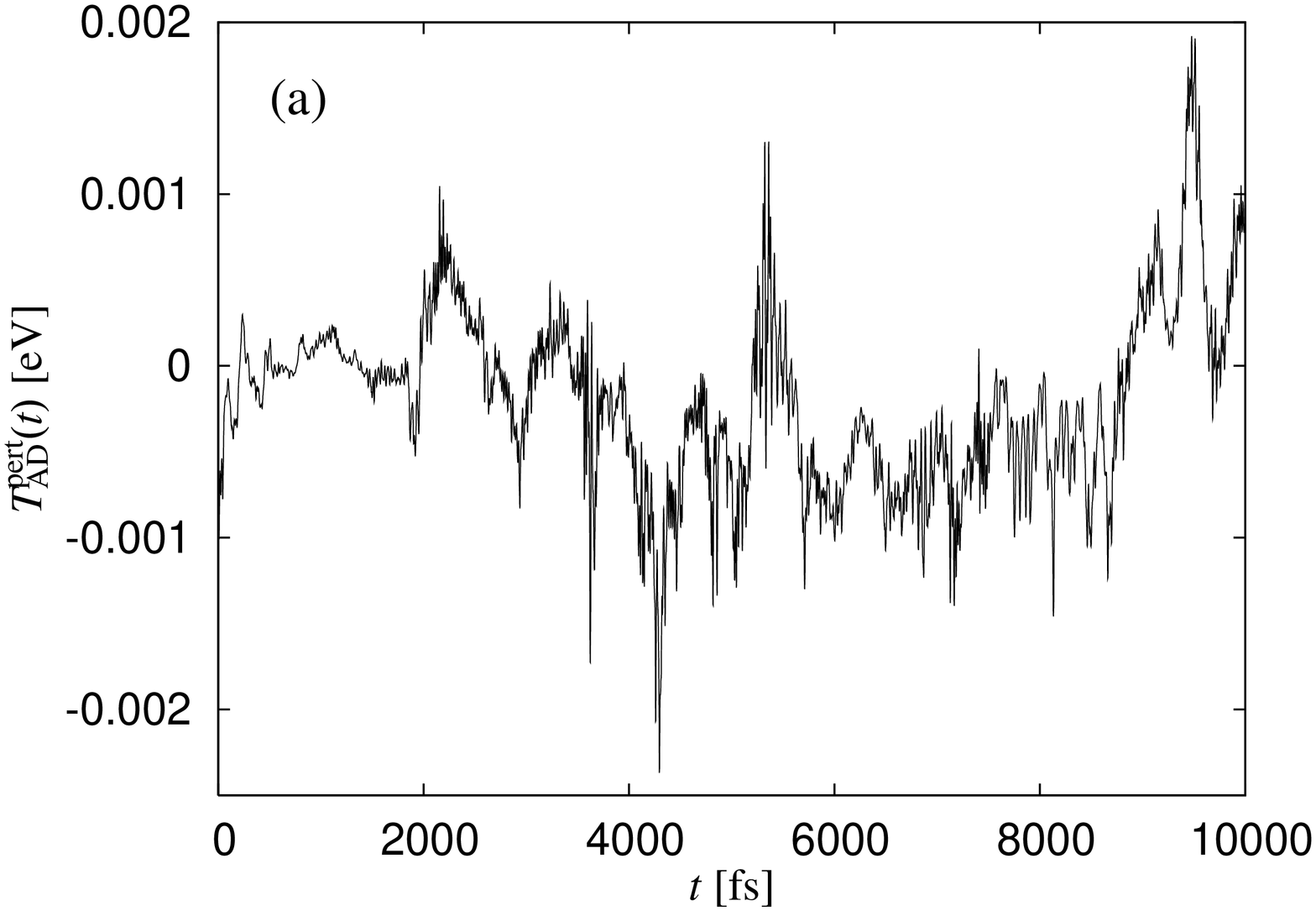}\\
\centering\includegraphics[width=3.8in, angle=-90]{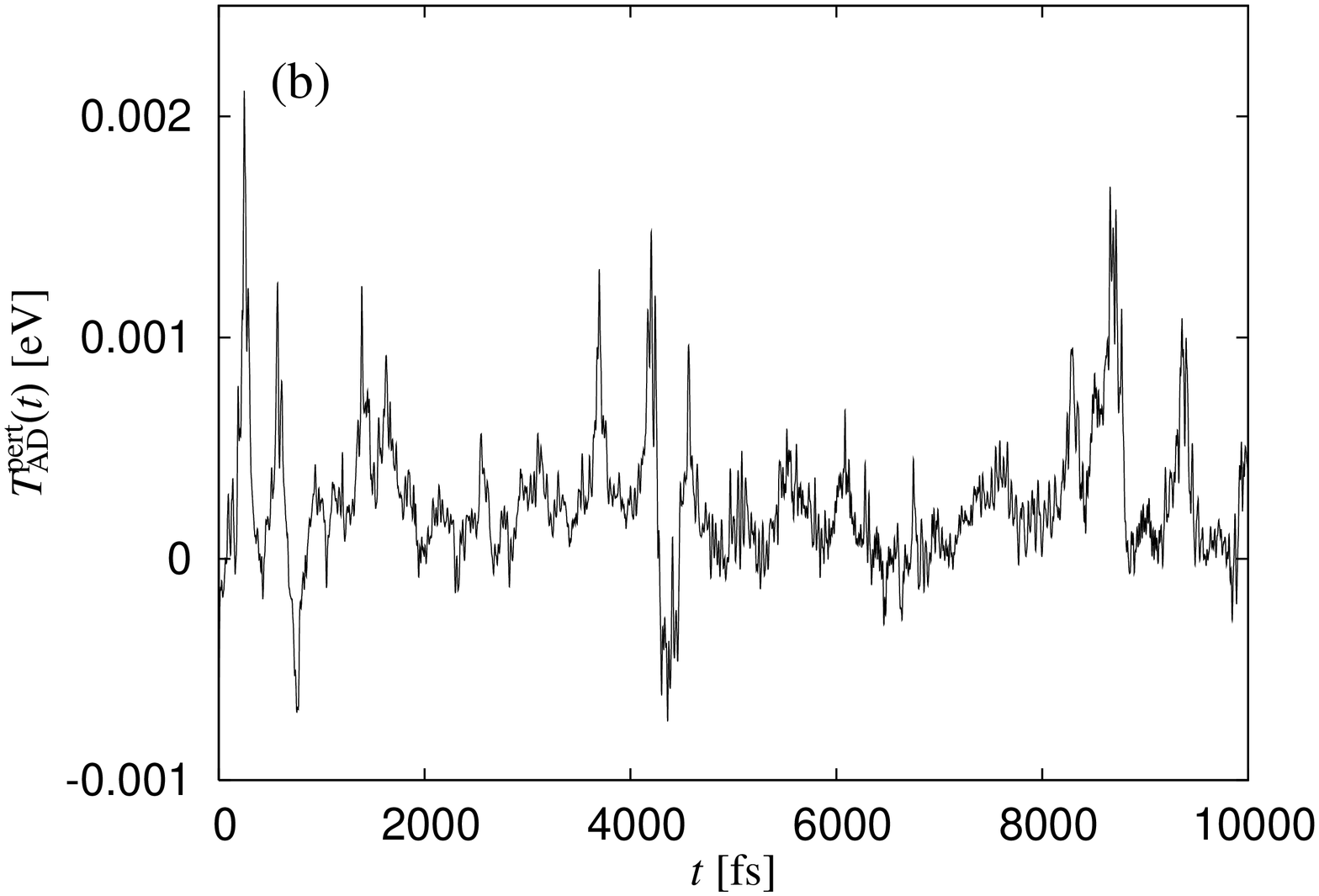}
\caption{\label{fig1}
$T_\mathrm{AD}^\mathrm{pert}(t)$ for (a) the CS from the LUMO of
anthracene to the LUMO of TCNE in PhCN, and (b) the CR from the LUMO of
TCNE to the HOMO of naphthalene in PhCN.
}
\end{figure}

\begin{figure}[!htbp]
\centering\includegraphics[width=3.8in, angle=-90]{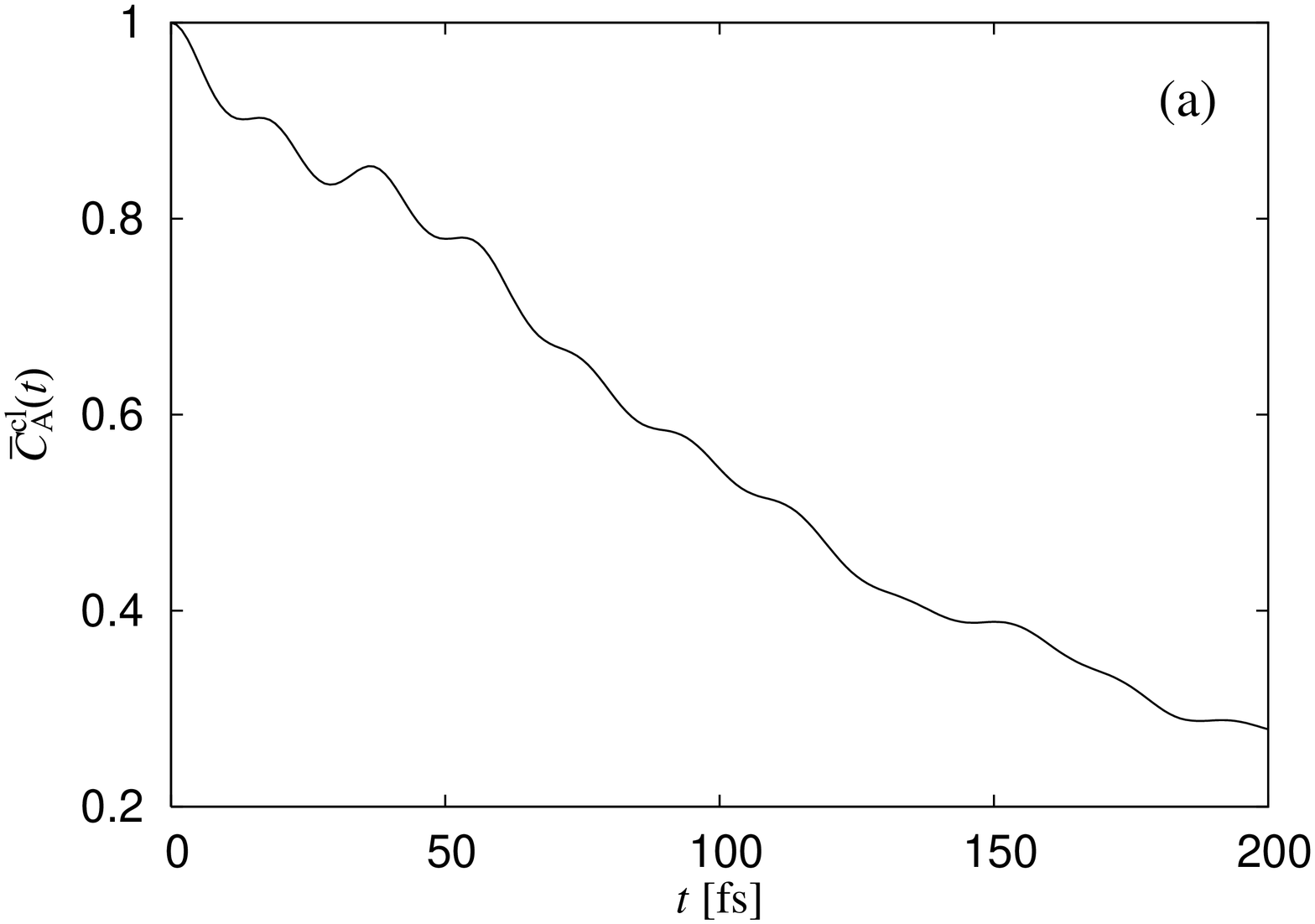}\\
\centering\includegraphics[width=3.8in, angle=-90]{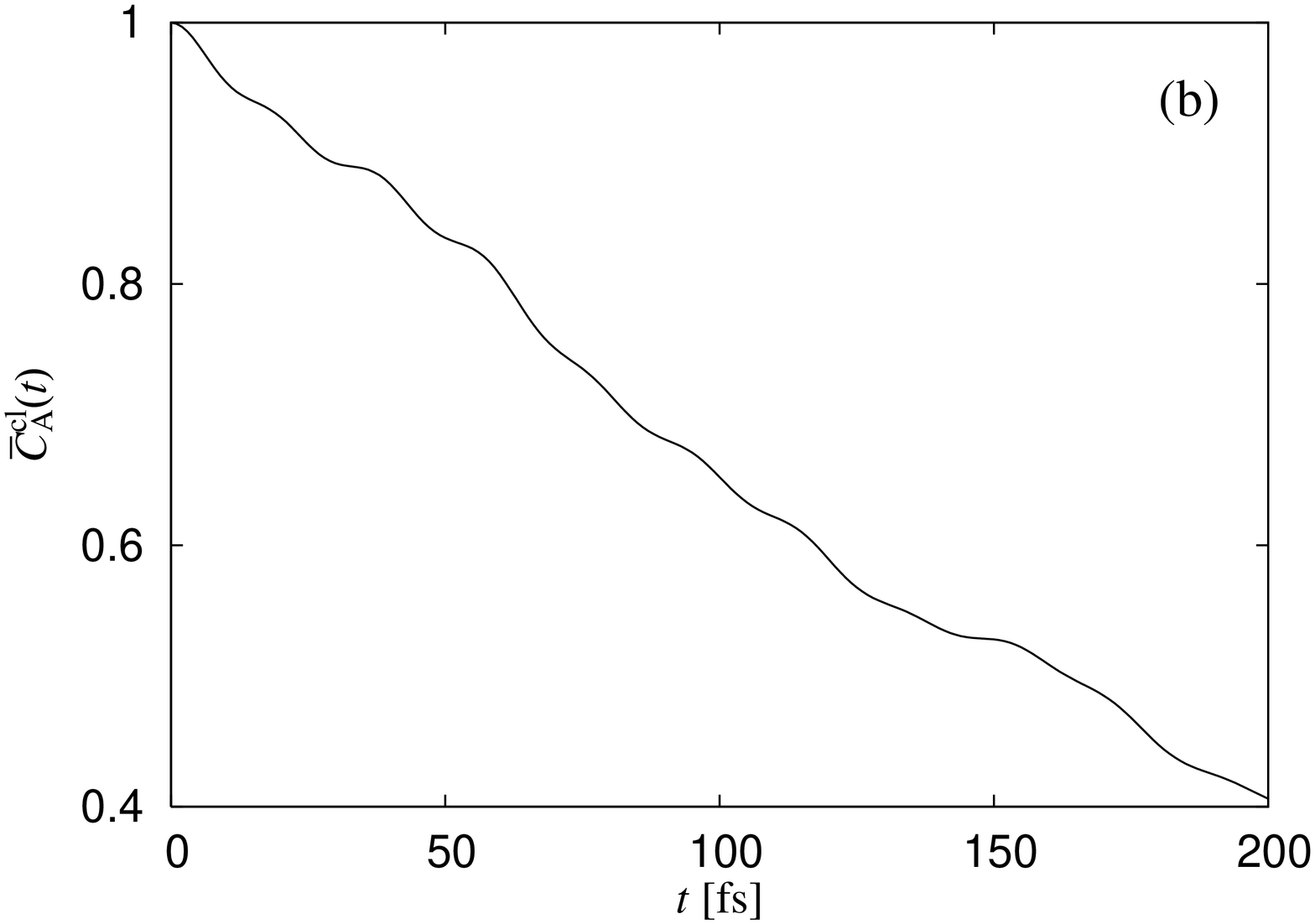}
\caption{\label{fig2}
$\bar{C}_\mathrm{A}^\mathrm{cl}(t)$ from
$T_\mathrm{AD}^\mathrm{pert}(t)$ for (a) the CS from the LUMO of
anthracene to the LUMO of TCNE in PhCN, and (b) the CR from the LUMO of
TCNE to the HOMO of naphthalene in PhCN.
}
\end{figure}

In Figs.~\ref{fig1} and \ref{fig2} we show the second-order perturbation
results for (a) the CS from the LUMO of anthracene to the LUMO of TCNE,
and (b) the CR from the LUMO of TCNE to the HOMO of naphthalene, which
are in the deep tunneling regime.

Figures \ref{fig1}(a) and (b) show the patterns of $T_\mathrm{AD}(t)$ in
second-order perturbation theory, denoted
$T_\mathrm{AD}^\mathrm{pert}(t)$, for portions of the MD trajectories
with anthracene and naphthalene, respectively, and the corresponding
time correlation functions $C_\mathrm{A}^\mathrm{cl}(t)$, normalized to their
initial values, are plotted in Figs.~\ref{fig2}(a) and (b).
We denote the normalized time correlation function by
$\bar{C}_\mathrm{A}^\mathrm{cl}(t) =
C_\mathrm{A}^\mathrm{cl}(t)/C_\mathrm{A}^\mathrm{cl}(0)$.

The effective correlation time $\tau_\mathrm{A}$ may be taken as the
time for which the function
$\bar{C}_\mathrm{A}^\mathrm{cl}(t) -
\bar{C}_\mathrm{A}^\mathrm{cl}(\infty)$
decays to $1/e$ of its initial value.
From Figs.~\ref{fig2}(a) and (b) we can see that $\tau_\mathrm{A}$ is
more than a few tens of femtoseconds in agreement with the previous
observations.\cite{Troisi_JACS04,Skourtis_PNAS05,Nishioka_JPCB05}
Using the value $\lambda = 0.902$ eV, from Table 2 of
Ref.~\onlinecite{Zimmt_JPCA03}, in Eq.~(\ref{tauFC}) yields
$\tau_\mathrm{FC} \approx 4.3$ fs.
Thus, $\tau_\mathrm{A}$ is too large to introduce significant dynamical
corrections to the ET rate for these processes.

\begin{figure}[!htbp]
\centering\includegraphics[width=3.8in, angle=-90]{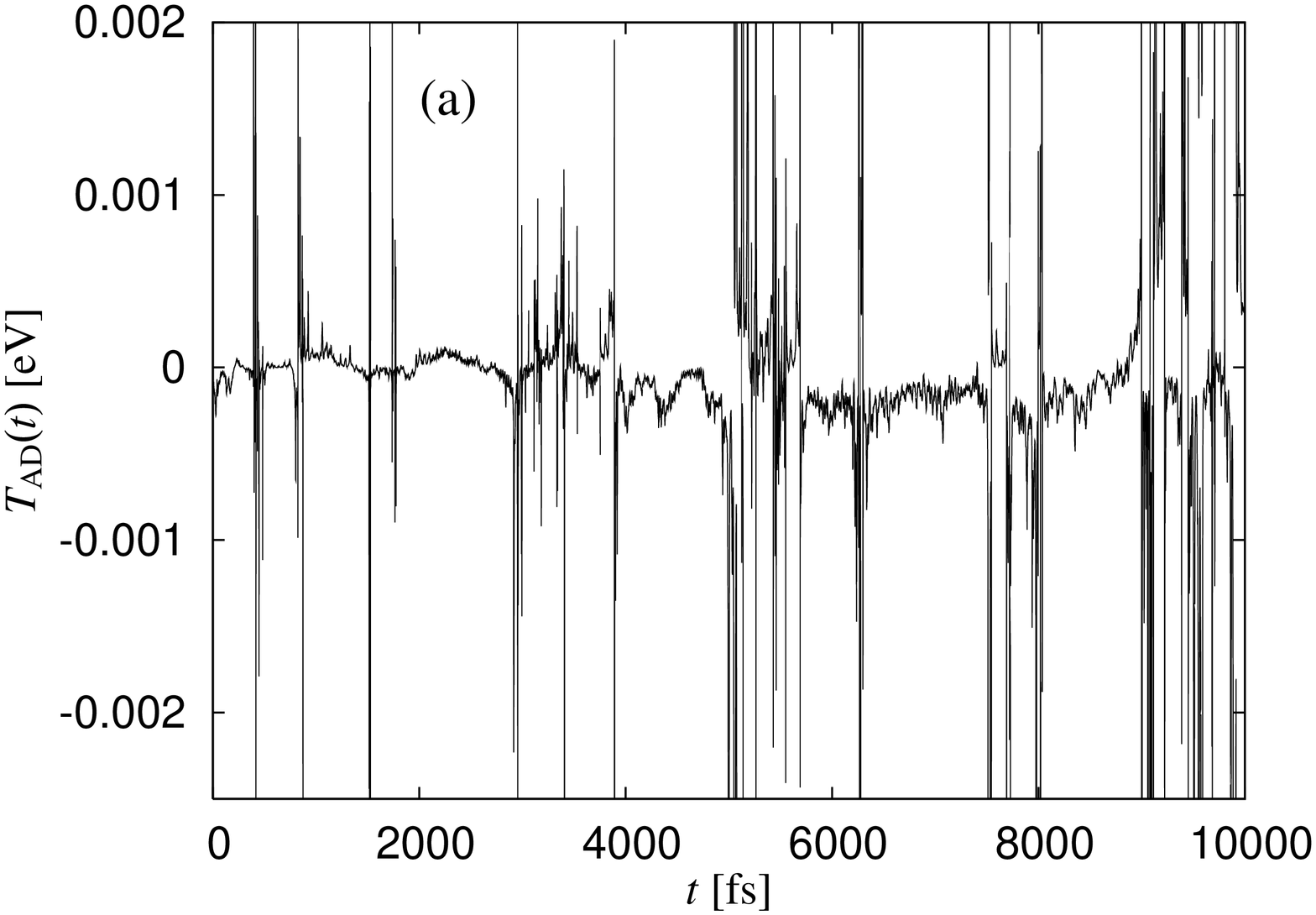}\\
\centering\includegraphics[width=3.8in, angle=-90]{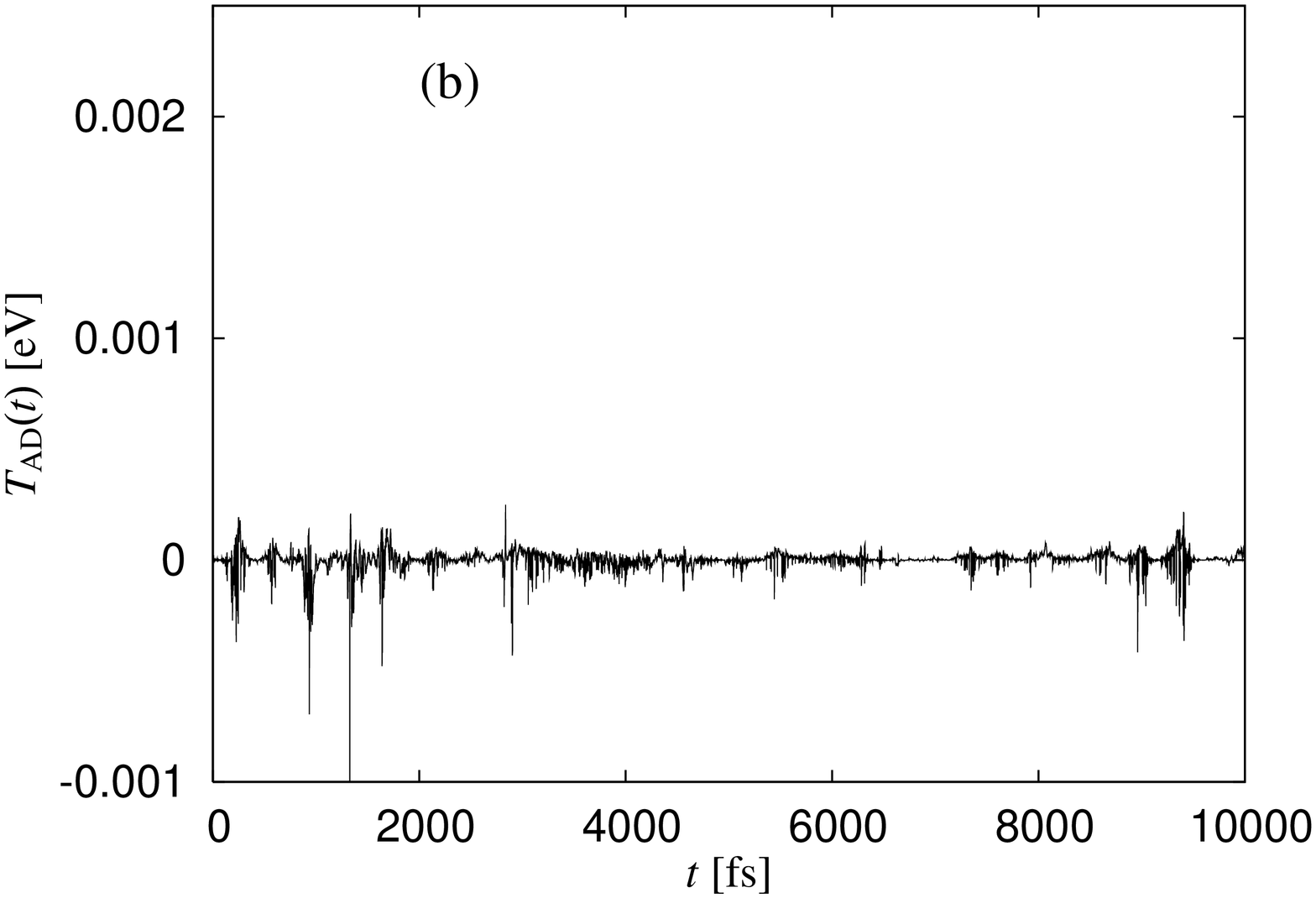}
\caption{\label{fig3}
$T_\mathrm{AD}(t)$ without level broadening for (a) the CS from the LUMO
of anthracene to the LUMO of TCNE in PhCN, and (b) the CR from the LUMO
of TCNE to the HOMO of naphthalene in PhCN.
}
\end{figure}

Figures \ref{fig3}(a) and (b) show the results of full $T$-matrix
calculations on the same systems.
Comparison of Figs.~\ref{fig1} and \ref{fig3} indicates that the full
$T$-matrix calculations gives a significant error, which mainly results
from the small difference between the initial electronic state energy
$E_i$ and the eigenenergy of the Hamiltonian $\hat{H}_\mathrm{S}$ in the
Green's function (\ref{GS}).
This error can be corrected by introducing the self-energy terms, which
describe the effect of the surroundings on the relevant system.
Thereby the free Green's function $\hat{G}_\mathrm{S}(E)$ in the
$T$-matrix is replaced by the effective reduced Green's function
(see Appendix \ref{A}).

The self-energy is a non-Hermitian matrix.
The anti-Hermitian component of the self-energy is responsible for the
broadening of the energy levels, while the Hermitian component can
conceptually be viewed as a correction to the system Hamiltonian
$\hat{H}_\mathrm{S}$.
For simplicity we ignore the mixing of the states of the relevant system
by the coupling to the surroundings.
In this case the self-energy becomes diagonal in the basis of
eigenstates of the system Hamiltonian $\hat{H}_\mathrm{S}$.
Furthermore, the energy dependence is disregarded, and all the diagonal
elements are set equal.
Consequently, the real part of the self-energy vanishes, and the
reservoir effect is characterized by a single level broadening
parameter, which we denote $\Gamma$.
Due to the level broadening term, $T_\mathrm{AD}$ is complex, and the
time correlation function $C_\mathrm{A}^\mathrm{cl}(t)$ is decomposed
into two autocorrelation functions of the real and imaginary parts of
$T_\mathrm{AD}$ as in Eq.~(\ref{CReclCImcl}).
Figures \ref{fig4} and \ref{fig5} show the consequences of including the
level broadening.

Figure \ref{fig4} summarizes a representative set of simulations of
$\mathrm{Re}T_\mathrm{AD}(t)$ (green line) and
$\mathrm{Im}T_\mathrm{AD}(t)$ (red line) for the CS from the LUMO of
anthracene to the LUMO of TCNE.
For (a), (b), (c), and (d), the level broadening is chosen as
(a) $\Gamma = 1$ eV, (b) $\Gamma = 0.1$ eV, (c) $\Gamma = 0.01$ eV,
(d) $\Gamma = 0.001$ eV.
For comparison $T_\mathrm{AD}^\mathrm{pert}(t)$ of Fig.~\ref{fig1} is
superimposed on each graph (black line).

Reasonably good agreement of the full $T$-matrix result with
$T_\mathrm{AD}^\mathrm{pert}(t)$ is obtained by choosing $\Gamma$ to be
$\simeq 0.01-0.1$ eV (Figs.~\ref{fig4}(b) and (c)), where the
contribution of $\mathrm{Im}T_\mathrm{AD}(t)$ is negligibly small
compared to that of $\mathrm{Re}T_\mathrm{AD}(t)$.
When $\Gamma$ is set larger or smaller (Figs.~\ref{fig4}(a) and (d)),
the contribution of $\mathrm{Im}T_\mathrm{AD}(t)$ becomes significant,
and the full $T$-matrix result apparently deviates from
$T_\mathrm{AD}^\mathrm{pert}(t)$.

We get similar behavior patterns for the CR from the LUMO of TCNE to the
HOMO of naphthalene as shown in Fig.~\ref{fig5}, though the deviation at
small level broadening becomes smaller due to the higher energy
separation between the D and B LUMO levels.

To analyze the origin of the deviation, we compare the full $T$-matrix
and the second-order perturbation results for the CS from the LUMO of
anthracene to the LUMO of TCNE, including the level broadening in both
calculations, in Fig.~\ref{fig6}.
The full $T$-matrix includes all possible intermediate
multiple-scattering processes and thus the D and A orbitals as
intermediate states.
On the other hand, the second-order perturbation theory exclude them
from the intermediate processes.
Consequently, the deviation of the full $T$-matrix result from the
second-order perturbation result arises due to the participation of the
D and A orbitals, particularly the initial electronic state, in the
intermediate processes.
In Fig.~\ref{fig6}, the gray line is the second-order perturbation
result, and the black line is the full $T$-matrix result.
Figures \ref{fig6}(c) and (d) clearly show that the deviation becomes
apparent when we use a smaller broadening parameter.
In Figs.~\ref{fig6}(a) and (b) both results are almost identical.
Similar results can be obtained for the CR from the LUMO of TCNE to the
HOMO of naphthalene (data not shown).
In conclusion, the deviation between the full $T$-matrix and the
second-order perturbation results at small level broadening can be
attributed to an underestimate of the level broadening for the initial
electronic state, while the deviation at large level broadening can be
attributed to an overestimate of the level broadening for the B states.
In the latter case significant differences do not appear between the
full $T$-matrix and the second-order perturbation calculations including
the level broadening because they both involve the B states for which
the large level broadening causes the error.

\section{\label{resonant}Resonant tunneling regime}
We now turn to the resonant tunneling regime.
In the previous section we have considered the CS from the LUMO of
anthracene to the LUMO of TCNE and the CR from the LUMO of TCNE to the
HOMO of naphthalene in PhCN solvent.
Here the LUMO energies of anthracene and TCNE and the HOMO energy of
naphthalene lie deep inside the solvent HOMO-LUMO gap.
To realize the resonant tunneling regime, we consider the CS from the
LUMO of naphthalene to the LUMO of TCNE in the same solvent, whereby the
LUMO energy of D is shifted into resonance with the solvent LUMO.
In the previous section we have seen that for $\Gamma \simeq 0.01-0.1$
eV we get the best results.
This estimation of the level broadening may also be used for the D, A,
and B states here, considering the similarities in chemical structure,
namely, cyano groups in TCNE and PhCN, and aromatic rings in anthracene,
naphthalene, and PhCN.
These similarities suggest similar interactions with the reservoir, and
thus we expect a similar level broadening.

Figure \ref{fig7} shows the patterns of $\mathrm{Re}T_\mathrm{AD}(t)$
(black line) and $\mathrm{Im}T_\mathrm{AD}(t)$ (gray line) with the
level broadening parameter chosen as above in (a), (b), (c), and (d),
respectively.
From this we can see that the $T$-matrix includes a substantial
contribution from the imaginary part.
In Fig.~\ref{fig8} the same data (black line) are shown in comparison
with the second-order perturbation results with the level broadening
(gray line).
This indicates that for the small level broadening the second-order
perturbation theory fails badly in the resonant tunneling regime.
In the following, we consider only the full $T$-matrix.
The black solid lines in Fig.~\ref{fig9} show the results for the
autocorrelation functions $C_\mathrm{Re}^\mathrm{cl}(t)$ and
$C_\mathrm{Im}^\mathrm{cl}(t)$, normalized to their initial values.
We denote the normalized autocorrelation functions by
$\bar{C}_\mathrm{Re}^\mathrm{cl}(t) =
C_\mathrm{Re}^\mathrm{cl}(t)/C_\mathrm{Re}^\mathrm{cl}(0)$ and
$\bar{C}_\mathrm{Im}^\mathrm{cl}(t) =
C_\mathrm{Im}^\mathrm{cl}(t)/C_\mathrm{Im}^\mathrm{cl}(0)$.
The level broadening parameter in (a)$-$(d) is chosen as above.
For (b), (c), and (d), $\bar{C}_\mathrm{Re}^\mathrm{cl}(t)$ exhibits
very fast initial decay and subsequent periodic damping with a period of
approximately $20$ fs.
In all cases, the oscillation frequencies are almost equal, while the
amplitudes of $\bar{C}_\mathrm{Re}^\mathrm{cl}(t)$ are larger than those
of $\bar{C}_\mathrm{Im}^\mathrm{cl}(t)$.
This feature can be understood by analyzing the dominant contributions
to $\bar{C}_\mathrm{Re}^\mathrm{cl}(t)$ and
$\bar{C}_\mathrm{Im}^\mathrm{cl}(t)$.
The gray solid lines in Fig.~\ref{fig9} show that for
$\bar{C}_\mathrm{Re}^\mathrm{cl}(t)$ and
$\bar{C}_\mathrm{Im}^\mathrm{cl}(t)$ in (b) and (c) and
$\bar{C}_\mathrm{Im}^\mathrm{cl}(t)$ in (d) the dominant contribution
comes from the two adjacent eigenstates of the system Hamiltonian
$\hat{H}_\mathrm{S}$, which are closest to the LUMO of D (naphthalene),
namely the initial electronic state $| \phi_i \rangle$, and the LUMO of
B (PhCN solvent).
The deviations in (a) can be explained by the large contributions from
the other eigenstates due to the large level broadening.
On the other hand, the deviation from
$\bar{C}_\mathrm{Re}^\mathrm{cl}(t)$ in (d) is more subtle.
We will return to this issue later on.

We denote the two adjacent eigenstates by $| \psi_+ \rangle$ and
$| \psi_- \rangle$ with the higher eigenenergy $E_+$ and the lower
eigenenergy $E_-$, respectively, their contribution to
$T_\mathrm{AD}(t)$ by $T_\mathrm{AD}^\mathrm{2s}(t)$, and the
contributions to $\bar{C}_\mathrm{Re}^\mathrm{cl}(t)$ and
$\bar{C}_\mathrm{Im}^\mathrm{cl}(t)$ by
$\bar{C}_\mathrm{Re}^\mathrm{cl2s}(t)$ and
$\bar{C}_\mathrm{Im}^\mathrm{cl2s}(t)$, respectively.
The two states $| \psi_+ \rangle$ and $| \psi_- \rangle$ alternately
fluctuate back and forth between the LUMOs of D and B, which are nearly
degenerate.
To demonstrate this more clearly, we consider a two-state system whose
Hamiltonian is written as a sum
$\hat{H}_\mathrm{S} = \hat{H}_0^\mathrm{2s} + \hat{V}$.
Here we assume that the eigenfunctions of $\hat{H}_0^\mathrm{2s}$ are
given by the initial electronic state $| \phi_i \rangle$ and the solvent
LUMO, denoted by $| \phi_\mathrm{B} \rangle$, with the corresponding
eigenvalues $E_i$ and $\epsilon_\mathrm{B}$, respectively.
These states are coupled by the perturbation $\hat{V}$.
For simplicity we assume that the two-state system is unaffected by the
coupling between the B and A orbitals $| \phi_\mathrm{B} \rangle$ and
$| \phi_\mathrm{A} \rangle$, although the perturbation $\hat{V}$
includes its contribution.
In this case, in the basis of the functions $| \phi_i \rangle$ and
$| \phi_\mathrm{B} \rangle$, $\hat{H}_\mathrm{S}$ is represented by the
matrix
\begin{equation}
 \label{2states}
 \hat{H}_\mathrm{S} =
 \begin{pmatrix}
            E_i&    V\, e^{- i \eta}\\
 V\, e^{i \eta}& \epsilon_\mathrm{B}
 \end{pmatrix}
 ,
\end{equation}
where we have denoted
$V e^{i\eta} = \langle \phi_\mathrm{B} | \hat{V} | \phi_i \rangle$ with
$V$ taken real and positive and the phase factor $e^{i\eta}$.
The eigenstates and eigenvalues of the Hamiltonian $\hat{H}_\mathrm{S}$
are denoted $| \tilde{\psi}_+ \rangle$ and $| \tilde{\psi}_- \rangle$,
and $\tilde{E}_+$ and $\tilde{E}_-$, respectively.
The eigenvalues are given by
\begin{equation}
 \label{secularSOL}
 \tilde{E}_\pm = \frac{E_i + \epsilon_\mathrm{B} \pm
 \sqrt{(E_i - \epsilon_\mathrm{B})^2 + 4 V^2}}{2}.
\end{equation}
Thus, $| \tilde{\psi}_+ \rangle$ remains above the initial electronic
state and the solvent LUMO while $| \tilde{\psi}_- \rangle$ remains
below them.
The eigenstates are expressed by
\begin{align}
 \label{Pstate}
 &| \tilde{\psi}_+ \rangle =
 \cos\theta\, e^{- i \eta / 2} | \phi_i \rangle
 + \sin\theta\, e^{i \eta / 2} | \phi_\mathrm{B} \rangle,
 \\
 \label{Mstate}
 &| \tilde{\psi}_- \rangle =
 - \sin\theta\, e^{- i \eta / 2} | \phi_i \rangle
 + \cos\theta\, e^{i \eta / 2} | \phi_\mathrm{B} \rangle,
\end{align}
where
\begin{align}
 \label{sin}
 &\sin\theta =
 \frac{X}{(1 + X^2)^{1/2}},
 \\
 \label{cos}
 &\cos\theta =
 \frac{1}{(1 + X^2)^{1/2}},
 \\
 \label{X}
 &X \equiv \frac{\tilde{E}_+ - E_i}{V}
 = \frac{V}{\tilde{E}_+ - \epsilon_\mathrm{B}}
 = \frac{V}{E_i - \tilde{E}_-}.
\end{align}
Here $V$ and the eigenstates and eigenvalues of $\hat{H}_0^\mathrm{2s}$
and $\hat{H}_\mathrm{S}$ depend on time via the solvent nuclear
coordinates.
Figure \ref{fig10} shows that $\tilde{E}_+$ (dark gray solid line) and
$\tilde{E}_-$ (gray solid line) are very good approximations to the
exact eigenenergies $E_+$ (black dashed line) and $E_-$
(black dotted line), respectively, and Fig.~\ref{fig11} shows that
$\sin^2\theta$ (dark gray solid line) and $\cos^2\theta$
(gray solid line) reproduce the population exchange of
$| \phi_i \rangle$ and $| \phi_\mathrm{B} \rangle$ between the exact
eigenstates $| \psi_+ \rangle$ and $| \psi_- \rangle$, described by
$|\langle \phi_i | \psi_- \rangle|^2$ (black dashed line) and
$|\langle \phi_i | \psi_+ \rangle|^2$ (black dotted line) or
$|\langle \phi_\mathrm{B} | \psi_+ \rangle|^2$ (black dashed line) and
$|\langle \phi_\mathrm{B} | \psi_- \rangle|^2$ (black dotted line).
These results allow us to approximate $T_\mathrm{AD}^\mathrm{2s}$ by
\begin{align}
 \label{TAD2sPM}
 T_\mathrm{AD}^\mathrm{2s}
 \approx\,
 &\langle \phi_\mathrm{A} | \hat{V} | \tilde{\psi}_+ \rangle
 \frac{1}{E_i - \tilde{E}_+ + i \Gamma / 2}
 \langle \tilde{\psi}_+ | \hat{V} | \phi_\mathrm{D} \rangle
 \nonumber
 \\
 &+ \langle \phi_\mathrm{A} | \hat{V} | \tilde{\psi}_- \rangle
 \frac{1}{E_i - \tilde{E}_- + i \Gamma / 2}
 \langle \tilde{\psi}_- | \hat{V} | \phi_\mathrm{D} \rangle.
\end{align}
Since by definition
$\langle \phi_i | \hat{V} | \phi_\mathrm{D} \rangle = 0$, we have
\begin{equation}
 \label{TAD2s}
 T_\mathrm{AD}^\mathrm{2s} = \mathrm{Re}T_\mathrm{AD}^\mathrm{2s}
 + i \mathrm{Im}T_\mathrm{AD}^\mathrm{2s},
\end{equation}
with
\begin{align}
 \label{ReTAD2s}
 \mathrm{Re}T_\mathrm{AD}^\mathrm{2s}
 &\approx \left[
 \frac{(E_i - \tilde{E}_+) \sin^2\theta}
 {(E_i - \tilde{E}_+)^2 + (\Gamma / 2)^2} +
 \frac{(E_i - \tilde{E}_-) \cos^2\theta}
 {(E_i - \tilde{E}_-)^2 + (\Gamma / 2)^2}
 \right]
 \langle \phi_\mathrm{A} | \hat{V} | \phi_\mathrm{B} \rangle
 \langle \phi_\mathrm{B} | \hat{V} | \phi_\mathrm{D} \rangle
 \nonumber
 \\
 &\equiv R_+ + R_-,
\end{align}
and with
\begin{align}
 \label{ImTAD2s}
 \mathrm{Im}T_\mathrm{AD}^\mathrm{2s}
 &\approx - \left[
 \frac{(\Gamma / 2) \sin^2\theta}
 {(E_i - \tilde{E}_+)^2 + (\Gamma / 2)^2} +
 \frac{(\Gamma / 2) \cos^2\theta}
 {(E_i - \tilde{E}_-)^2 + (\Gamma / 2)^2}
 \right]
 \langle \phi_\mathrm{A} | \hat{V} | \phi_\mathrm{B} \rangle
 \langle \phi_\mathrm{B} | \hat{V} | \phi_\mathrm{D} \rangle
 \nonumber
 \\
 &\equiv I_+ + I_-.
\end{align}
Here, $R_+$ and $R_-$ denote the contributions of the states
$| \tilde{\psi}_+ \rangle$ and $| \tilde{\psi}_- \rangle$, respectively,
to the real part $\mathrm{Re}T_\mathrm{AD}^\mathrm{2s}$, and $I_+$ and
$I_-$ denote the contributions of the states $| \tilde{\psi}_+ \rangle$
and $| \tilde{\psi}_- \rangle$, respectively, to the imaginary part
$\mathrm{Im}T_\mathrm{AD}^\mathrm{2s}$.
Since $\tilde{E}_+ > E_i > \tilde{E}_-$, $R_+$ and $R_-$ have opposite
signs while $I_+$ and $I_-$ have the same sign.
The black dashed lines in Fig.~\ref{fig9} show this approximation
results for $\bar{C}_\mathrm{Re}^\mathrm{cl2s}(t)$ and
$\bar{C}_\mathrm{Im}^\mathrm{cl2s}(t)$.
These plots show very good agreement with the exact results
(gray solid lines) except for (d).
The deviations are exposed due to the small level broadening.
According to Eqs.~(\ref{secularSOL}), (\ref{ReTAD2s}), and
(\ref{ImTAD2s}), $| \tilde{\psi}_+ \rangle$ and
$| \tilde{\psi}_- \rangle$, which are nearly identical to
$| \psi_+ \rangle$ and $| \psi_- \rangle$, respectively, contribute to
$\mathrm{Re}T_\mathrm{AD}^\mathrm{2s}$ destructively, while they
contribute to $\mathrm{Im}T_\mathrm{AD}^\mathrm{2s}$ constructively.
Thus, $\mathrm{Re}T_\mathrm{AD}$ is more influenced by the other states
than $\mathrm{Im}T_\mathrm{AD}$.
This is the reason why the deviation of
$\bar{C}_\mathrm{Re}^\mathrm{cl2s}(t)$ from
$\bar{C}_\mathrm{Re}^\mathrm{cl}(t)$ is much larger than that of
$\bar{C}_\mathrm{Im}^\mathrm{cl2s}(t)$ from
$\bar{C}_\mathrm{Im}^\mathrm{cl}(t)$.

For the small level broadenings $\Gamma \simeq 0.001-0.1$ eV, the
timescale for the initial decay of $\bar{C}_\mathrm{Re}^\mathrm{cl}(t)$
reaches several femtoseconds and thus may be fast enough to introduce
significant dynamical corrections to the ET rate expressions.
On the other hand, if the level broadening is as large as $1$ eV, the
dynamical non-Condon effects will be insignificant.
The rapid initial decay of $\bar{C}_\mathrm{Re}^\mathrm{cl}(t)$ is
followed by the periodic behavior with much the same period.
The fluctuations of $\mathrm{Re}T_\mathrm{AD}$ and
$\mathrm{Im}T_\mathrm{AD}$ are caused by the solvent motions.
From the timescale viewpoint, we infer that the periodic behaviors of
the autocorrelation functions are caused by bond stretching vibrations
of PhCN.
The higher frequency components in the initial decay should be created
by the coupling between various vibrational modes.

The ET rate can be calculated using Eq.~(\ref{kAD5}).
For
$\tilde{C}_\mathrm{A}^\mathrm{cl}(\epsilon / \hslash)
/ C_\mathrm{A}^\mathrm{cl}(0)$
we use the Fourier transforms of $\bar{C}_\mathrm{Re}^\mathrm{cl}(t)$
and $\bar{C}_\mathrm{Im}^\mathrm{cl}(t)$ in Fig.~\ref{fig9}.
Figure \ref{fig10} shows the energy gap dependence of
$k_{\mathrm{A} \leftarrow \mathrm{D}}$ for the level broadening chosen
as above in (a), (b), (c), and (d), respectively.
The gray solid, black solid, and black dashed lines are the results for
the quantum correction factors of Eqs.~(\ref{Qst}), (\ref{Qh}), and
(\ref{Qsc}), respectively.
As expected, for the small level broadenings $\Gamma \simeq 0.001-0.1$
eV, the dynamical non-Condon effects cause the substantial enhancement
of the ET rate in the activated inverted region.
This enhancement becomes larger with decreasing level broadening.
It also depends on the quantum correction factor used.
The quantum correction factor of Eq.~(\ref{Qh}) generates larger
enhancement than that of Eq.~(\ref{Qst}).
On the other hand, that of Eq.~(\ref{Qsc}) causes extraordinarily large
enhancement of the ET rate regardless of the level broadening because
this approximation badly overestimates the value of
$\tilde{C}_\mathrm{A}(\omega)$ for high frequencies.

\section{\label{con}Conclusion}
We have investigated the significance of the dynamical non-Condon effect
on the ET rate in the coherent resonant tunneling regime.
The time-dependent Fermi's golden rule expression for the ET rate has
been generalized to deal with the resonance in the $T$-matrix framework.
The dynamical effect is thereby expressed in terms of the time
correlation function of the non-Hermitian $T$-matrix.
This time correlation function can be separated into the autocorrelation
and the cross-correlation part.
The former satisfies detailed balance, whereas the latter does not
satisfy it.
When the shortest timescale for the change of the time correlation
function is comparable to the decay time of the time-dependent
Franck-Condon factor, the dynamical non-Condon effects become
significant.

In the semiclassical approximation the Fourier transform of the quantum
time correlation function is replaced by the classical time correlation
function multiplied by the quantum correction factor.
In this case the cross-correlation part is disregarded.
The classical time correlation function has been evaluated using the
combined MD/QC simulations.
As an example of the resonant tunneling we have considered the ET from
the LUMO of naphthalene to the LUMO of TCNE in PhCN.
For simplicity the D and A molecules were held fixed during the
simulations.
Experimentally, similar situations can be created by linked D-A systems
such as C-clamp molecules.

We have introduced the level broadening parameter to include the effect
of the surroundings on the relevant system.
Disregarding the level broadening leads to unacceptable results even in
the deep tunneling regime.
The parameter values have been estimated for the CS from the LUMO of
anthracene to the LUMO of TCNE and the CR from the LUMO of TCNE to the
HOMO of naphthalene in PhCN, which are in the deep tunneling regime.
Considering the similarities in chemical structure, we expect a similar
level broadening for the above-mentioned resonant tunneling system.
By comparing the full $T$-matrix results with the second-order
perturbation results, the best parameter values are found to be
$\Gamma \simeq 0.01-0.1$ eV.
For such level broadenings, the timescale for the initial decay of the
time correlation function reaches several femtoseconds, and the ET rate
can be significantly enhanced by the dynamical non-Condon effects.

\appendix\section{}\label{A}
We consider a Hamiltonian written as a sum
\begin{equation}
 \label{H}
 \hat{H} = \hat{H}_\mathrm{S} + \hat{H}_\mathrm{R} +
 \hat{V}_\mathrm{SR},
\end{equation}
where $\hat{H}_\mathrm{S}$ is the Hamiltonian of the relevant system,
given by the Fock matrix (\ref{FOCK}), $\hat{H}_\mathrm{R}$ is the
Hamiltonian of the surroundings, and $\hat{V}_\mathrm{SR}$ is the
interaction between them.
Let $\{ | s \rangle \}$ and $\{ | r \rangle \}$ be the orthonormal sets
of eigenstates of $\hat{H}_\mathrm{S}$ and $\hat{H}_\mathrm{R}$,
respectively.
Then $\hat{H}_\mathrm{S}$, $\hat{H}_\mathrm{R}$, and
$\hat{V}_\mathrm{SR}$ may be written in the representation as
\begin{align}
 \label{HSA}
 &\hat{H}_\mathrm{S} = \sum_s E_s | s \rangle \langle s |,
 \\
 \label{HR}
 &\hat{H}_\mathrm{R} = \sum_r E_r | r \rangle \langle r |,
 \\
 \label{VSR}
 &\hat{V}_\mathrm{SR} = \sum_{s,r} (V_{sr} | s \rangle \langle r | +
 V_{rs} | r \rangle \langle s |),
\end{align}
where we denote $V_{sr} = \langle s | \hat{V}_\mathrm{SR} | r \rangle$,
etc.
The retarded Green's function of the total Hamiltonian is given by
\begin{equation}
 \label{G}
 \hat{G}(E) = \frac{1}{E - \hat{H} + i 0^+}.
\end{equation}

We focus on the relevant subspace of the total system and attempt to
characterize the effect of the surroundings on the relevant system.
The free retarded Green's function of the relevant system is given by
Eq.~(\ref{GS}).
Then the Dyson equation is expressed as
\begin{equation}
 \label{DYSON}
 \hat{G}(E) = \hat{G}_\mathrm{S}(E) +
 \hat{G}_\mathrm{S}(E)\, \hat{V}_\mathrm{SR}\, \hat{G}(E).
\end{equation}
Using the resolution of the identity operator
\begin{equation}
 \label{identity}
 \sum_s | s \rangle \langle s | + \sum_r | r \rangle \langle r | = 1
\end{equation}
leads to
\begin{align}
 \label{DGs's}
 &G_{s's}(E) = (G_\mathrm{S})_{ss}(E)\, \delta_{s',s} +
 (G_\mathrm{S})_{s's'}(E) \sum_r V_{s'r} G_{rs}(E),
 \\
 \label{DGrs}
 &G_{rs}(E) = (G_\mathrm{S})_{rr}(E) \sum_{s''} V_{rs''} G_{s''s}(E),
\end{align}
where
$G_{s's}(E) = \langle s' | \hat{G}(E) | s \rangle$,
$(G_\mathrm{S})_{ss}(E) =
\langle s | \hat{G}_\mathrm{S}(E) | s \rangle$,
etc.
Inserting Eq.~(\ref{DGrs}) into Eq.~(\ref{DGs's}) it follows that
\begin{equation}
 \label{Gs''s}
 \sum_{s''} \left[(E - E_{s'} + i 0^+) \delta_{s',s''} -
 B_{s's''}(E)\right] G_{s''s}(E) = \delta_{s',s},
\end{equation}
where $B_{s's''}(E)$ are the self-energy matrix elements given by
\begin{equation}
 \label{Bs's''}
 B_{s's''}(E) = \sum_r \frac{V_{s'r} V_{rs''}}{E - E_r + i 0^+}.
\end{equation}
This defines the effective reduced Green's function for the relevant
subspace.
Note that the self-energy matrix is non-Hermitian due to the
infinitesimal term $i 0^+$.
For simplicity we assume that the mixing of the states $| s \rangle$ by
the coupling to the reservoir can be disregarded, namely we employ the
diagonal approximation to the self-energy matrix.
Thereby the self-energy matrix elements read
\begin{equation}
 \label{Bss'}
 B_{ss'}(E) = B_s(E)\, \delta_{s,s'},
\end{equation}
with
\begin{equation}
 \label{Bs1}
 B_s(E) = \sum_r \frac{|V_{sr}|^2}{E - E_r + i 0^+},
\end{equation}
and thus the reduced Green's function becomes a diagonal matrix, that
is,
\begin{equation}
 \label{Gs's}
 G_{s's}(E) = \frac{1}{(E - E_s + i 0^+) - B_s(E)} \delta_{s',s}.
\end{equation}
The imaginary part of the self-energy is equal to the broadening of the
density of states in the relevant system, while the real part
corresponds to the shifts in the energy levels $E_s$.

Furthermore, we assume that the set of states $| r \rangle$ constitutes
a continuum of states.
In this case the summation over $r$ can be replaced by the integral
\begin{equation}
 \label{int}
 \sum_r \rightarrow \int_{-\infty}^\infty dE_r \rho_\mathrm{R}(E_r),
\end{equation}
where $\rho_\mathrm{R}(E)$ denotes the reservoir density of states.
Equation (\ref{Bs1}) now takes the form
\begin{equation}
 \label{Bs2}
 B_s(E) = \int_{-\infty}^\infty dE_r
 \frac{\overline{(|V_{sr}|^2)_{E_r}}\,
 \rho_\mathrm{R}(E_r)}{E - E_r + i 0^+}
 = \frac{1}{2 \pi} \int_{-\infty}^\infty dE_r
 \frac{\Gamma_s(E_r)}{E - E_r + i 0^+},
\end{equation}
where $\overline{(|V_{sr}|^2)_E}$ is the average of the squared
coupling over all continuum levels $r$ that have energy $E$, defined by
$\overline{(|V_{sr}|^2)_E} =
\sum_r |V_{sr}|^2 \delta(E - E_r) / \sum_r \delta(E - E_r) =
(\rho_\mathrm{R}(E))^{-1} \sum_r |V_{sr}|^2 \delta(E - E_r)$,
and where
\begin{equation}
 \label{GAMMA}
 \Gamma_s(E) \equiv
 2 \pi\, \overline{(|V_{sr}|^2)_E}\, \rho_\mathrm{R}(E).
\end{equation}
In the simplest case where $\Gamma_s(E)$ does not depend on $E$ and $s$,
we get
\begin{equation}
 \label{Bs3}
 B_s(E) = \frac{\Gamma}{2 \pi} \int_{-\infty}^\infty dE_r
 \frac{1}{E - E_r + i 0^+}
 = - i \frac{\Gamma}{2}.
\end{equation}
In this case the effective reduced Green's function can be expressed in
the form
\begin{equation}
 \label{Geff}
 \hat{G}(E) = \frac{1}{E - \hat{H}_\mathrm{S} + i \Gamma / 2}.
\end{equation}
Note that the infinitesimal term $i 0^+$ in Eq.~(\ref{Gs's}) can be
disregarded relative to $i \Gamma / 2$.

\newpage

\newpage

\newpage

\newpage

\begin{figure}[!htbp]
\centering\includegraphics[width=3.8in, angle=-90]{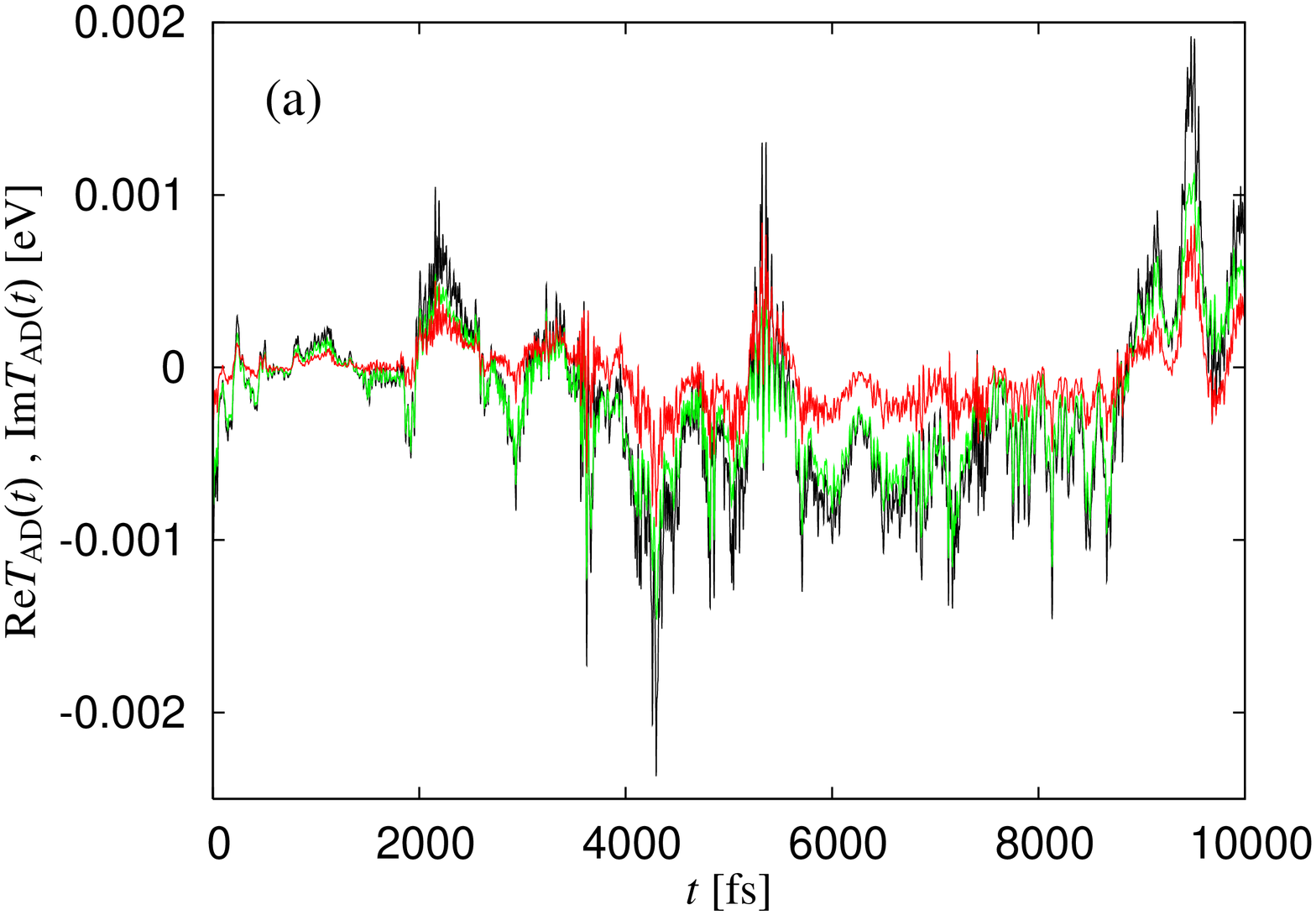}\\
\centering\includegraphics[width=3.8in, angle=-90]{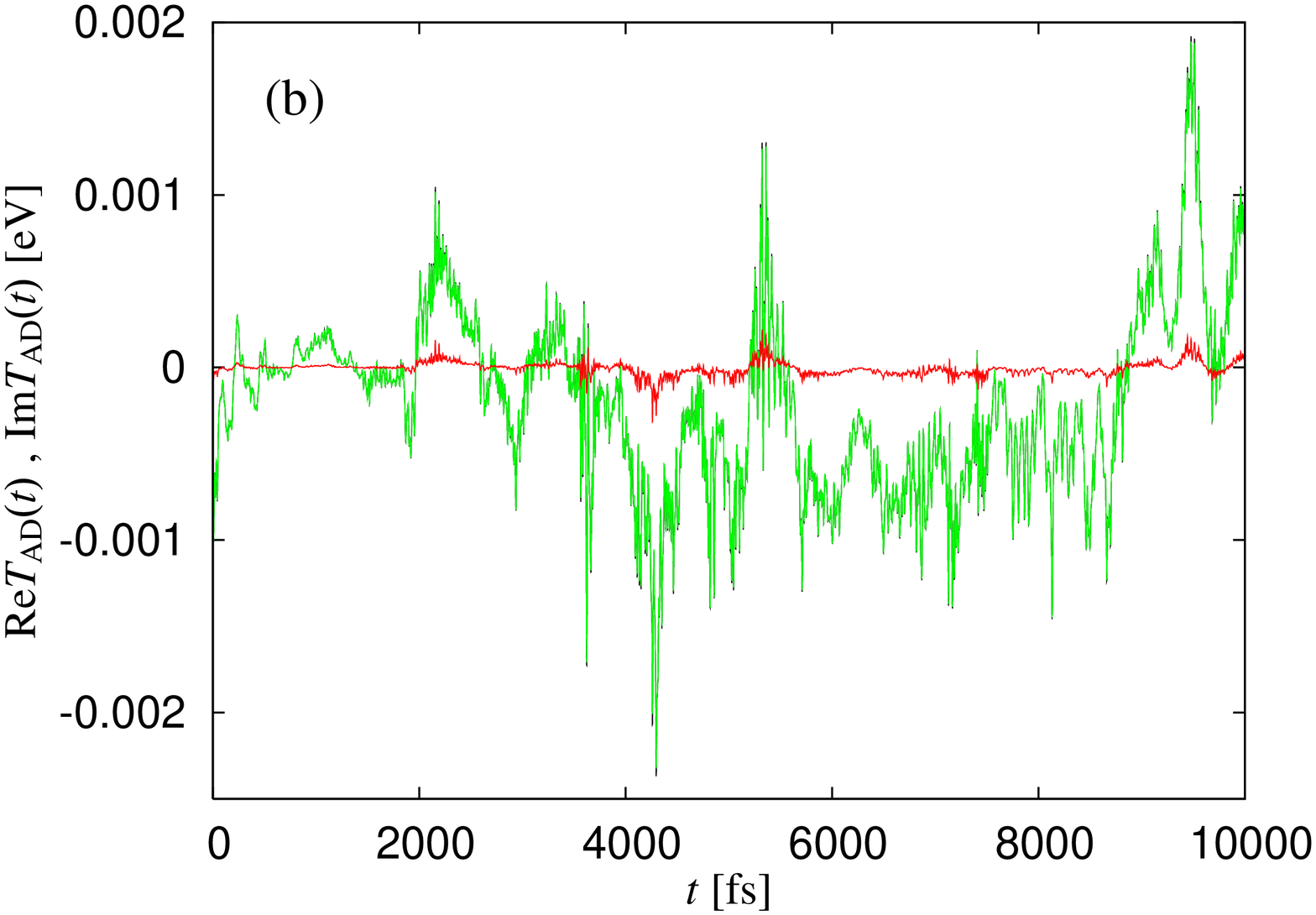}
\end{figure}

\newpage

\begin{figure}[!htbp]
\centering\includegraphics[width=3.8in, angle=-90]{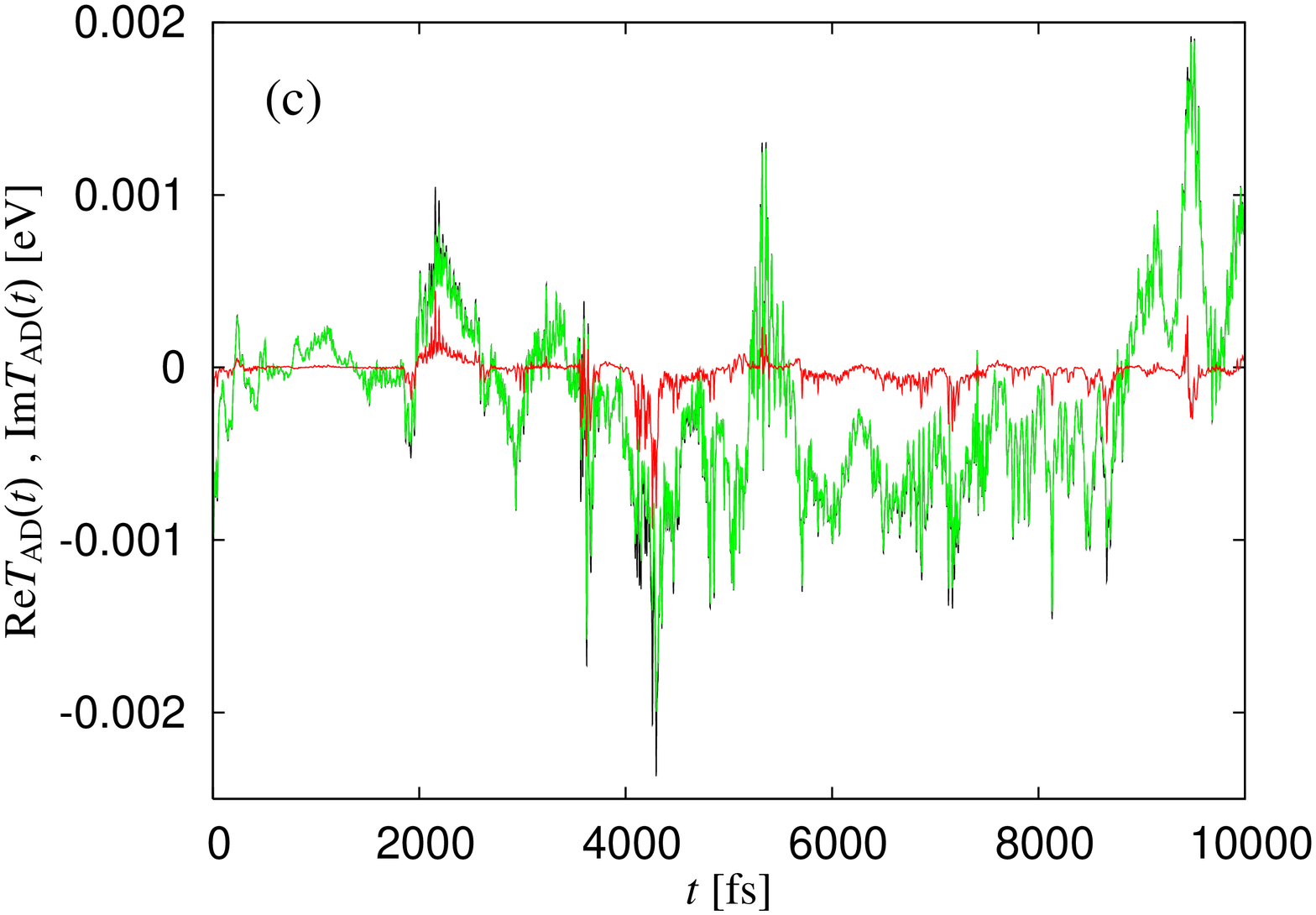}\\
\centering\includegraphics[width=3.8in, angle=-90]{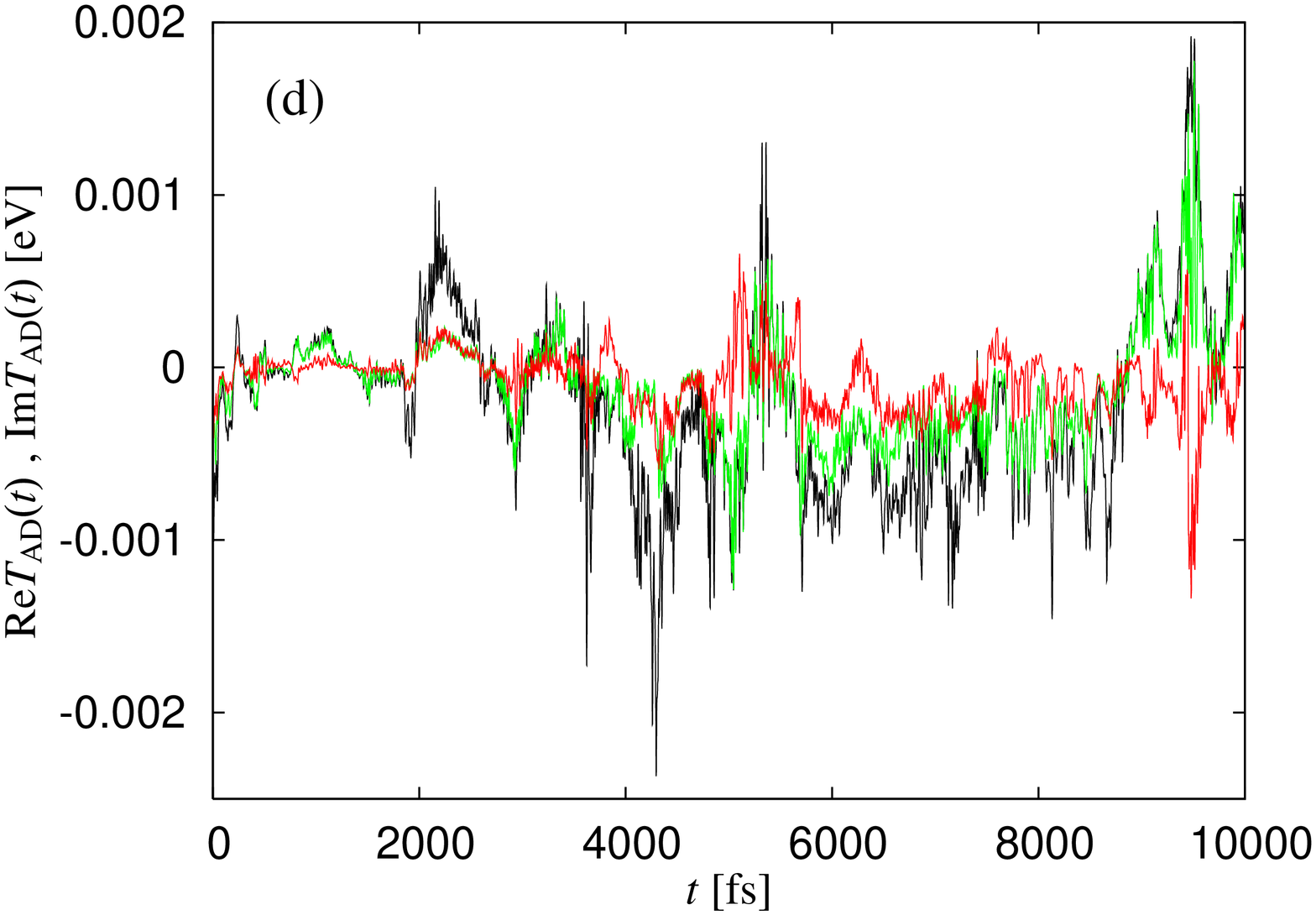}
\caption{\label{fig4}
$\mathrm{Re}T_\mathrm{AD}(t)$ (green line),
$\mathrm{Im}T_\mathrm{AD}(t)$ (red line) with level broadening, and
$T_\mathrm{AD}^\mathrm{pert}(t)$ (black line) without level broadening
for the CS from the LUMO of anthracene to the LUMO of TCNE in PhCN.
The level broadening is chosen as
(a) $\Gamma = 1$ eV, (b) $\Gamma = 0.1$ eV, (c) $\Gamma = 0.01$ eV,
(d) $\Gamma = 0.001$ eV.
}
\end{figure}

\newpage

\begin{figure}[!htbp]
\centering\includegraphics[width=3.8in, angle=-90]{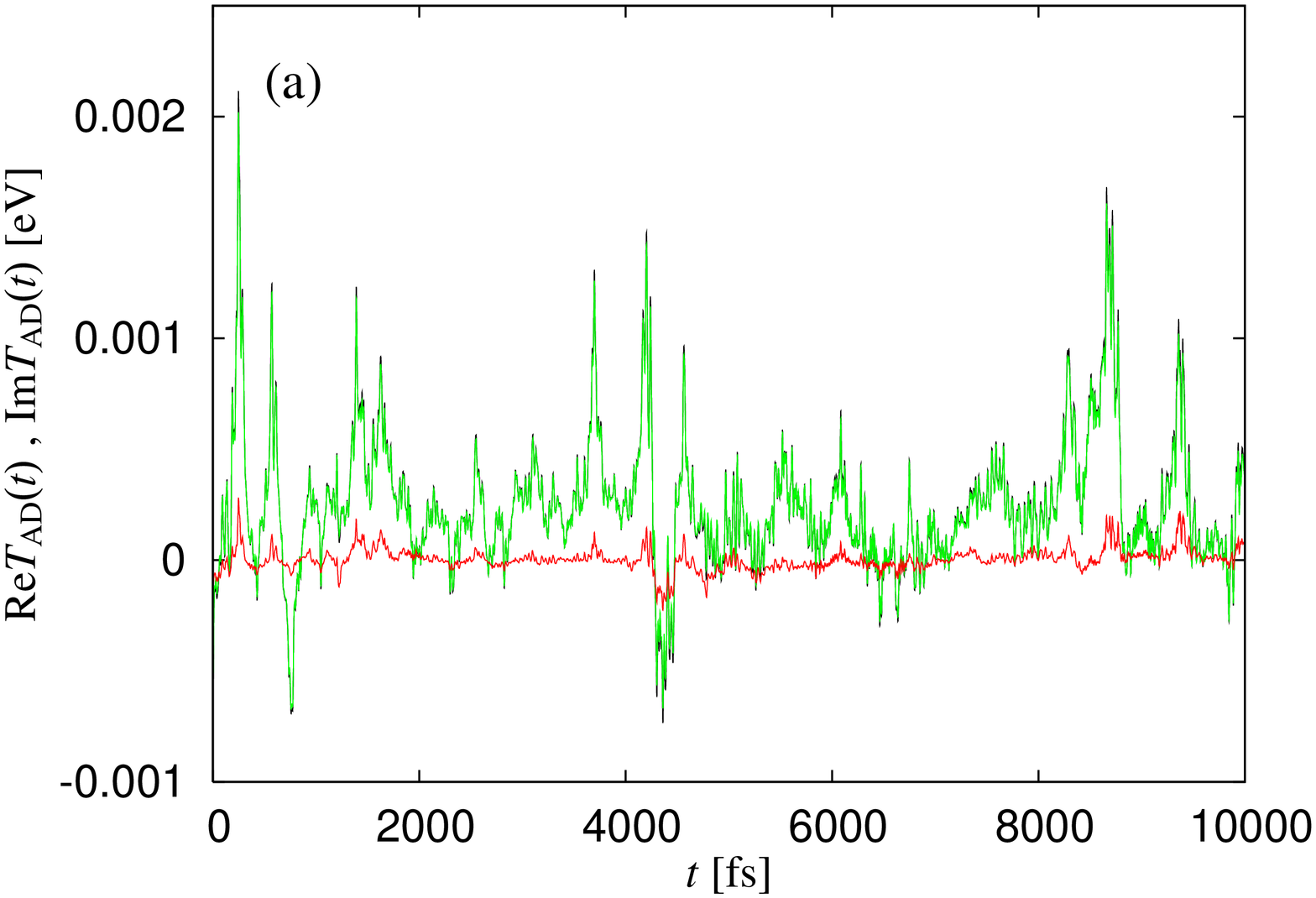}\\
\centering\includegraphics[width=3.8in, angle=-90]{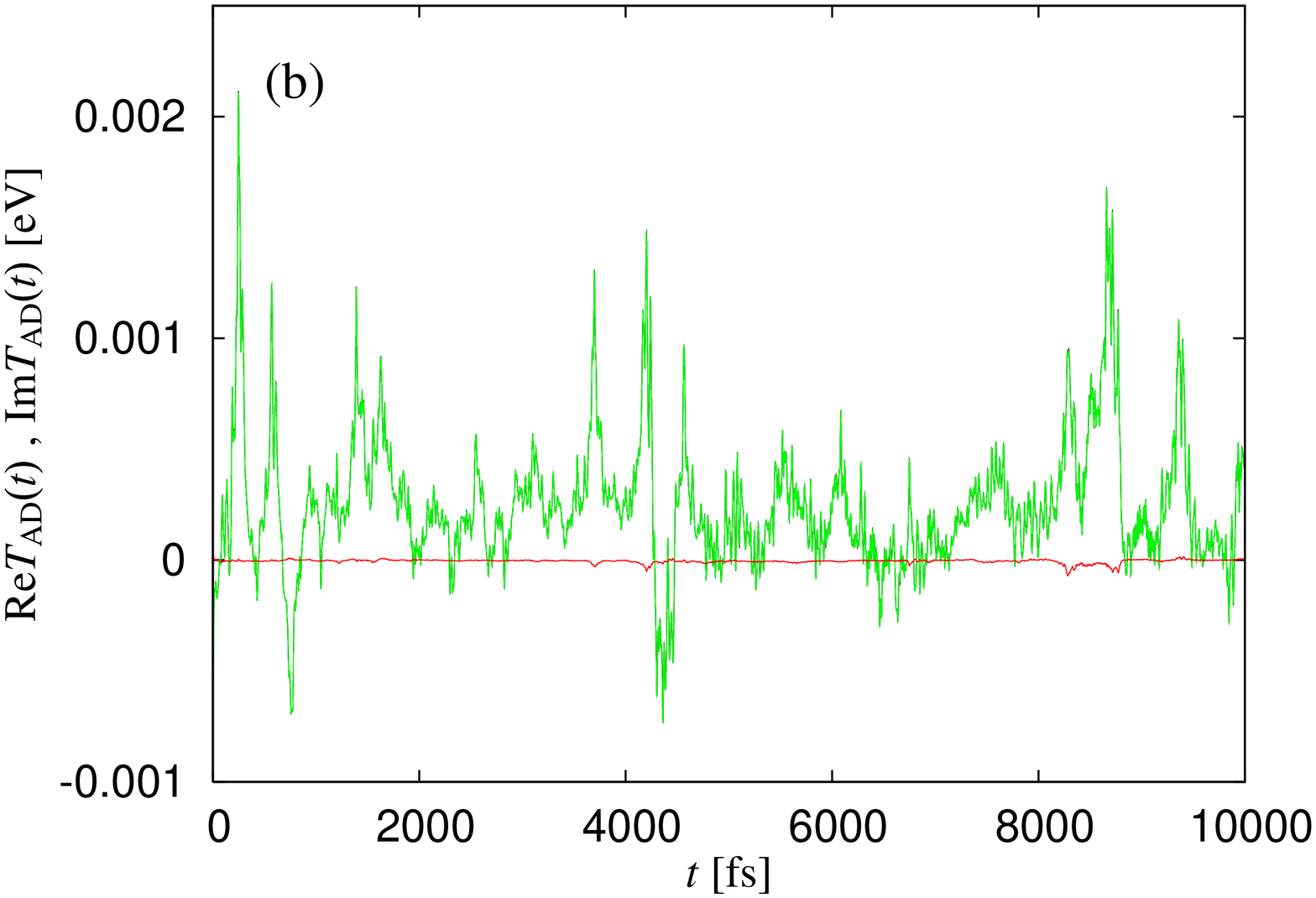}
\end{figure}

\newpage

\begin{figure}[!htbp]
\centering\includegraphics[width=3.8in, angle=-90]{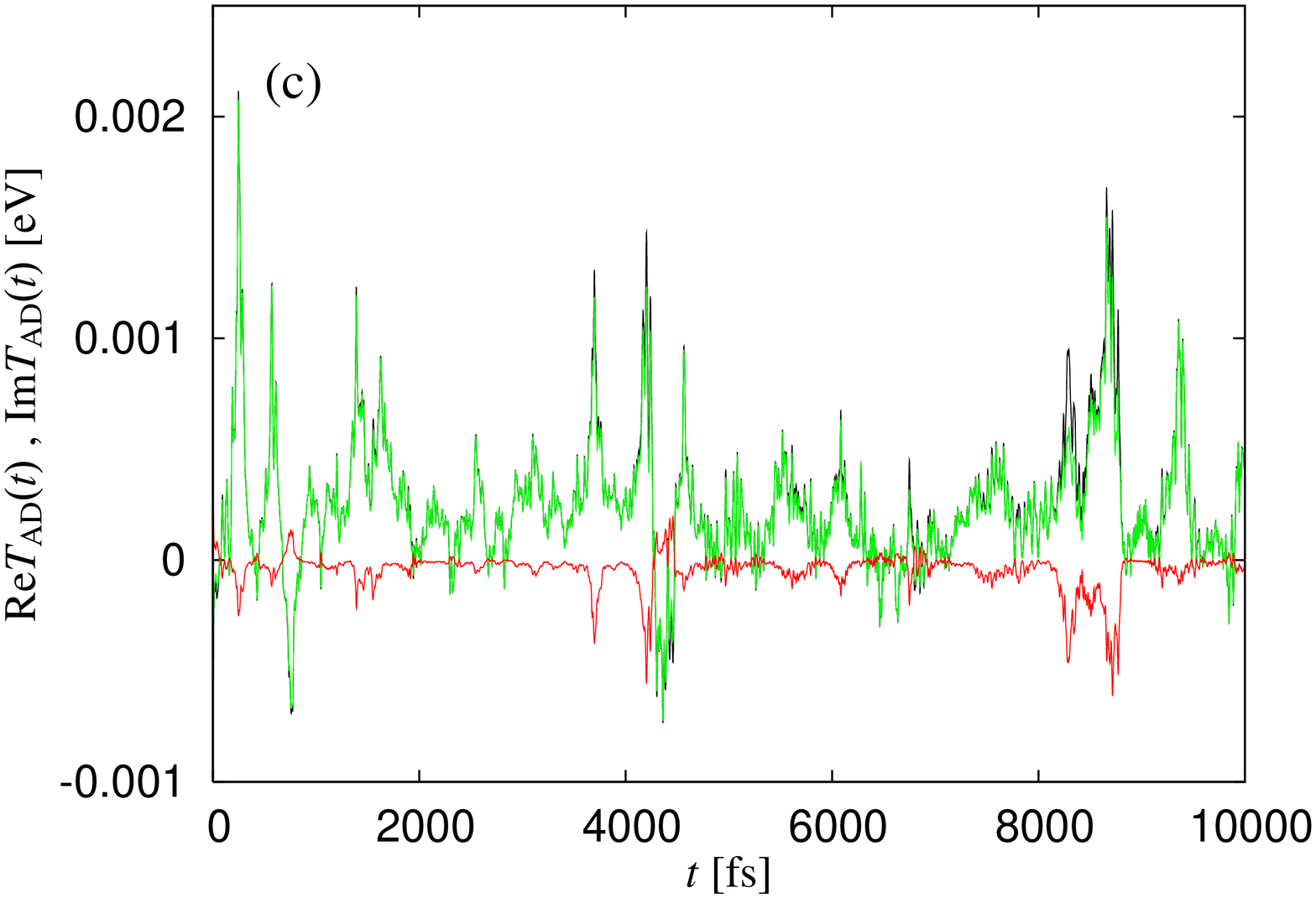}\\
\centering\includegraphics[width=3.8in, angle=-90]{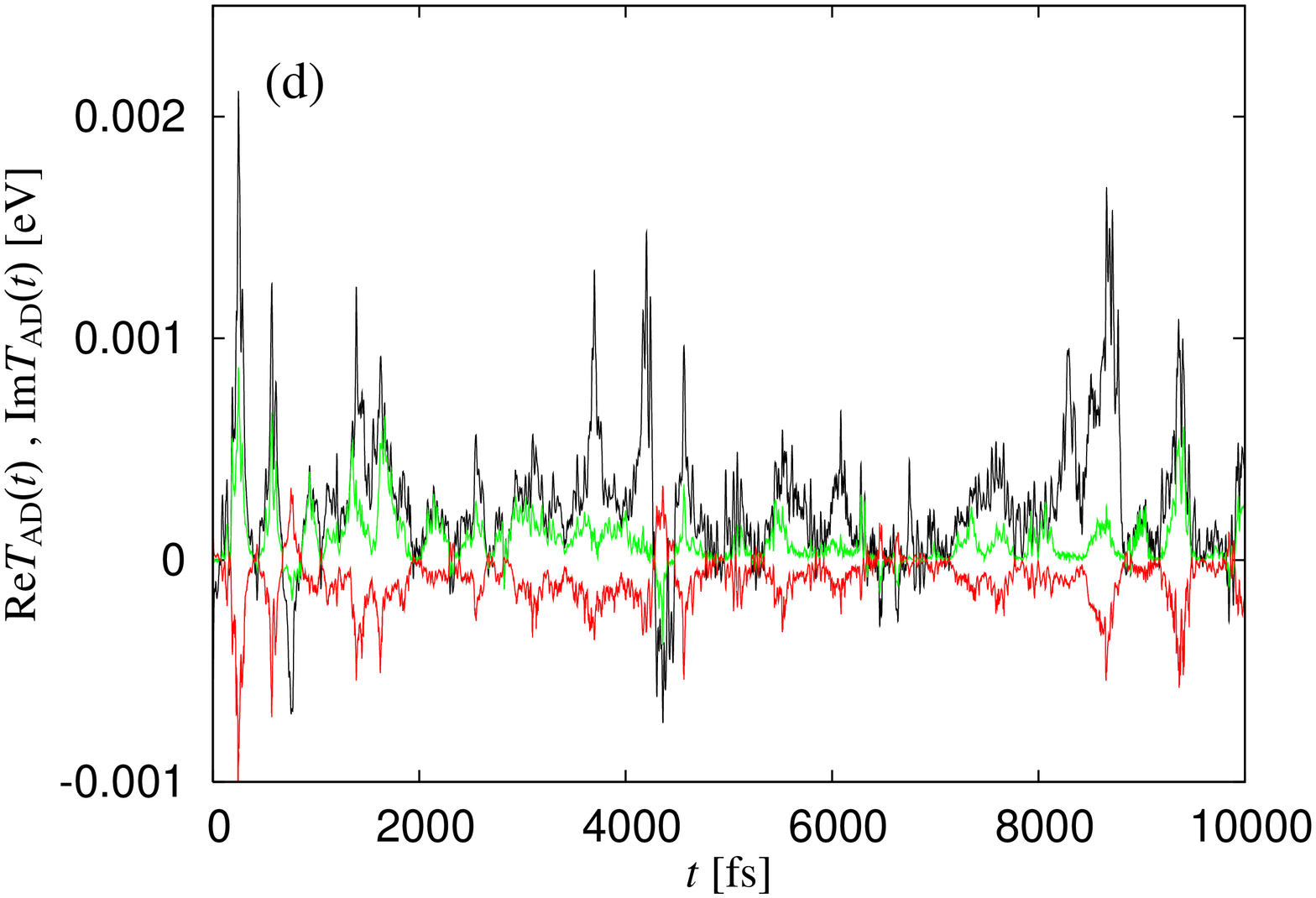}
\caption{\label{fig5}
$\mathrm{Re}T_\mathrm{AD}(t)$ (green line),
$\mathrm{Im}T_\mathrm{AD}(t)$ (red line) with level broadening, and
$T_\mathrm{AD}^\mathrm{pert}(t)$ (black line) without level broadening
for the CR from the LUMO of TCNE to the HOMO of naphthalene in PhCN.
The level broadening is chosen as
(a) $\Gamma = 1$ eV, (b) $\Gamma = 0.1$ eV, (c) $\Gamma = 0.01$ eV,
(d) $\Gamma = 0.001$ eV.
}
\end{figure}

\newpage

\begin{figure}[!htbp]
\centering\includegraphics[width=3.8in, angle=-90]{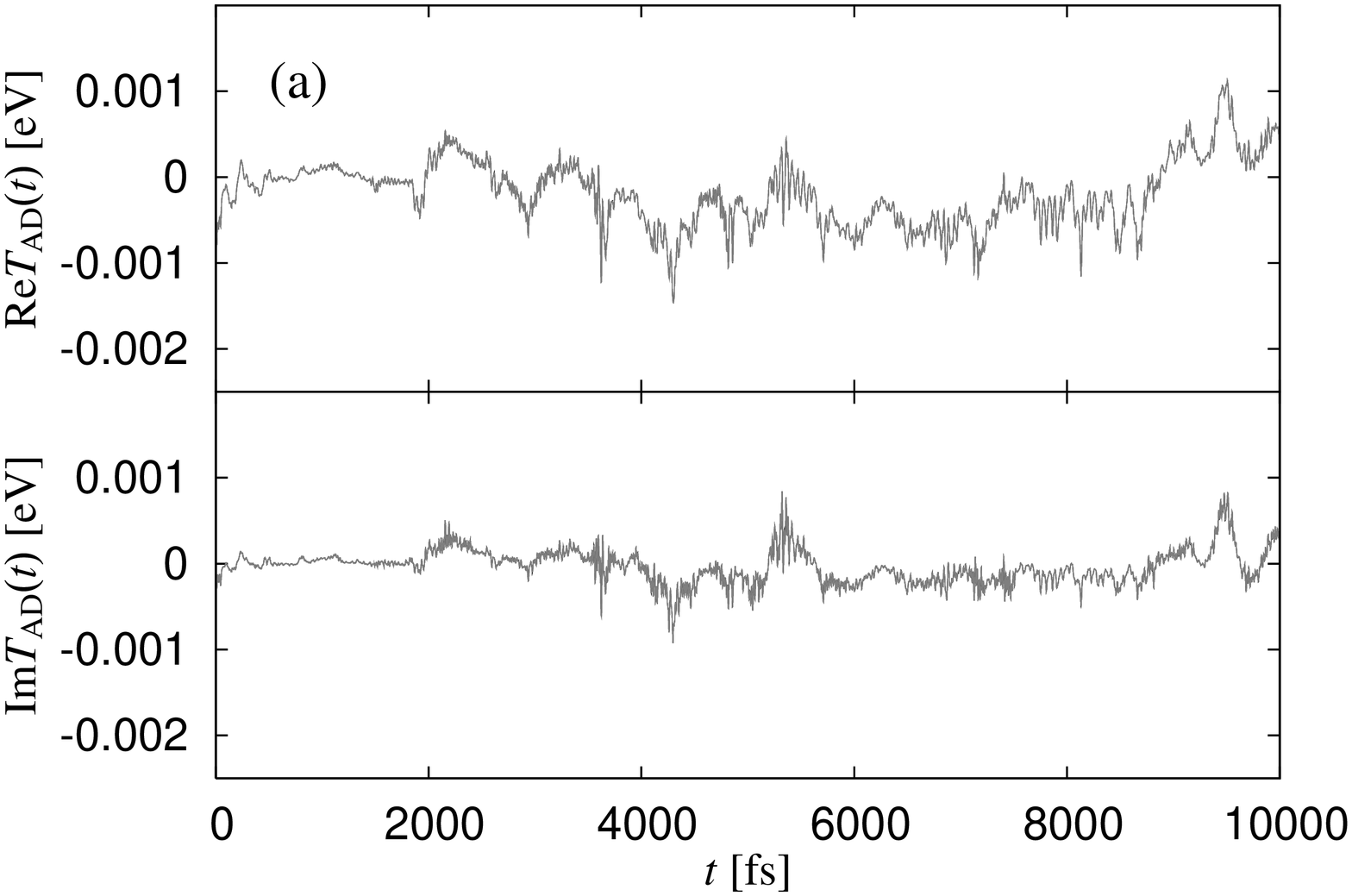}\\
\centering\includegraphics[width=3.8in, angle=-90]{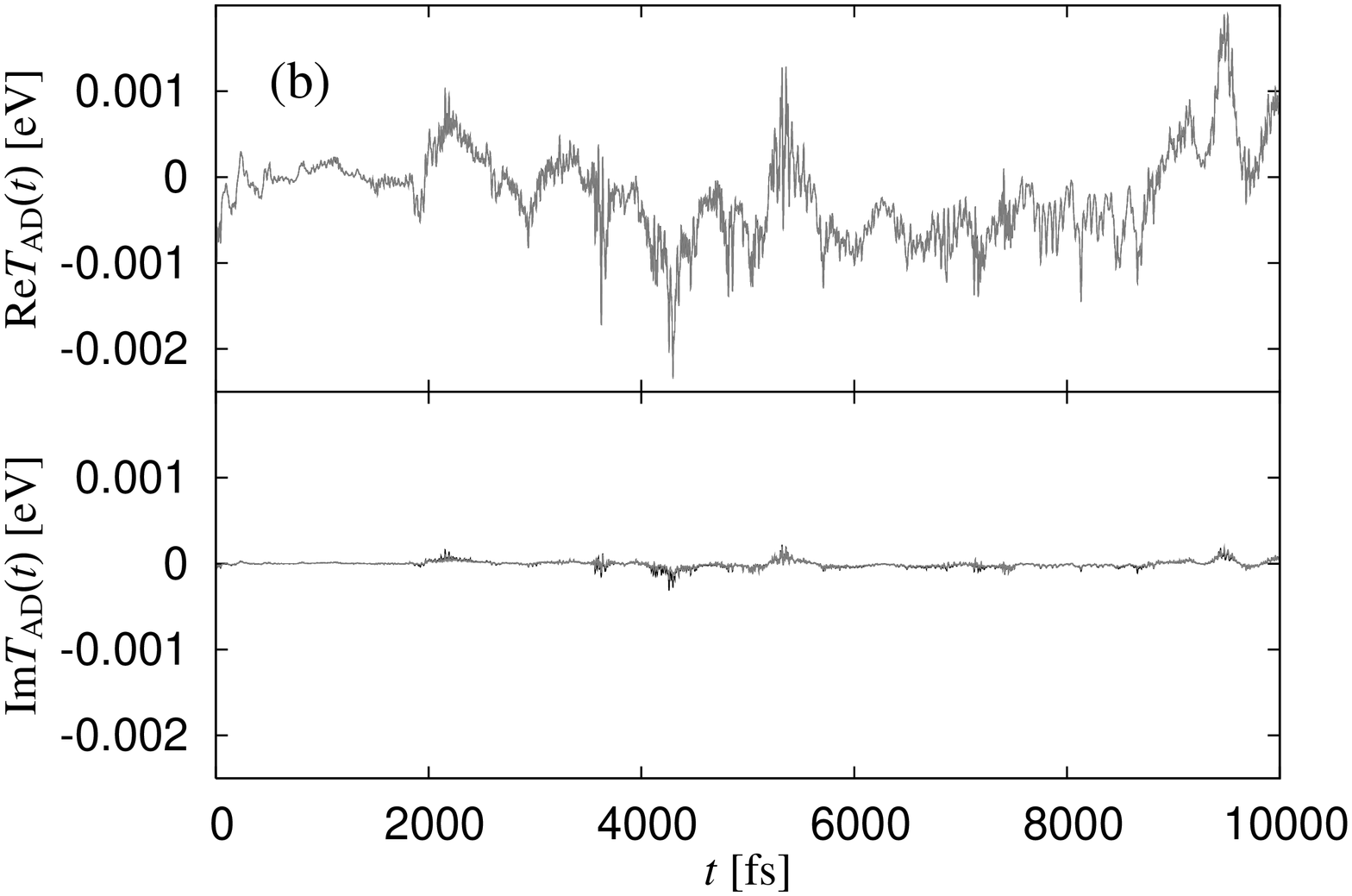}
\end{figure}

\newpage

\begin{figure}[!htbp]
\centering\includegraphics[width=3.8in, angle=-90]{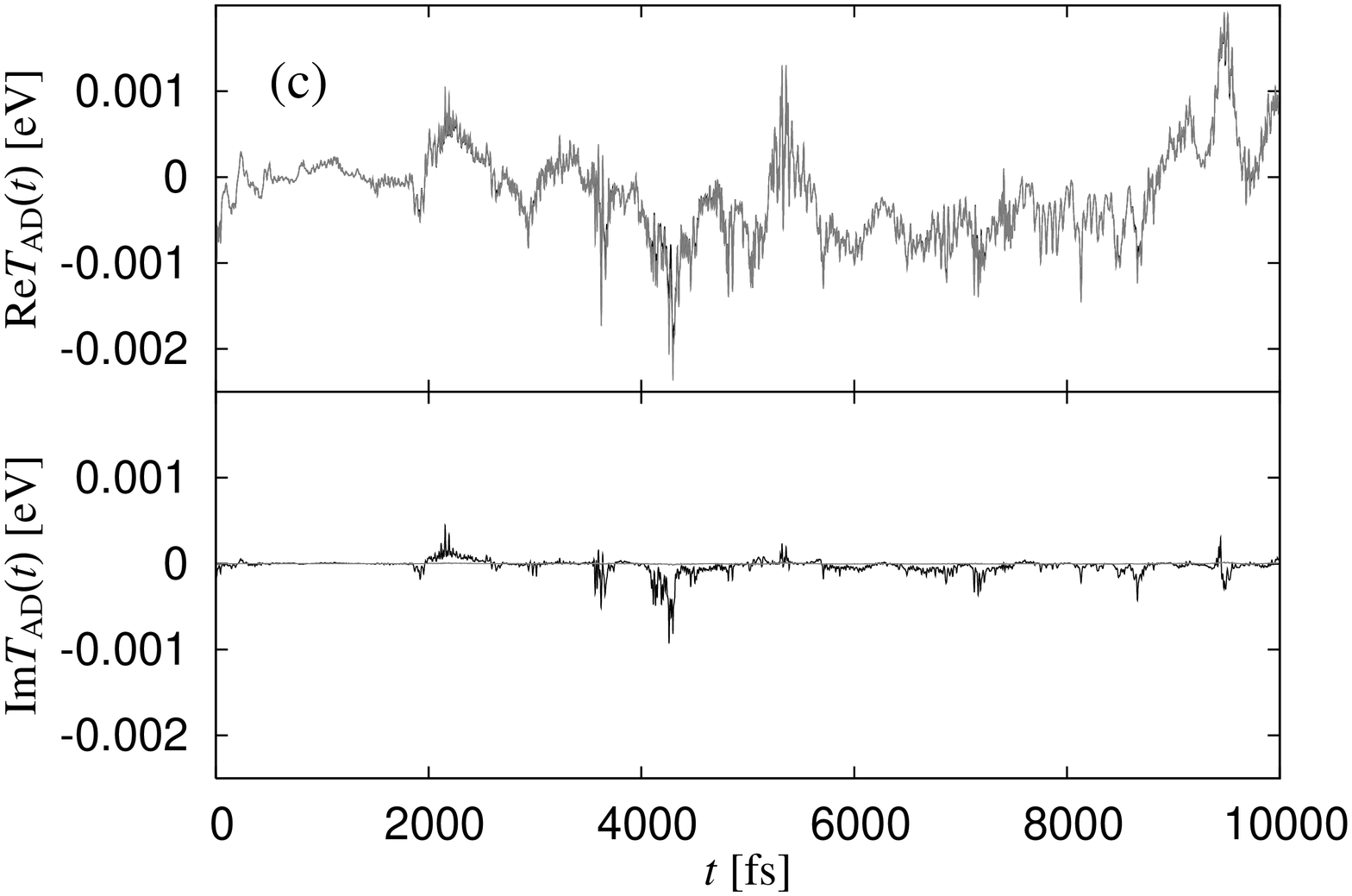}\\
\centering\includegraphics[width=3.8in, angle=-90]{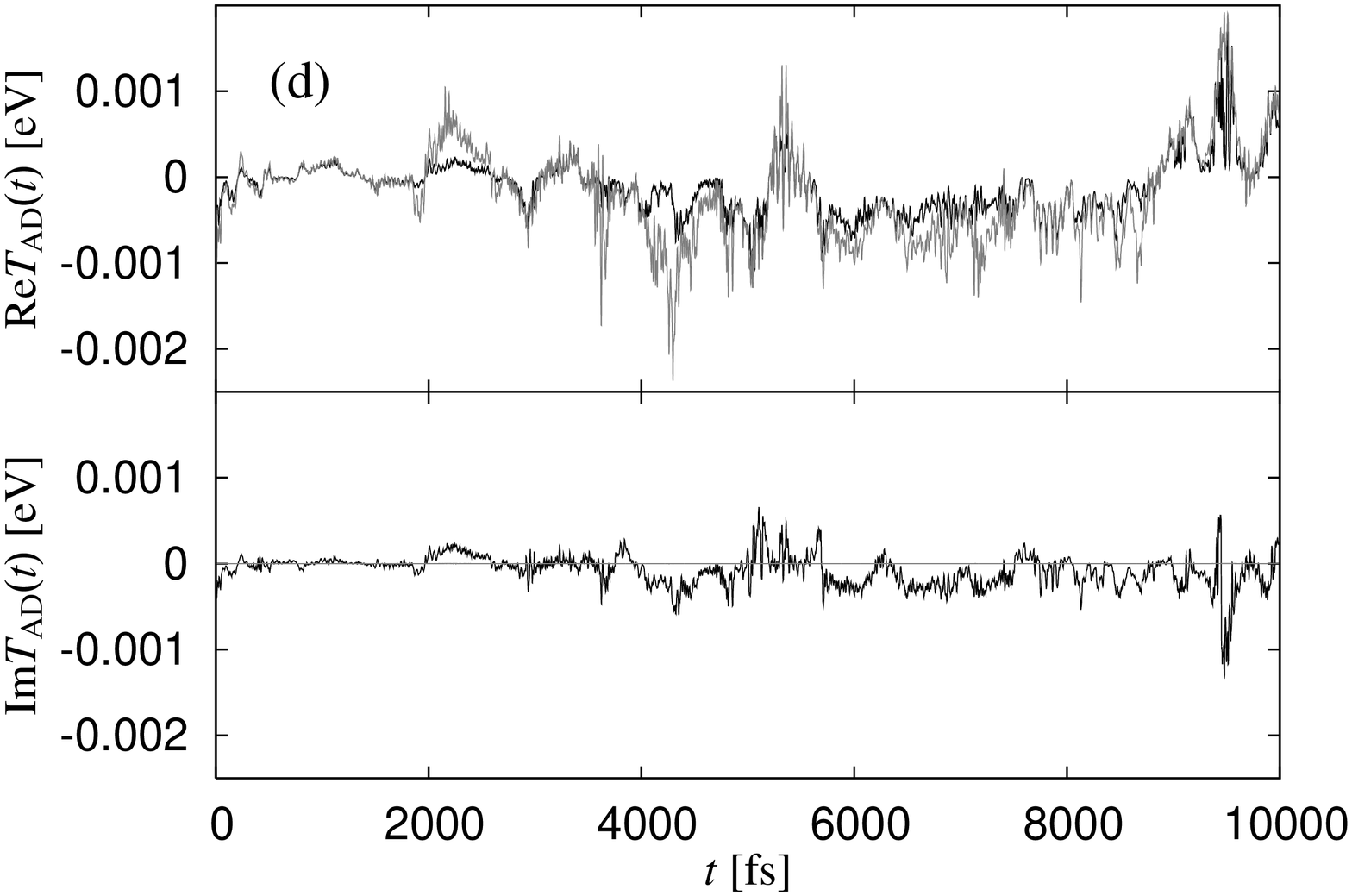}
\caption{\label{fig6}
$\mathrm{Re}T_\mathrm{AD}(t)$ and $\mathrm{Im}T_\mathrm{AD}(t)$ with
level broadening for the CS from the LUMO of anthracene to the LUMO of
TCNE in PhCN.
The gray line is the second-order perturbation result, and the black
line is the full $T$-matrix result.
The level broadening is chosen as
(a) $\Gamma = 1$ eV, (b) $\Gamma = 0.1$ eV, (c) $\Gamma = 0.01$ eV,
(d) $\Gamma = 0.001$ eV.
}
\end{figure}

\newpage

\begin{figure}[!htbp]
\centering\includegraphics[width=3.8in, angle=-90]{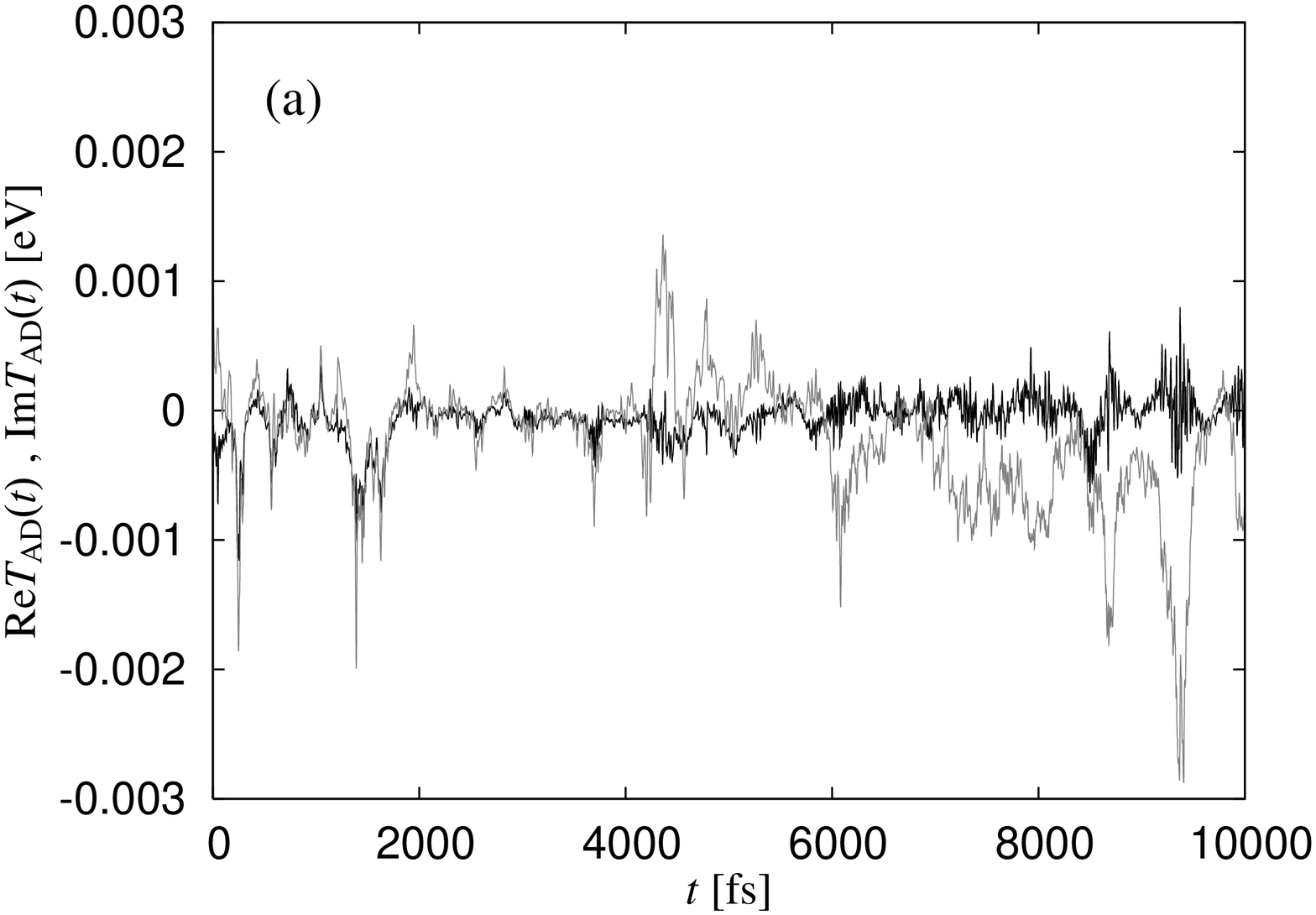}\\
\centering\includegraphics[width=3.8in, angle=-90]{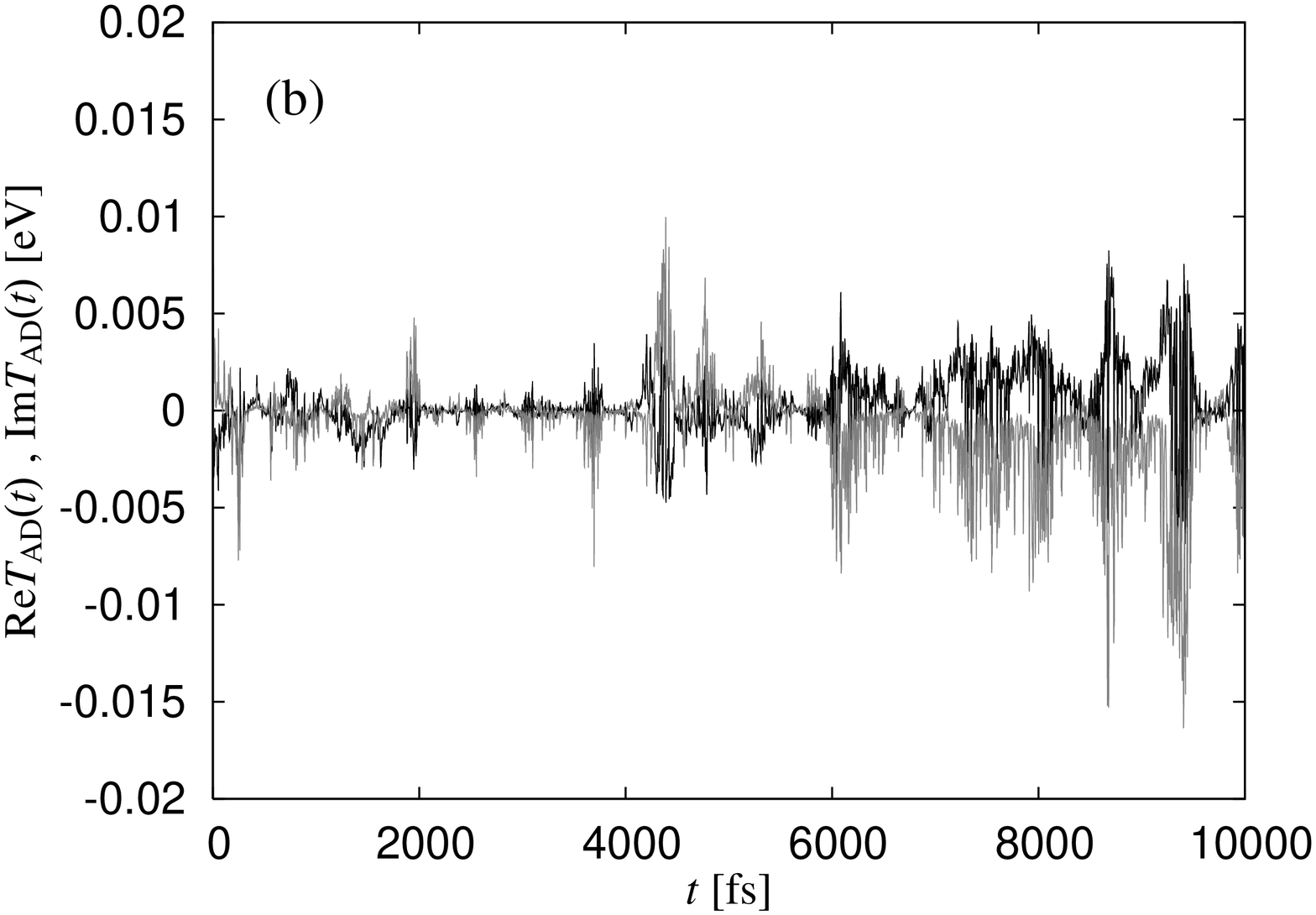}
\end{figure}

\begin{figure}[!htbp]
\centering\includegraphics[width=3.8in, angle=-90]{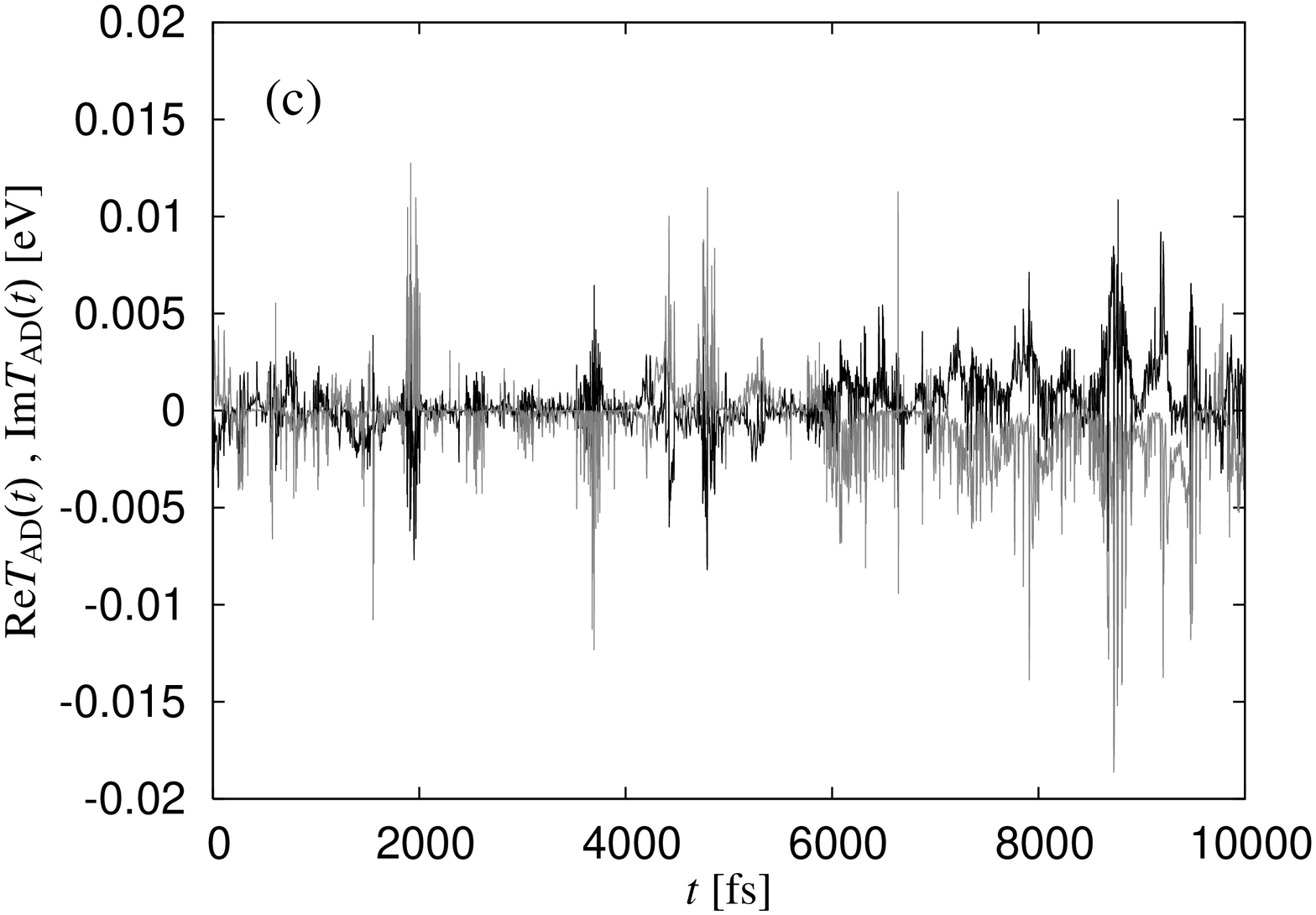}\\
\centering\includegraphics[width=3.8in, angle=-90]{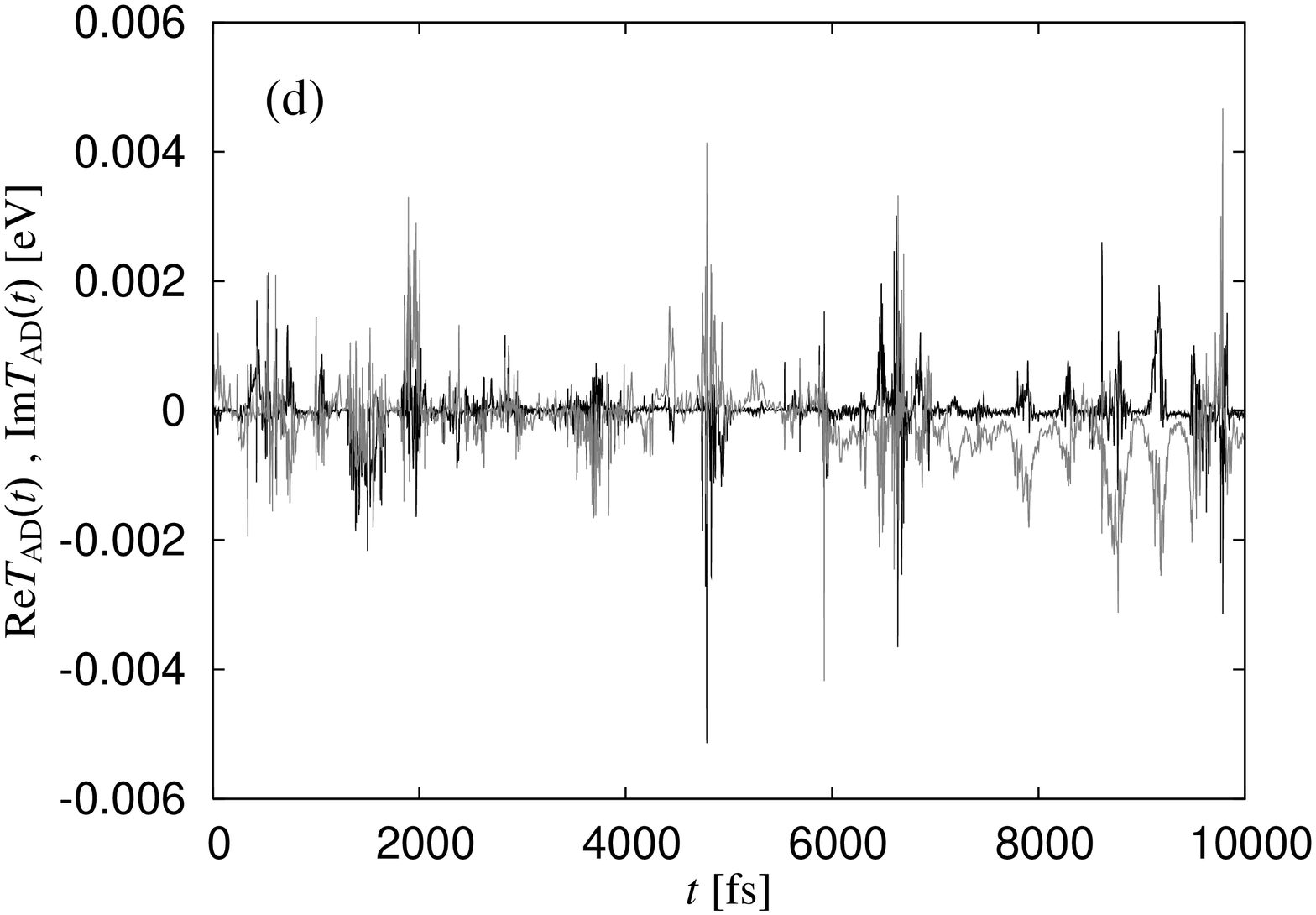}
\caption{\label{fig7}
$\mathrm{Re}T_\mathrm{AD}(t)$ (black line) and
$\mathrm{Im}T_\mathrm{AD}(t)$ (gray line) with level broadening for the
CS from the LUMO of naphthalene to the LUMO of TCNE in PhCN.
The level broadening is chosen as
(a) $\Gamma = 1$ eV, (b) $\Gamma = 0.1$ eV, (c) $\Gamma = 0.01$ eV,
(d) $\Gamma = 0.001$ eV.
}
\end{figure}

\newpage

\begin{figure}[!htbp]
\centering\includegraphics[width=3.8in, angle=-90]{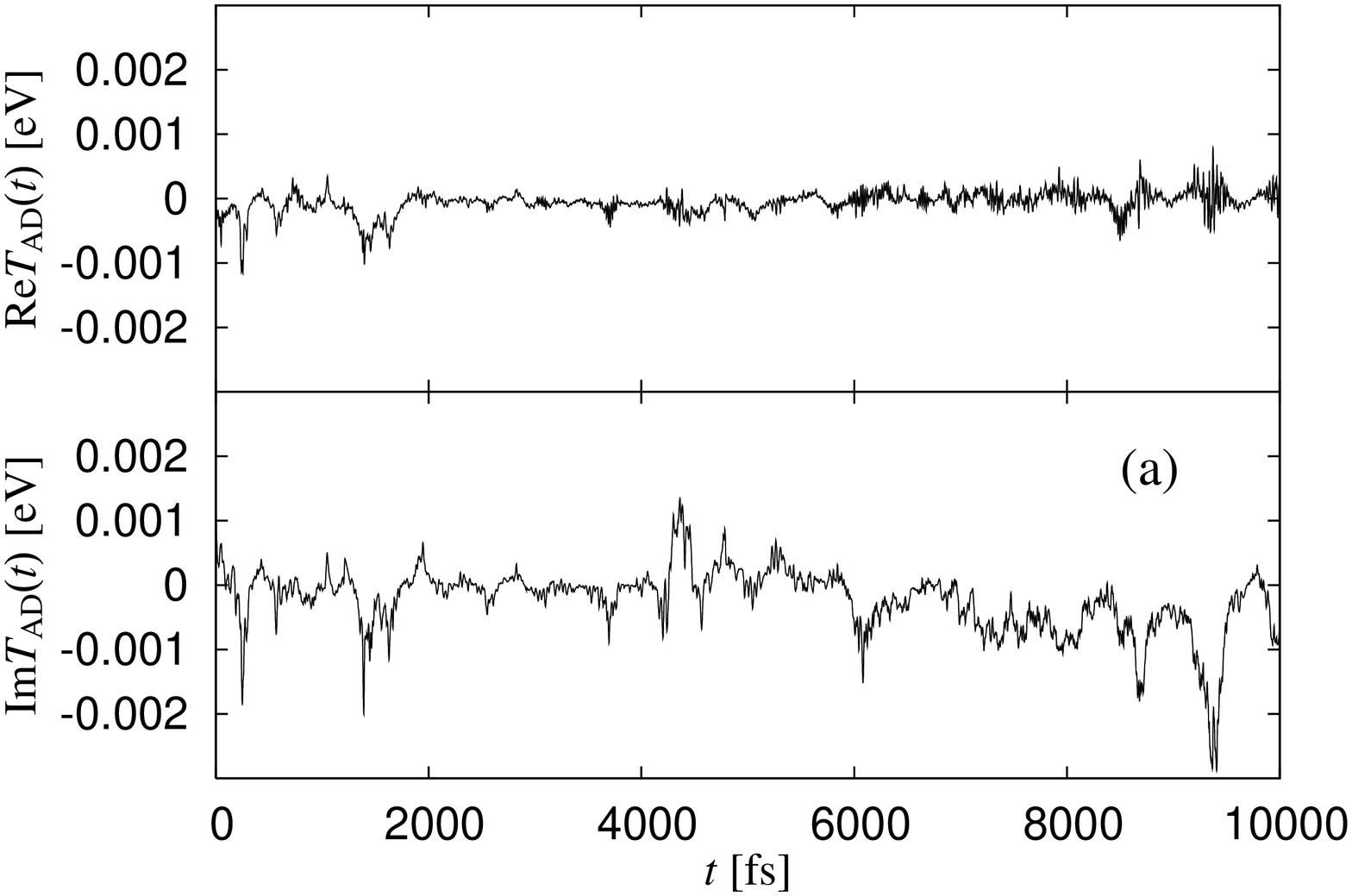}\\
\centering\includegraphics[width=3.8in, angle=-90]{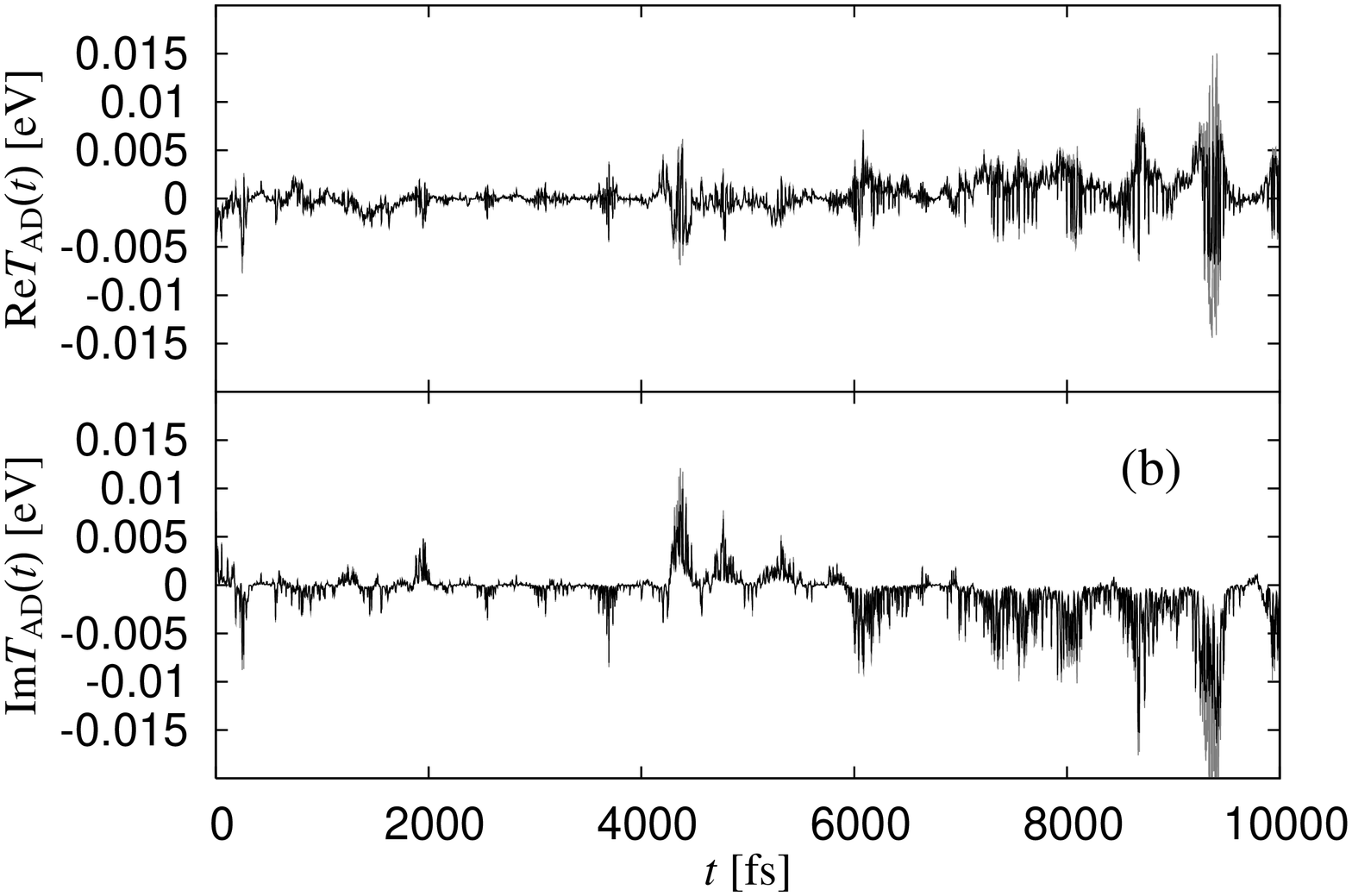}
\end{figure}

\newpage

\begin{figure}[!htbp]
\centering\includegraphics[width=3.8in, angle=-90]{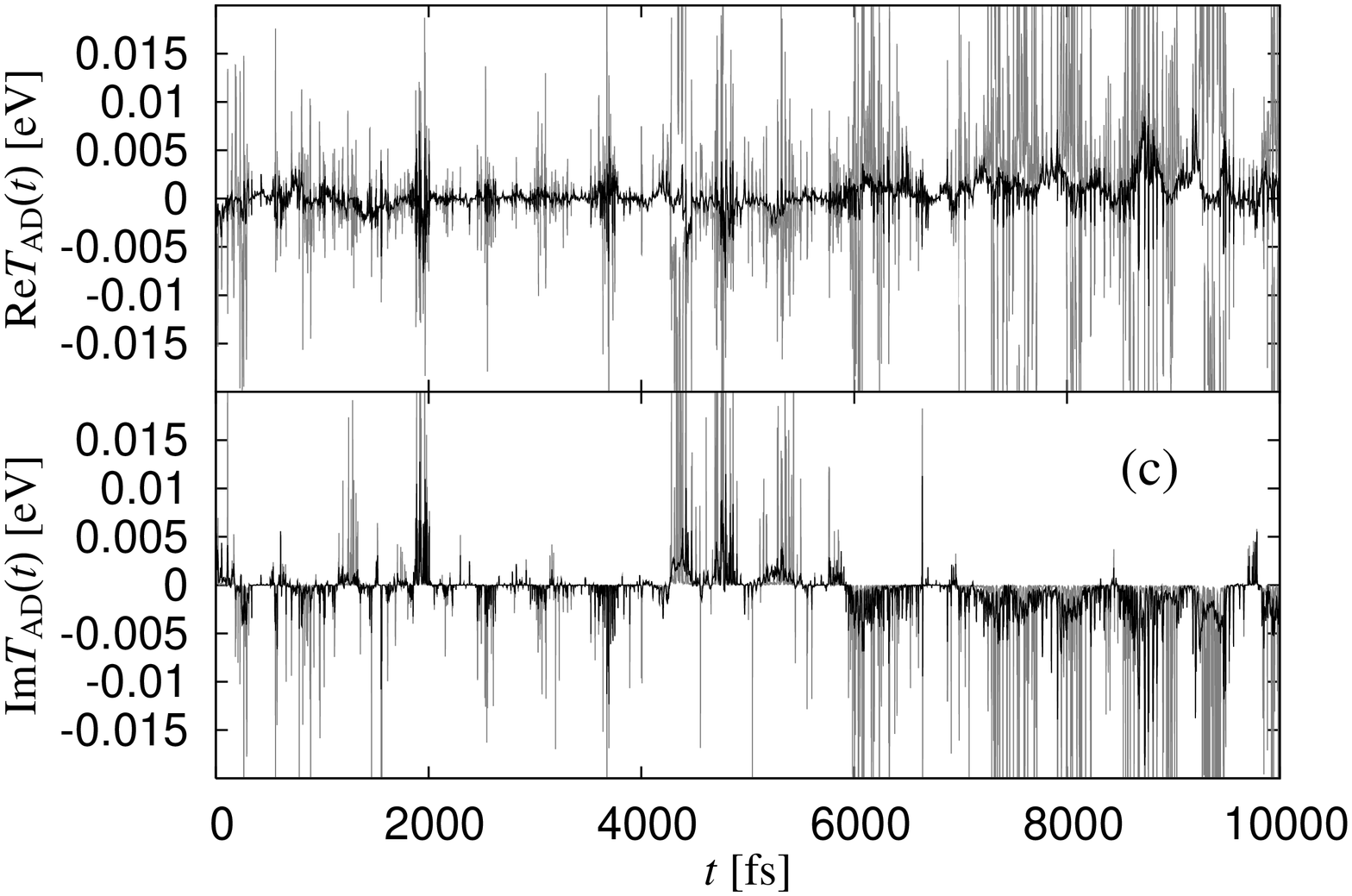}\\
\centering\includegraphics[width=3.8in, angle=-90]{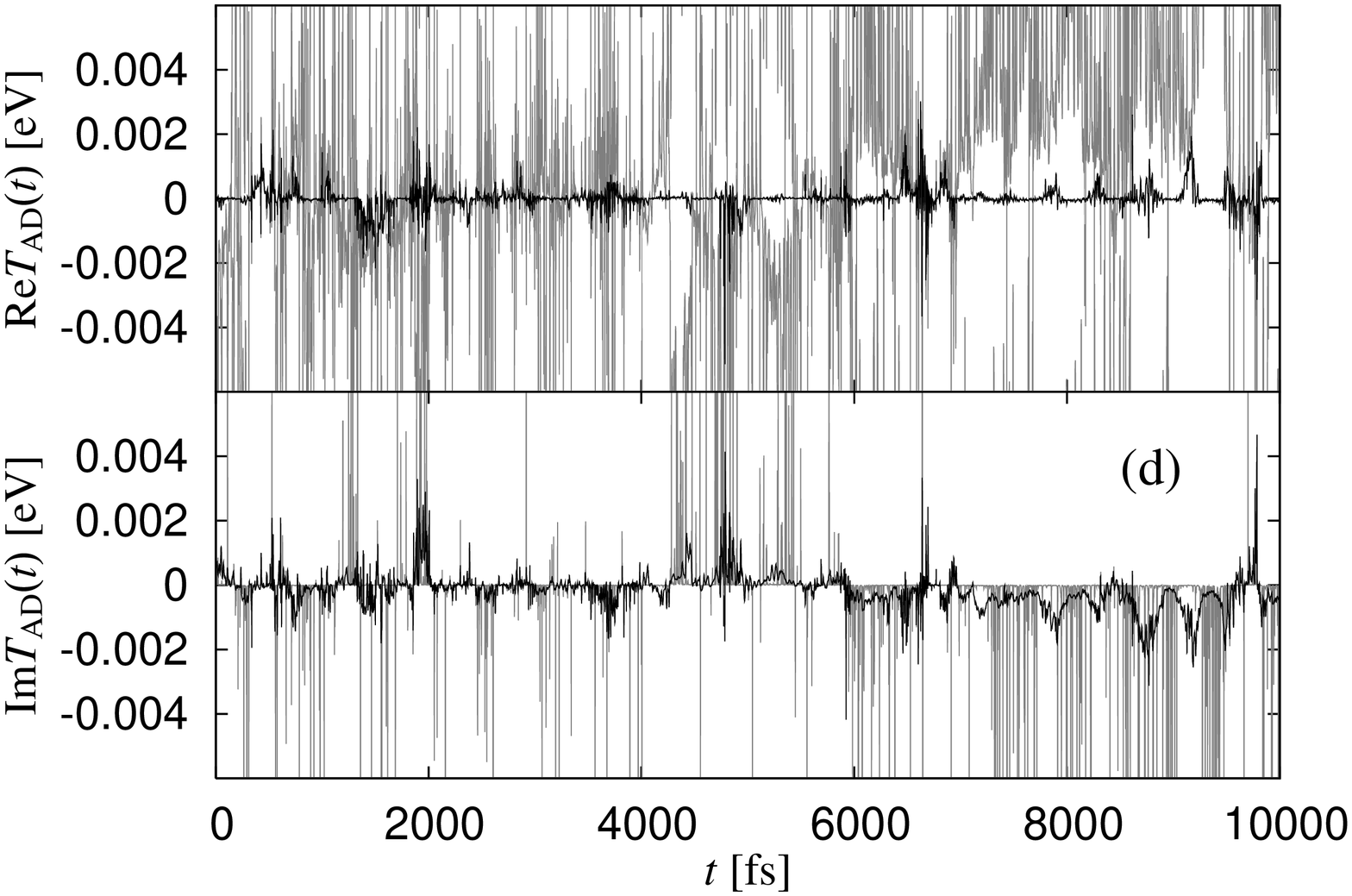}
\caption{\label{fig8}
$\mathrm{Re}T_\mathrm{AD}(t)$ and $\mathrm{Im}T_\mathrm{AD}(t)$ with
level broadening for the CS from the LUMO of naphthalene to the LUMO of
TCNE in PhCN.
The gray line is the second-order perturbation result, and the black
line is the full $T$-matrix result.
The level broadening is chosen as
(a) $\Gamma = 1$ eV, (b) $\Gamma = 0.1$ eV, (c) $\Gamma = 0.01$ eV,
(d) $\Gamma = 0.001$ eV.
}
\end{figure}

\newpage

\begin{figure}[!htbp]
\centering\includegraphics[width=3.8in, angle=-90]{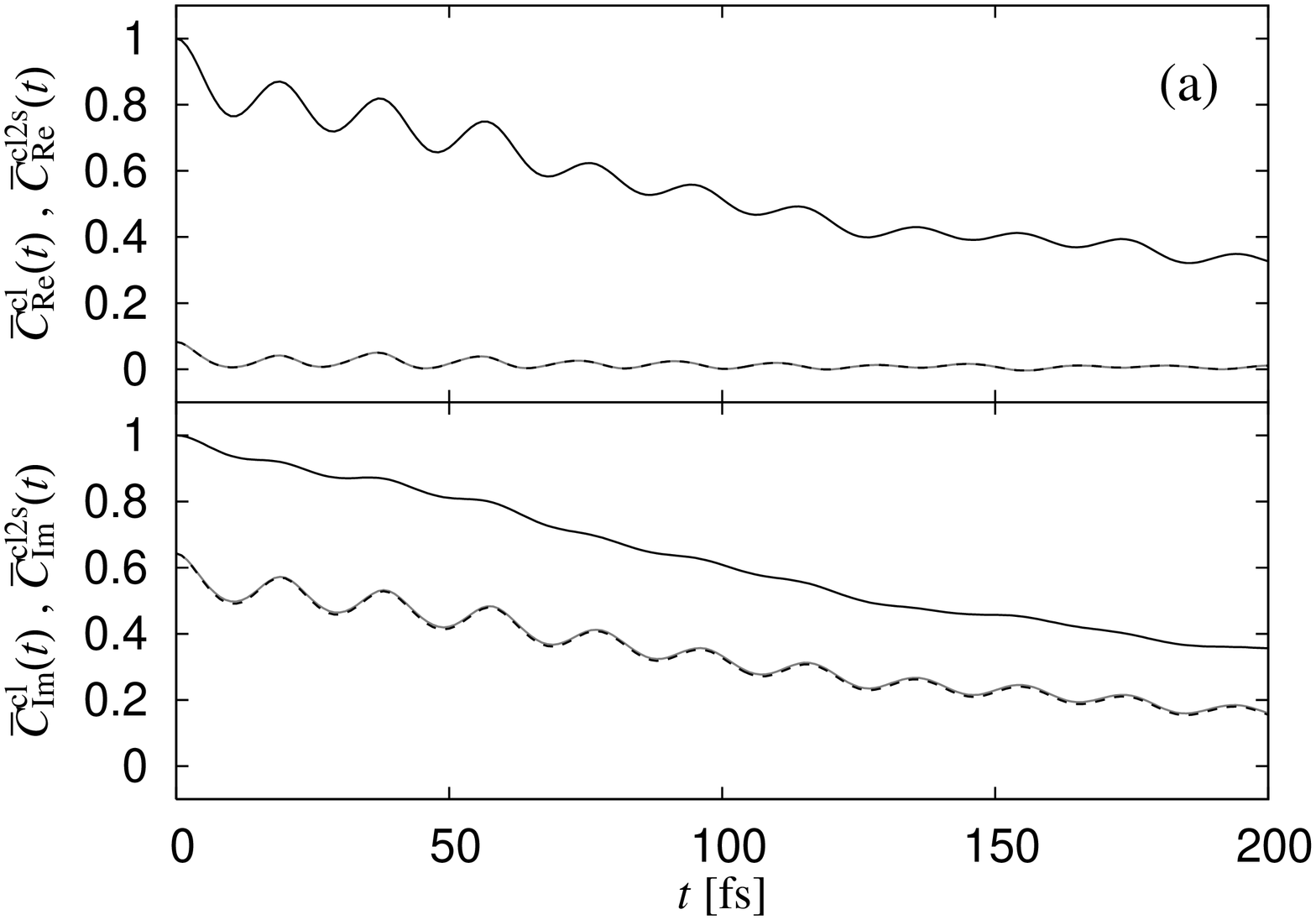}\\
\centering\includegraphics[width=3.8in, angle=-90]{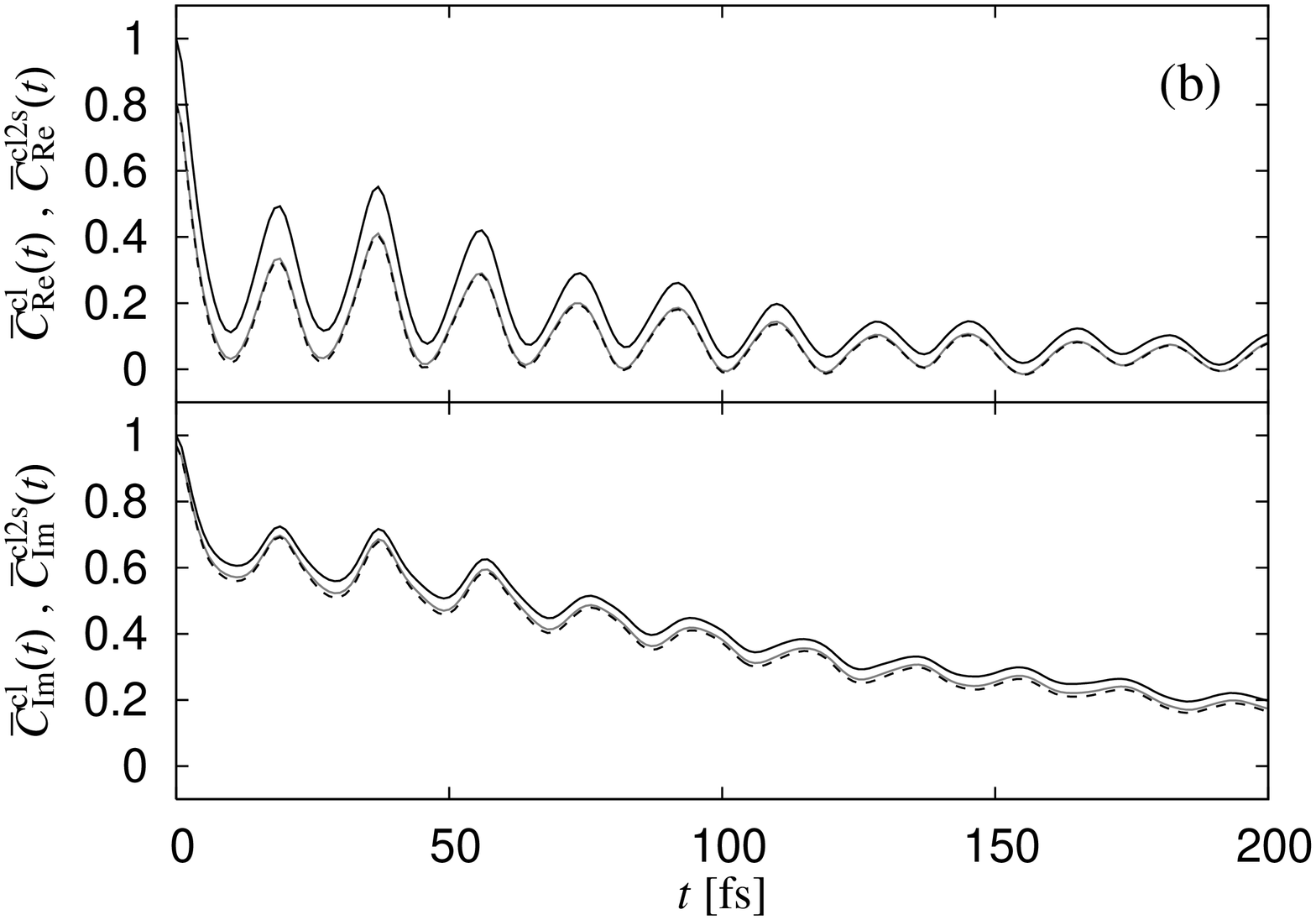}
\end{figure}

\newpage

\begin{figure}[!htbp]
\centering\includegraphics[width=3.8in, angle=-90]{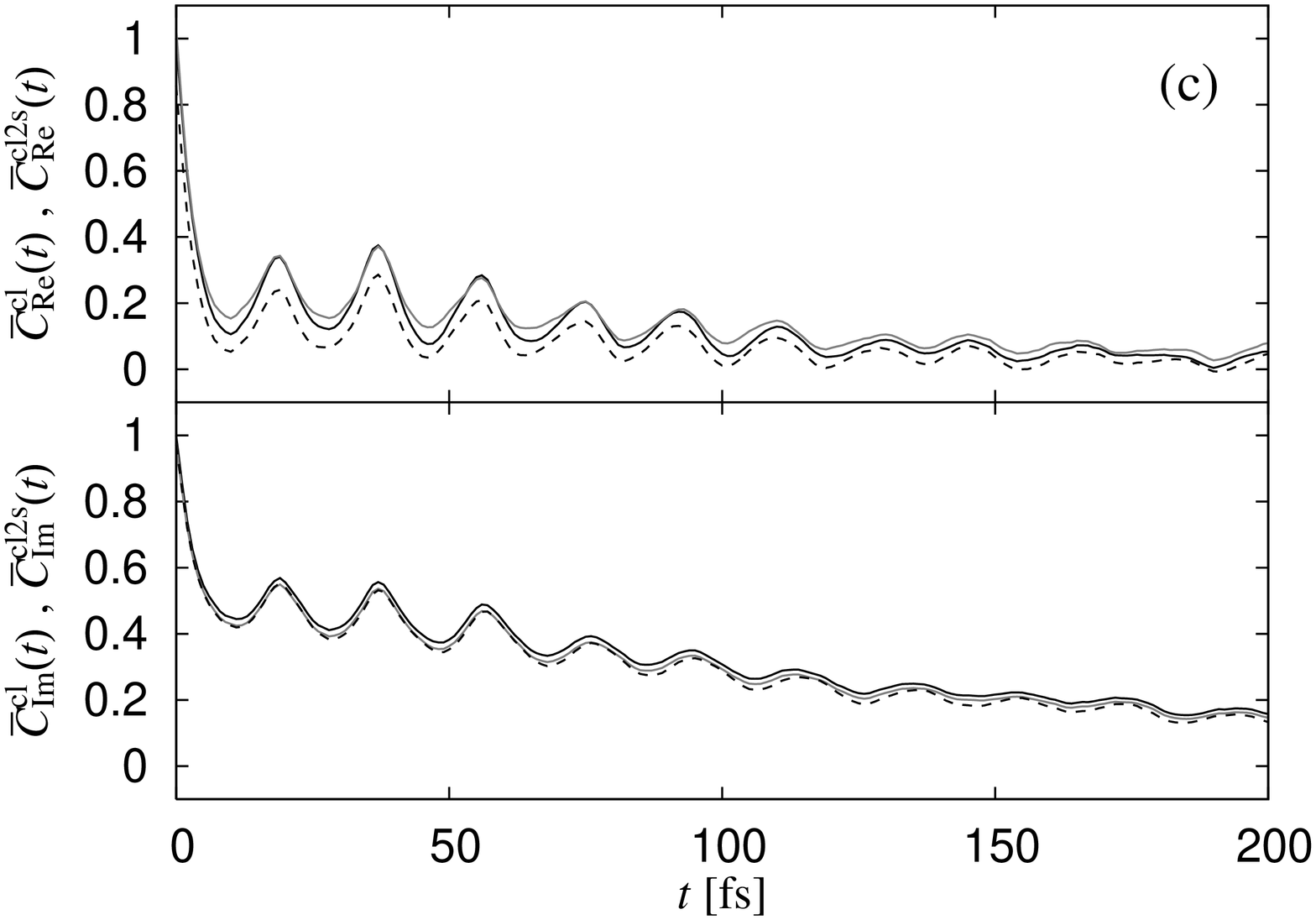}\\
\centering\includegraphics[width=3.8in, angle=-90]{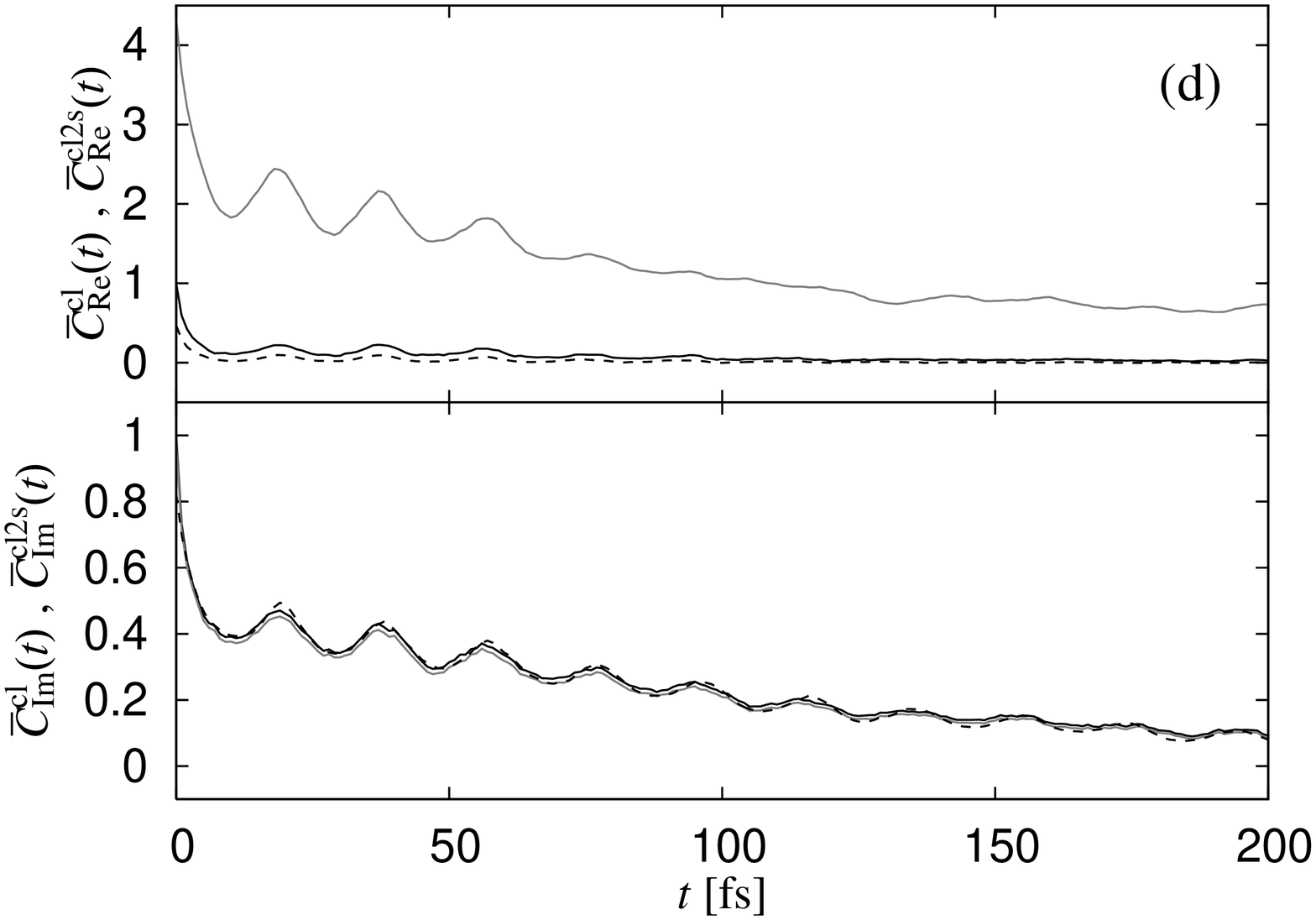}
\caption{\label{fig9}
$\bar{C}_\mathrm{Re}^\mathrm{cl}(t)$ (black solid line),
$\bar{C}_\mathrm{Re}^\mathrm{cl2s}(t)$ (gray solid line),
$\bar{C}_\mathrm{Im}^\mathrm{cl}(t)$ (black solid line), and
$\bar{C}_\mathrm{Im}^\mathrm{cl2s}(t)$ (gray solid line) for the CS from
the LUMO of naphthalene to the LUMO of TCNE in PhCN.
The black dashed lines are the approximation results for
$\bar{C}_\mathrm{Re}^\mathrm{cl2s}(t)$ and
$\bar{C}_\mathrm{Im}^\mathrm{cl2s}(t)$ using Eqs.~(\ref{ReTAD2s}) and
(\ref{ImTAD2s}).
The level broadening is chosen as
(a) $\Gamma = 1$ eV, (b) $\Gamma = 0.1$ eV, (c) $\Gamma = 0.01$ eV,
(d) $\Gamma = 0.001$ eV.
}
\end{figure}

\newpage

\begin{figure}[!htbp]
\centering\includegraphics[width=3.8in, angle=-90]{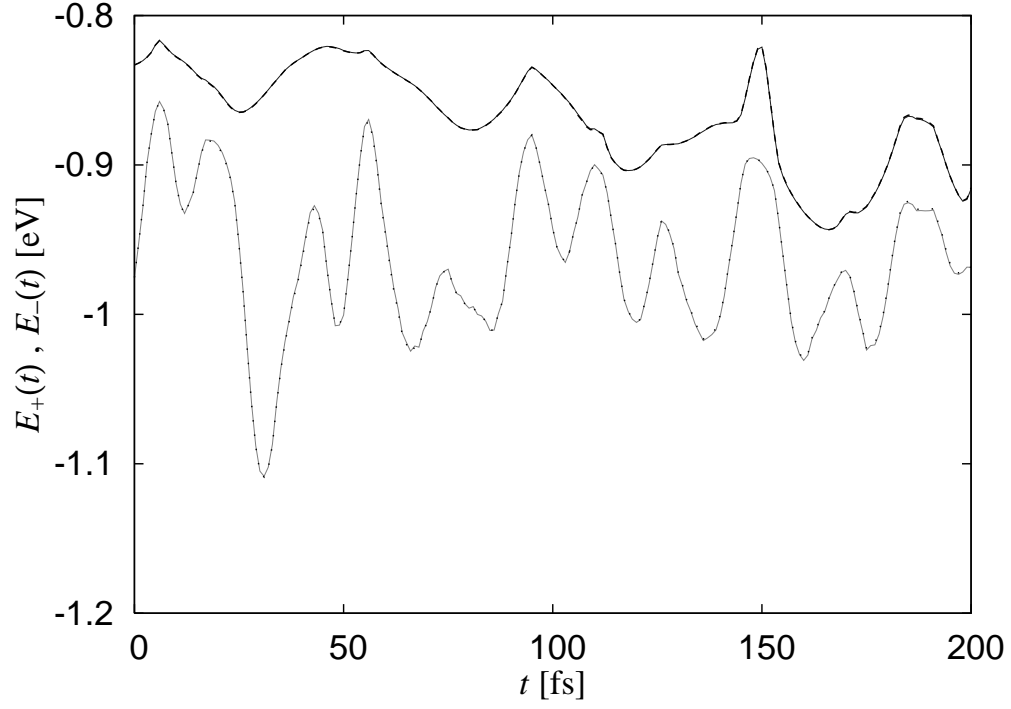}
\caption{\label{fig10}
$\tilde{E}_+(t)$ (dark gray solid line) in comparison with $E_+(t)$
(black dashed line), and $\tilde{E}_-(t)$ (gray solid line) in
comparison with $E_-(t)$ (black dotted line).
}
\end{figure}

\begin{figure}[!htbp]
\centering\includegraphics[width=3.8in, angle=-90]{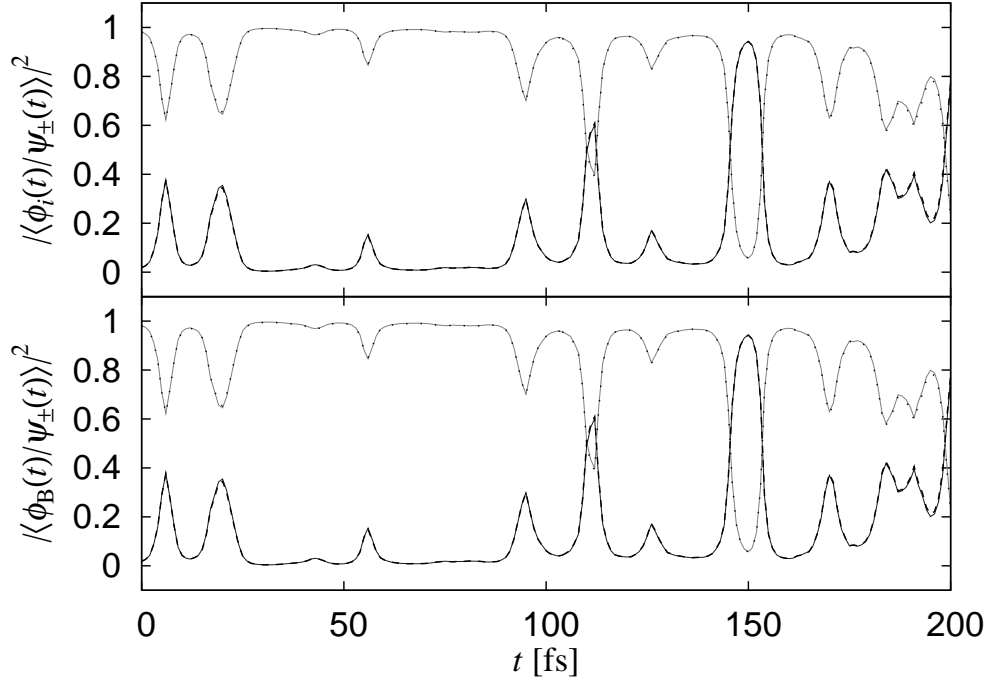}
\caption{\label{fig11}
Time dependence of $\sin^2\theta$ (dark gray solid line) in comparison
with $|\langle \phi_i | \psi_- \rangle|^2$ (black dashed line) and
$|\langle \phi_\mathrm{B} | \psi_+ \rangle|^2$ (black dashed line), and
$\cos^2\theta$ (gray solid line) in comparison with
$|\langle \phi_i | \psi_+ \rangle|^2$ (black dotted line) and
$|\langle \phi_\mathrm{B} | \psi_- \rangle|^2$ (black dotted line).
}
\end{figure}

\newpage

\begin{figure}[!htbp]
\centering\includegraphics[width=3.8in, angle=-90]{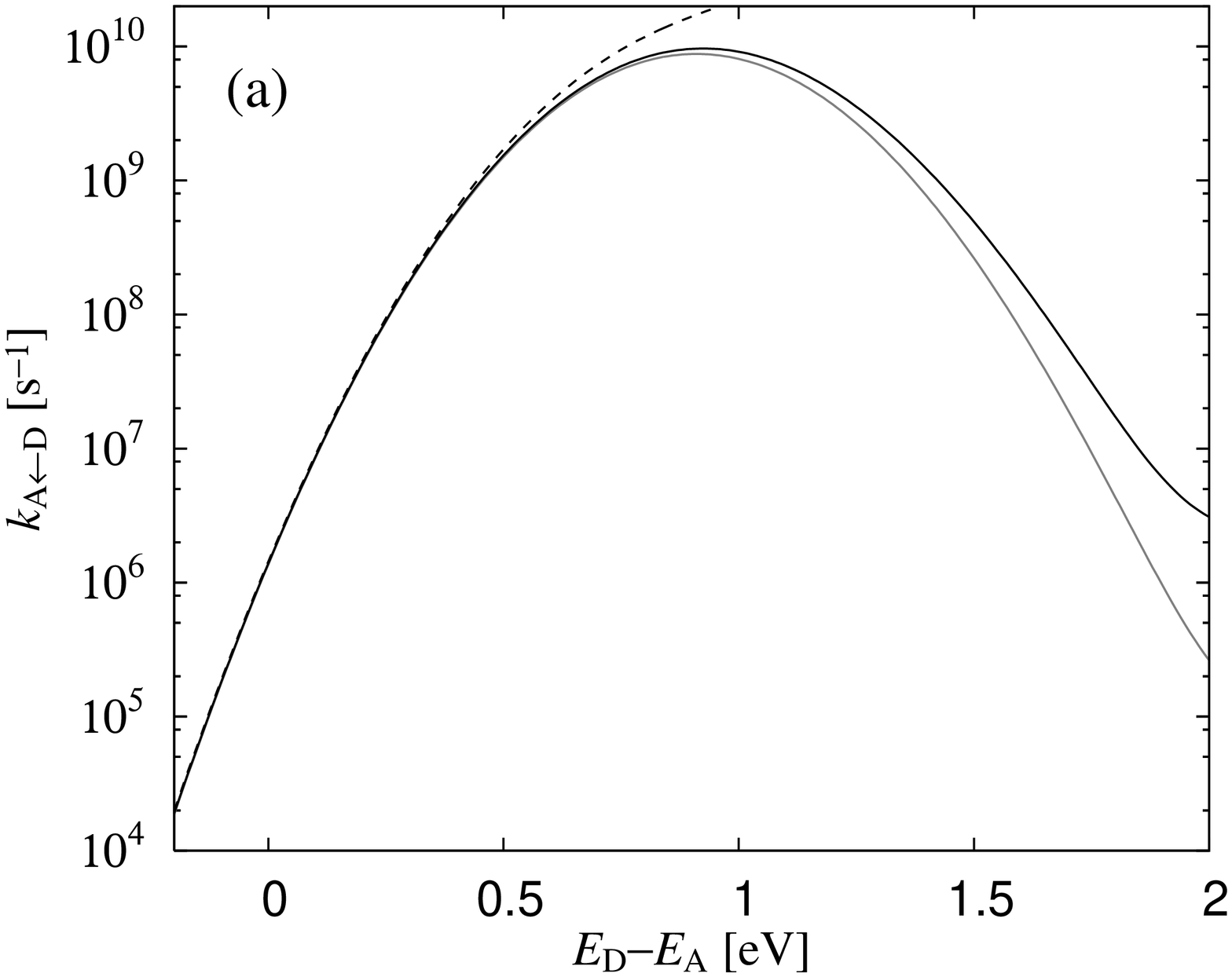}\\
\centering\includegraphics[width=3.8in, angle=-90]{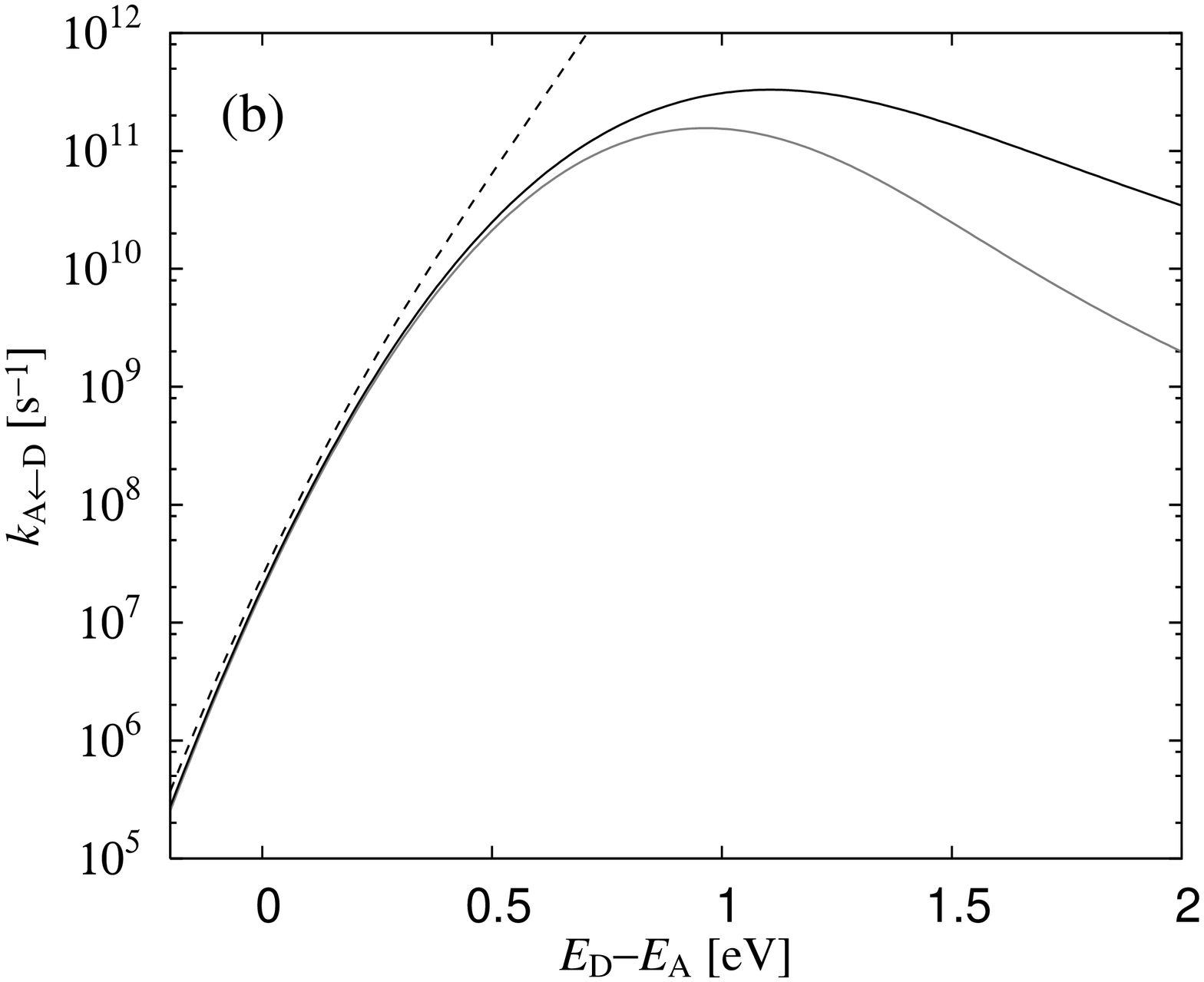}
\end{figure}

\newpage

\begin{figure}[!htbp]
\centering\includegraphics[width=3.8in, angle=-90]{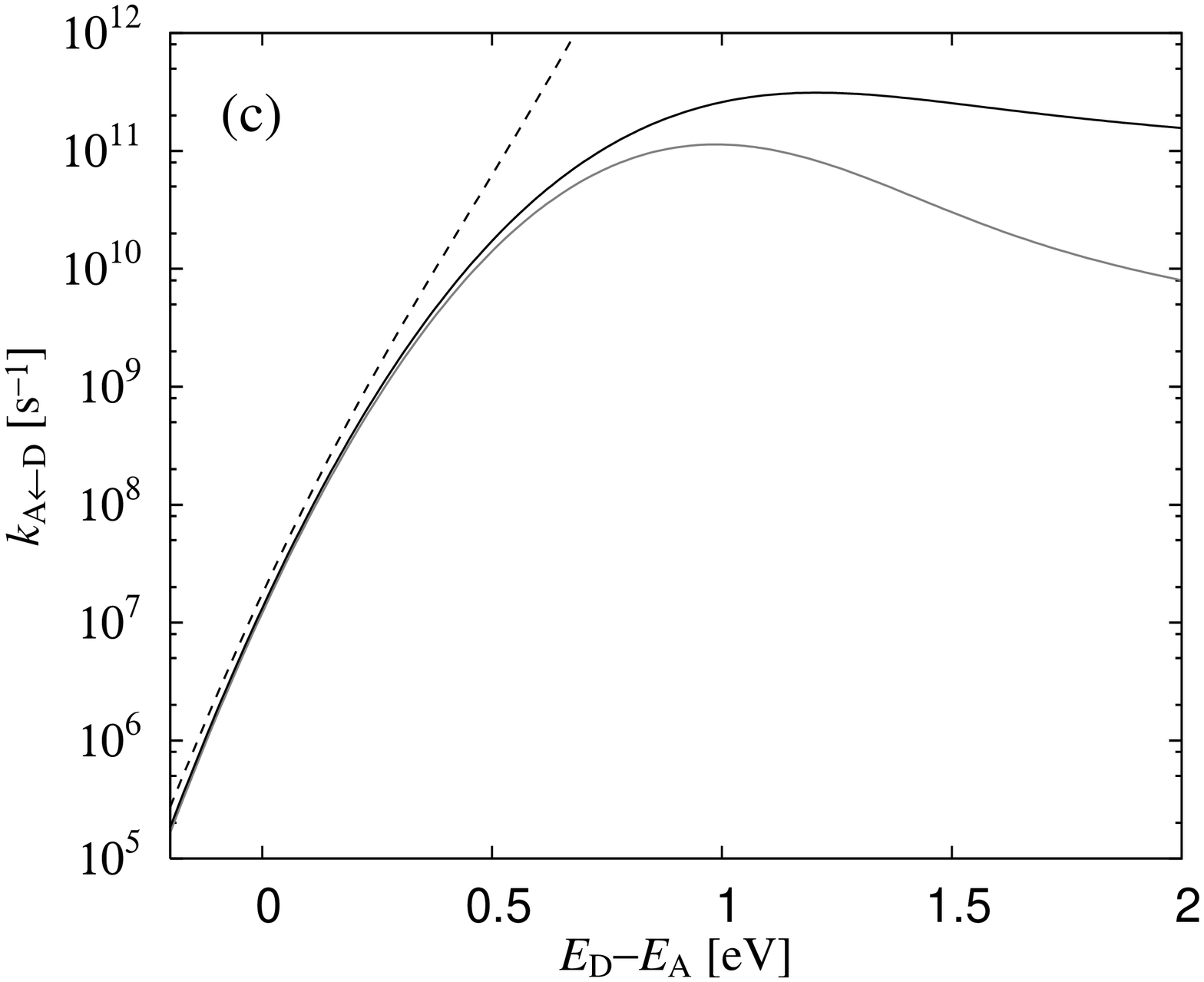}\\
\centering\includegraphics[width=3.8in, angle=-90]{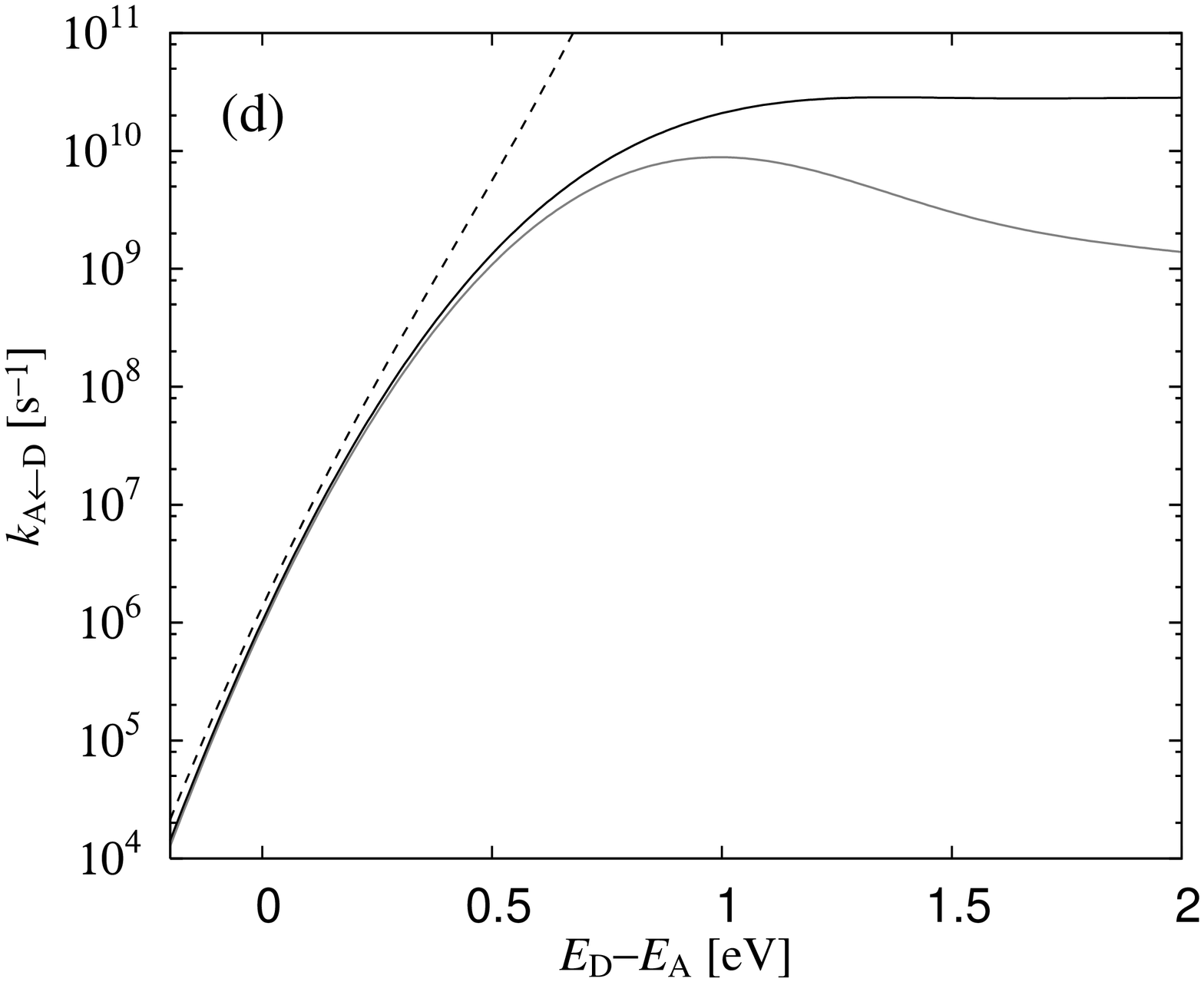}
\caption{\label{fig12}
Energy gap dependence of the ET rate.
The gray solid, black solid, and black dashed lines are the results for
the quantum correction factors of Eqs.~(\ref{Qst}), (\ref{Qh}), and
(\ref{Qsc}), respectively.
The level broadening is chosen as
(a) $\Gamma = 1$ eV, (b) $\Gamma = 0.1$ eV, (c) $\Gamma = 0.01$ eV,
(d) $\Gamma = 0.001$ eV.
}
\end{figure}

\end{document}